\documentclass[onecolumn,tighten]{aastex63}
\hypersetup{linkcolor=red,citecolor=blue,filecolor=cyan,urlcolor=magenta}
\usepackage{amsmath}
\usepackage{natbib}
\usepackage{color}

\newcommand{\msun}{\mbox{$M_{\odot}$}}

\newcommand{\aco}{$a_{Co}$}
\newcommand{\vecaco}{${\mathbf{a_{Co}}}$}
\newcommand{\vecacf}{${\mathbf{a_{Cf}}}$}

\newcommand{\vecag}{$\mathbf{a_{\mathrm g}}$}
\newcommand{\gaia}{{\it Gaia}}
\newcommand{\OT}{{\em OT\,}}
\newcommand{\Fm}{{\em F19m\,}}
\newcommand{\LM}{{\em LM10\,}}
\newcommand{\LMm}{{\em LMm\,}}
\newcommand{\tev}{$t_{\rm ev}$}

\newcommand{\kms}{\ensuremath{\,\mathrm{km\ s}^{-1}}}
\newcommand{\kmskpc}{\ensuremath{\,\mathrm{{km\ s}^{-1}\ {kpc}^{-1}}}}

\defcitealias{valluri_etal_10}{V10}
\defcitealias{valluri_etal_12}{V12}
\defcitealias{deb_etal_08}{D08}
\defcitealias{BT08}{BT}
\defcitealias{bailin_steinmetz_04}{BS04}

\received{16, Sept, 2020}
\revised{8, February, 2021}
\accepted{xxx, 2021}
\date{\today}
\submitjournal{ApJ}

\shorttitle{Figure rotation of the Milky Way Halo}
\shortauthors{Valluri, Price-Whelan \& Snyder}

\begin{document}

\title{%
Detecting the Figure Rotation of Dark Matter Halos with Tidal Streams}

\correspondingauthor{Monica Valluri}
\email{mvalluri@umich.edu}

\author[0000-0002-6257-2341]{Monica Valluri}
\affiliation{Department of Astronomy, University of Michigan,
1085 S.\ University Ave., Ann Arbor, MI 48109, USA}

\author[0000-0003-0872-7098]{Adrian~M.~Price-Whelan}
\affiliation{Center for Computational Astrophysics, Flatiron Institute, 162 Fifth Ave, New York, NY 10010, USA}

\author{Sarah J. Snyder}
\affiliation{Department of Astronomy, University of Michigan,
1085 S.\ University Ave., Ann Arbor, MI 48109, USA}

\begin{abstract}
The dark matter halos that surround Milky Way-like galaxies in cosmological simulations are, to first order, triaxial. Nearly 30 years ago it was  predicted that such triaxial dark matter halos should exhibit steady figure rotation or tumbling motions for durations of several gigayears. The angular frequency of figure rotation predicted by cosmological simulations is described by a log-normal distribution of pattern speed  $\Omega_p$  with a median value  $0.15\,h\,\kmskpc$ ($\sim  0.15\,h\,{\rm rad}\,{\rm Gyr}^{-1}$ $\sim 9^{\circ}\,h\,{\rm Gyr}^{-1}$)  and a width of 0.83$h\kmskpc$. These pattern speeds are so small that they have generally been considered both unimportant and undetectable. In this work we show that even extremely slow figure rotation can significantly alter the structure of extended stellar streams produced by the tidal disruption of satellites in the Milky Way halo. We simulate the behavior of a Sagittarius-like polar tidal stream in triaxial dark matter halos with different shapes, when the halos are rotated about the three principal axes.  For pattern speeds typical of cosmological halos we demonstrate, for the first time, that a Sagittarius-like tidal stream would be altered to a degree that is detectable even with current observations.  This discovery will potentially allow for a future measurement of figure rotation of the Milky Way's dark matter halo, perhaps enabling the first evidence of this relatively unexplored prediction of $\Lambda$CDM. 
\end{abstract}
 
\keywords{%
stars: kinematics and dynamics, 
Galaxy: kinematics and dynamics;
Galaxy: fundamental parameters;
Galaxy: halo;
Galaxy: formation;
Galaxy: evolution;
(Cosmology): dark matter;
Methods: numerical
}

\section{Introduction}
\label{intro}

The Milky Way (MW) is an important laboratory for dark matter (DM) science. A robust prediction of cosmological simulations containing collisionless DM (and no baryons) is that halos are triaxial (with short/long axis ratio $ \sim 0.6$ and intermediate/long axis  ratio $\sim 0.8$), with axis ratios that are almost independent of radius \citep{dubinski_carlberg_91,jing_suto_00}.  
The  dissipative collapse of cold baryonic gas and the formation of stellar disks alter halo shapes making them  oblate or nearly spherical within the inner one-third of the virial radius, but allowing them to remain triaxial at intermediate radii and prolate at large radii \citep{kazantzidis_etal_04_shapes, deb_etal_08,  zemp_etal_12}. Cosmological simulations with Warm Dark Matter ~\citep[WDM, sterile neutrinos,][]{bose_frenk_16_WDM} and Self Interacting Dark Matter (SIDM)~\citep{peter_13_SIDM} also predict triaxial DM halos, although there are small but quantifiable differences in the radial variation in axis ratios and the degree of triaxiality. 

Since triaxial DM halos form via hierarchical mergers they generally have angular momentum, primarily due to the relative orbital angular momentum of the progenitor halos involved in the merger, but also due to the internal streaming motions within the halos. Since DM halos are triaxial, this angular momentum can manifest either as streaming motions of individual particles or as tumbling (figure rotation) of the entire triaxial halo, or both. $\Lambda$CDM cosmological $N$-body simulations predict that $\sim$90\% of dark matter halos are significantly triaxial and have measurable figure rotation \citep{dubinski_92,bailin_steinmetz_04,bryan_cress_07}. 
The pattern speed ($\Omega_p$) of figure rotation for halos from dark matter-only simulations follows a log normal distribution centered on  $0.148h\kmskpc$ with a width of 0.83$\kmskpc$~\footnote{$\Omega_p=0.148h\kmskpc=8.47^{\circ}h{\rm Gyr}^{-1}=30.4h\mu {\rm arcsec~yr}^{-1}$;$1\kmskpc \simeq 1 {\rm rad~Gyr}^{-1}$.}. \citet[][hereafter BS04]{bailin_steinmetz_04} find that the axis about which the figure rotates aligns fairly well with the halo minor axis in 85\% of the halos and with the major axis in the remaining 15\% of the halos.  The study by \citet{bryan_cress_07}  found that only  small fraction of halos (5/222) showed coherent rotation over 5~Gyr but  when rotation was measured over 1~Gyr most halos showed figure rotation with log normal distributed pattern speeds, with median and width similar to those found by \citepalias{bailin_steinmetz_04}. Since rotation is induced by torques from companions, the duration of steady rotation is expected to depend on the interaction and merger history of a galaxy. \citetalias{bailin_steinmetz_04} also found that for CDM halos $\Omega_p$ is correlated with the cosmological halo spin parameter\footnote{The halo spin parameter $\lambda= J|E|^{1/2}G^{-1}M^{-5/2}$ where $J, |E|$ and $M$ are the angular momentum, total energy and total mass of the halo.} $\lambda$  \citep{peebles_69}, but is independent of
halo mass. 

Valluri, Hofer et al. (in prep) have measured the pattern speed of figure rotation of DM  within 100~kpc of the center of disk galaxies in the Illustris suite of simulations \citep{vogelsberger_etal_14}. They find that most halos shows steady (coherent) figure rotation with  $\Omega_p \sim 0.15-0.6\kmskpc \sim 9^{\circ}-35^{\circ} \mathrm{Gyr}^{-1}$ over durations of $\sim 3-4$~Gyr.  Steady figure rotation for DM halos with baryons was not necessarily expected. Unlike DM-only simulations where the halos are strongly triaxial with nearly constant axis ratios as a function of radius,  DM halos in simulations with baryons have radially varying shapes: oblate at small radii, triaxial at intermediate radii and prolate at large radii \citep{kazantzidis_etal_04_shapes, debattista_etal_08, zemp_etal_12, Chua_2019}. In addition the presence of a dissipative baryonic component, which (in disk galaxies) is demonstrably rotating, is expected to absorb much of the  angular momentum from hierarchical mergers. We are unaware of any works that measure the pattern speeds of figure rotation of DM halos in WDM or SIDM cosmological simulations (with or without baryons). However, since halos in these simulations are triaxial \citep{bose_frenk_16_WDM,peter_13_SIDM} and have halo spin parameters $\lambda$ comparable to their CDM counterparts, it is reasonable to expect them to also have figure rotation (although future studies are need to measure the distribution of their pattern speeds).

Despite having been first predicted from cosmological simulations nearly 30 years ago \citep{dubinski_92}, few methods to measure the figure rotation of DM halos have been proposed, and it  has never been measured. Figure rotation was suggested as a mechanism to explain  the ``anomalous dust lanes'' in triaxial elliptical galaxies \citep{vanalbada_etal_82}. It was also suggested as a possible mechanism for driving spiral structure and warps in extremely extended HI disks. For instance the HI disk of NGC~2915~  is 30 times larger than its optical disk and shows a strong bisymmetric spiral feature seen only in the HI gas. Since this galaxy has no nearby companions that could have triggered the spiral features
\citep{bureau_etal_99, dubinski_chakrabarty_09, chakrabarty_dubinski_11}, it was proposed that the spiral was triggered by figure rotation of the DM halo. However  simulations show that in order for figure rotation to account for the observed features of NGC~2915, the DM halo would have to have a pattern speed  
$\Omega_p \sim 4-8$ \kmskpc, 25-50 times larger than median value predicted by cosmological simulations \citep{bekki_freeman_02,masset_bureau_03}. These simulations also  showed that production of a spiral feature also required the rotation axis of the halo to be significantly misaligned with the disk. These extreme requirements make it unlikely that halo figure rotation has triggered the spiral structure in the extended HI disk in NGC~2915. While it is still unclear how the extended spiral structure in the HI disk of NGC~2915 is generated, it cannot be the result of figure rotation of the dark matter halo. 

To our knowledge, no method for measuring extremely small figure rotation of a dark matter halo has ever  been proposed. {\em In this work we propose the first plausible method for measuring figure rotation of the MW halo that can be tested with current and future \gaia\ data.}  A definitive measurement of coherent figure rotation of the DM halo of the MW and/or other galaxies would be strong evidence of the particle nature of dark matter. In alternative theories such as MOND \citep{MOND,milgrom_2019}, dark matter does not exist, rather it is a modification in either gravity or Newton's second law at low acceleration scales, that mimics a dark component in galaxies. In MOND (and most other similar theories) it is only the baryons that produce the gravitational force. An unambiguous measurement of halo figure rotation would, therefore, be a  validation of dark matter models. In a potential theory like MOND, a disk galaxy like the MW cannot produce a triaxial potential that rotates independently of the disk potential.  While the MW does have a triaxial central bar of scale length $\sim 3-5$~kpc that comprises nearly 2/3rd the total stellar mass of the disk, its pattern speed, $\Omega_p \sim 40-50 \kmskpc$  \citep{bland_hawthorn_gerhard_16} is about 300 times larger than the pattern speeds predicted for DM halos. 

In the past decade numerous coherent tidal streams have been detected in the Milky Way halo. Since tidal streams consist of a large number of stars on similar orbits they are excellent tracers of the  gravitational potential of the Galaxy \citep{johnston_etal_99}. Since figure rotation induces an additional ``centrifugal potential'' \citep[herafter BT]{BT08} it  alters both the trajectory of a satellite and the morphology and kinematics of its tidal stream.  In principal, therefore, streams should be sensitive to figure rotation. The Sagittarius tidal stream (here after Sgr stream) \citep{mateo_etal_96, mateo_etal_98_sgrstream,majewski_etal_03,majewski_etal_04,carlin_etal_11} is the most prominent, coherent known stream in the MW halo, and indeed in the local universe. It has been used by numerous authors to probe the MW potential and is considered a  prototype of the dynamical tidally-induced evolution of satellites \citep[for a review see][]{law_majewski_16}. 

The effects of figure rotation are most easily seen in its effects on the morphology of orbits. However the orbital periods of  halo stars are so long that individual orbits and the effects of halo figure rotation on them are unobservable. Tidal streams are good proxies for the orbits of their progenitors and are frequently used to determine the properties of dark matter halos, such as their shapes \citep{johnston_etal_99,eyre_binney_09}.  If the halo of the MW is triaxial and it does rotate, the Sgr stream with a pericenter radius of 20~kpc and an apocenter radius of 100~kpc from the Galactic center \citep{belokurov_etal_14,hernitschek_etal_17}  extending over  $\sim 500^\circ$ on the sky is likely to be an ideal probe of figure rotation.  In this paper we explore how the rotation of a triaxial dark matter halo would alter the structure of a Sgr-like tidal stream. We do not attempt to either constrain the pattern speed of figure rotation, or other parameters of the Galactic potential. We simply describe the nature of the pseudo forces (Coriolis and centrifugal) resulting from figure rotation about three different axes and demonstrate how they would alter such a stream.

At present there is no consensus on the shape of the MW's dark matter halo. Despite two decades of efforts to use the spatial and velocity distributions of stars in the Sgr  stream to determine the shape of the halo \citep{johnston_etal_99, helmi_04, johnston_etal_05,law_majewski_10, deg_widrow_12, carlin_etal_12,dierickx_loeb_17a,dierickx_loeb_17b,fardal_etal_19}, current measurements include spherical, oblate, prolate and triaxial shapes. Until recently, no model satisfactorily reproduced all the observed features of the Sgr stream  such as  the large ratio of the trailing apocenter radius ($\sim 100$~kpc) to the leading apocenter distance ($\sim 50$~kpc),  the relatively small angle of $95^{\circ}$ between the leading and trailing apocenters \citep{belokurov_etal_14}, and the ``bifurcations'' in the Sgr stream \citep{fellhauer_etal_06,   penarrubia_etal_10, koposov_etal_12, slater_etal_13}.  Although the model of \citet{law_majewski_10} does very well to describe pre-2014 data \citep[also see][]{deg_widrow_12}, it requires an oblate-triaxial halo with the disk perpendicular to the intermediate axis of the halo, an orientation that is violently unstable  \citep{debattista_etal_13}. To match the angle between the leading and trailing apocenters a halo with a shallow central radial density profile with an extended flat core, not predicted by cosmological simulations, is needed \citep{belokurov_etal_14,fardal_etal_19}. In addition it has been shown that the LMC may significantly perturb the stream \citep{vera-ciro_helmi_13, gomez_besla_15} and alter the DM distribution of the halo by producing a wake \citep{garavito-camargo_etal_19}. {After the submission of this paper \citet{Vasiliev_tango} presented the most comprehensive model to date for the Sgr stream. Unlike most previous models, the Sgr dwarf galaxy in this model is tidally disrupted in the time dependent dark matter halo of the MW as it is being dynamically altered by the gravitational interaction with the LMC. In this model the shape of the dark matter halo not only varies with radius, it is also time dependent. They argue that the time dependence is necessary to explain the observed  morphology and 3D kinematics of the Sgr stream. We will return to a discussion of this Sgr-MW-LMC model in Section~\ref{sec:conclusions}.}

The effects of figure rotation on the orbit of a progenitor depend strongly on the shape of the DM halo. Since this is uncertain we simply adopt four plausible models with disk, bulge, and halo mass distributions motivated by previous work. We explore a small range of pattern speeds and use an evolution time \tev, for the Sgr-stream of 4~Gyr. This is larger than \tev$ \sim 2.3-2.9$~Gyr preferred by other authors \citep{fardal_etal_19, vasiliev_belokurov_20,Vasiliev_tango}, but short enough that it is reasonable to assume that the halo has maintained a constant pattern speed over this timescale.  Although some authors \citep{laporte_etal_18} claim, based on dynamical features in the Milky Way's stellar disk, that the Sgr stream has been evolving for at least 6~Gyr, we do not consider such a long timescale since it is unlikely that DM halos maintain a coherent pattern speed for such a long duration. In reality the progenitor of Sgr was probably at least $5\times 10^{10}$\msun\, and its initial infall probably started $\sim$10 Gyr ago from a much larger initial distance \citep{jiang_binney_00,  gibbons_etal_17}. Our experiments with  1.5~Gyr~$<$ \tev $<$~8~Gyr show that for the progenitor mass selected here  \tev $ \lesssim 3$~Gyr do not produce streams that are long enough to match the observations. We also found that \tev$ \gtrsim 6$~Gyr in the rotating potential result in tidal streams that are more strongly perturbed (less coherent) than streams in stationary potentials evolved over the same duration. We believe this is due to the fact that orbits in rotating potentials are on average more likely to be chaotic than in stationary potentials \citep{deibel_etal_11} and it is this increased chaoticity that results in less coherent streams \citep{price_whelan_etal_16}. For these reasons we limit our study to \tev$= 4$~Gyr. 

The objectives of this paper  are (a) to demonstrate that figure rotation of a (moderately) triaxial halo can demonstrably alter the morphology and kinematics of a tidal stream in ways that are already measurable with existing data, (b)  to highlight the features that would be most likely to distinguish a rotating halo from a static one. 

 In Section~\ref{sec:method} we describe the set up of our test-particle simulations. In Section~\ref{sec:orbits} we describe some general principles governing the behavior of orbits and tidal streams in triaxial halos subjected to figure rotation. We also show that the magnitude of the Coriolis force on a Sgr-like stream is a significant fraction of the gravitational force even for a small pattern speed. In Section~\ref{sec:results} we present our simulations of Sgr-like streams and make some comparisons with observations. In Section~\ref{sec:conclusions} we summarize our results and discuss the implications of this work and future directions.

\section{Simulations and Numerical Methods}
\label{sec:method}

{We use the Gala package \citep{gala:JOSS, gala} to explore orbits in a triaxial potential subject to rotation about each of three principle axes (Section~\ref{sec:orbits_triaxial}). The exploration of orbits in Section~\ref{sec:orbits_triaxial} is carried out in a potential that does not contain a disk or bulge. }

We use the same package to simulate  tidal streams in Milky-Way-like potentials (Sections~\ref{sec:Sgr-stream} \& \ref{sec:results}). Streams are generated by simulating the orbital evolution of  stars once they have been tidally stripped from the progenitor satellite. We use the ``particle-spray'' stream generation method \citep{fardal_etal_15} that assumes that stars are lost from the L1 and L2 Lagrange points of the progenitor at a uniform rate  \citep[e.g.,][] {Kupper_etal_12}. Once stars escape from the progenitor they experience the gravitational potential of both the progenitor and the Galaxy. In the current work we do not consider the effects of dynamical friction from the DM halo on the progenitor, even though the Sgr dwarf progenitor was probably massive enough to experience significant dynamical friction \citep[e.g.,][]{fardal_etal_19}. The ``particle-spray''  method used produces stream stars drawn from  an initial distribution function that depends on the potential (and mass) of the progenitor, such that more massive satellites produce dynamically hotter streams. While they are not as accurate as N-body simulations, the advantage of test particle simulation is that they allow for a rapid exploration of the parameter space (e.g.  Galactic potential parameters, halo shape, pattern speed and axis of figure rotation).

When discussing the Sgr stream we use a Galactocentric coordinate system that is right handed with the $x$-axis coincident with the Galactic $X$-axis,  $y$-axis parallel to the direction of the velocity of the LSR (parallel to the Galactic $Y$-axis) and $z$-axis perpendicular to the disk plane with the sun located at located at $(-8.122, 0, 0.0208)$~kpc (the default Galactocentric parameters in Astropy v4.0).  The Sgr dwarf progenitor has a mass of $6\times 10^8$\msun\,  \citep{law_majewski_10}, which is lower than some recent estimates \citep[$10^{10}-10^{11}$\msun,][]{laporte_etal_18}.  The progenitor potential is modeled as a spherical Plummer model with a core radius of 0.65~kpc and does not include a separate dark matter component. This is a lower  progenitor mass than recent estimates ($5\times 10^{10}$\msun) but since the mass of the progenitor does not change with time in ``particle-spray'' models, it produces streams with a somewhat closer visual appearance to the observed stream. For most of our models we use the present day phase-space coordinates for the Sgr dwarf obtained with  \gaia DR2 \citep{gaiacollab_helmi_18,vasiliev_belokurov_20}   ($x=17.2$~kpc, $y= 2.5$~kpc,  $z=-6.4$~kpc, $v_x=237.9~\kms, v_y= -24.3~\kms, v_z=209.0~\kms$). Starting at this position we evolved the orbit backwards in time for  4~Gyrs to determine the initial position and velocity of the progenitor.

All of  our Milky Way models use a composite halo+disk+bulge potential. We use three different halo shapes with  $a, b, c$ defined as the semi-axis lengths of the density (not potential) model along the Galactocentric $x,y,z$ axes respectively (in this work $a, b, c$ are not aligned with the long, intermediate and short axes of the halo, since these change from model to model).

The first Galactic model considered is the one found by \citet{law_majewski_10} (referred to hereafter as the \LM\ model)\footnote{The LMPotential in Gala}. This model has a logarithmic halo potential with rotation velocity set such that the total circular velocity $v_c = 220\kms$  at 8~kpc and  semi-axis lengths (relative to the longest axis) of {\em density profile} of $a=0.44, b=1, c=0.97$. The \LM model has a Miyamoto-Nagai disk \citep{miyamoto_nagai_75} with mass $M_{\rm disk} = 1\times10^{11}\msun$, radial scale length of $6.5$~kpc, and and vertical scale length of $0.26$~kpc. It contains a spherical Hernquist bulge with mass $M_{\rm b}= 3.4\times 10^{10}\msun$ and radial scale length $0.7$~kpc resulting in a somewhat deeper potential than in the other three models. The deeper potential results in simulations that are unable to produce a trailing stream with apocenter radius of $\sim 100$kpc  as observed by \citep{belokurov_etal_14,hernitschek_etal_17}. Therefore we only use it to illustrate that the effect of figure rotation in this deeper potential are similar to the effects in a shallower potential of a similar shape (the \LMm\, model below).
 
The other three models have triaxial dark matter halos with radial density profiles of {Navarro-Frenk-White  form (NFW) \citep{NFW} with mass distribution stratified in concentric ellipsoidal shells of parametrized by,
\begin{equation}
m^2 = (\frac{x^2}{a^2}+\frac{y^2}{b^2}+\frac{z^2}{c^2})
\end{equation}
where  $a >b>c$, are the semi-axis length. } The halo is defined by a circular velocity $v_c = 162\kms$ and a scale radius $r_s =  28$~kpc. The latter is somewhat larger than predicted by  $\Lambda$CDM simulations for the halo mass - concentration relationship for MW mass halos \citep{dutton_maccio_14, klypin_yepes_16}, but is consistent with estimates based on the local escape speed measurements from \gaia~DR2 \citep{hattori_etal_18}. The value of $r_s$ used here is  significantly smaller than  $r_s = 68$~kpc estimated by  \citet{fardal_etal_19}. The Gala package uses the  formulation of \citep{lee_suto_03} to compute the potential from the ellipsoidally stratified mass density distribution.

These three models use a Miyamoto-Nagai disk \citep{miyamoto_nagai_75} with $M_{\rm disk} = 6\times 10^{10}\msun$, a radial scale length of 3~kpc and vertical scale length 0.26~kpc. Instead of a box-peanut bulge/bar we use a spherical Hernquist bulge \citep{hernquist_90} with mass $M_{\rm b} = 6\times10^9$\msun\ and radial bulge scale length 0.7~kpc. These values were chosen since they produce streams that give a reasonably good match to many Sgr stream observations, including the larger apocenter radius of the trailing arm. We note that all our streams produce a slightly larger leading apocenter radius than observed, probably because we do not include dynamical friction, and possibly because the reflex motion of the MW center-of-mass due to the gravitational interaction with the LMC is not account for  \citep[ see,][]{Vasiliev_tango}.

\begin{figure*}
\begin{center}
\includegraphics[angle=0,trim=120. 0. 100. 0., clip, width=0.495\textwidth]{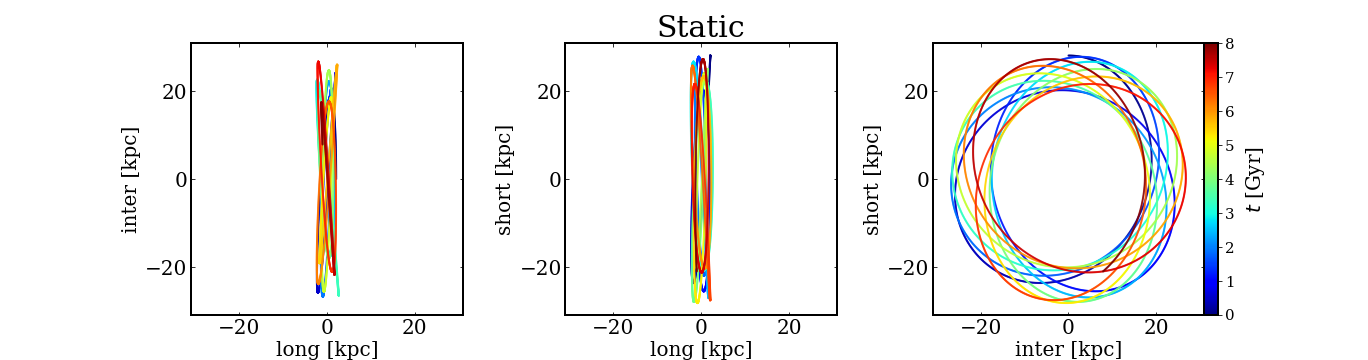} \\
\hspace{-0.25cm}\includegraphics[angle=0,trim=120. 0. 100. 0., clip, width=0.495\textwidth]{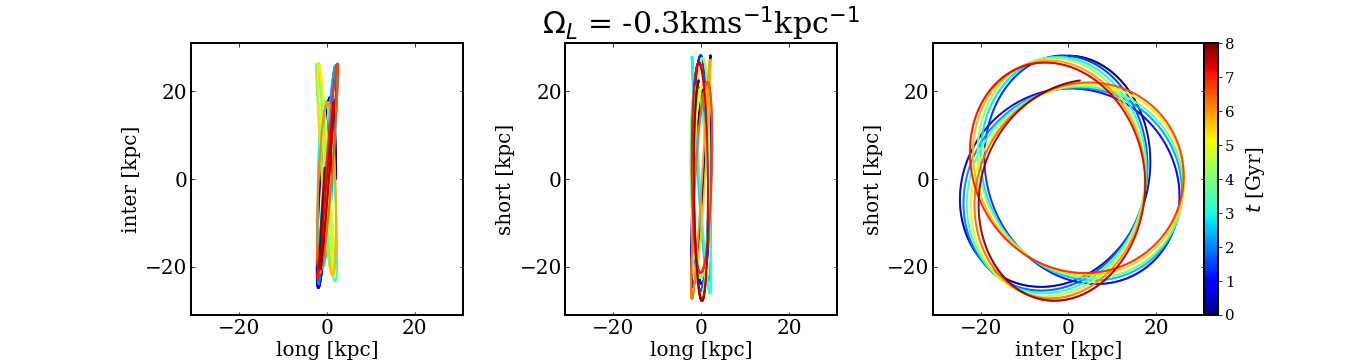} 
\includegraphics[angle=0,trim=120. 0. 100. 0., clip, width=0.495\textwidth]{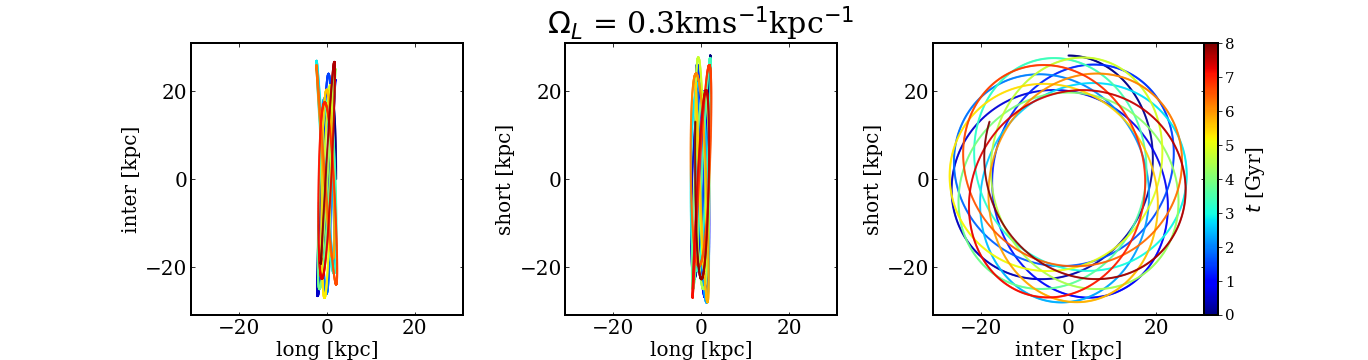} \\
\hspace{-0.25cm}\includegraphics[angle=0,trim=120. 0. 100. 0., clip, width=0.495\textwidth]{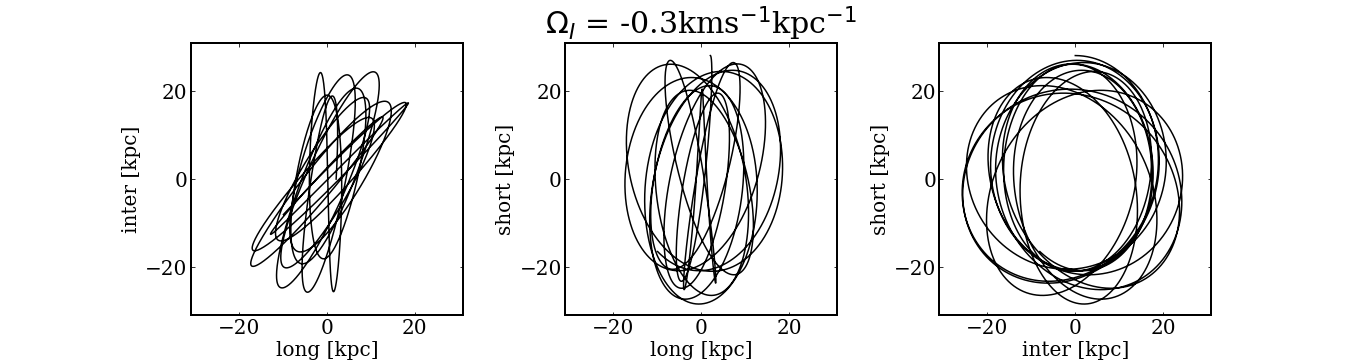} 
\includegraphics[angle=0,trim=120. 0. 100. 0., clip, width=0.495\textwidth]{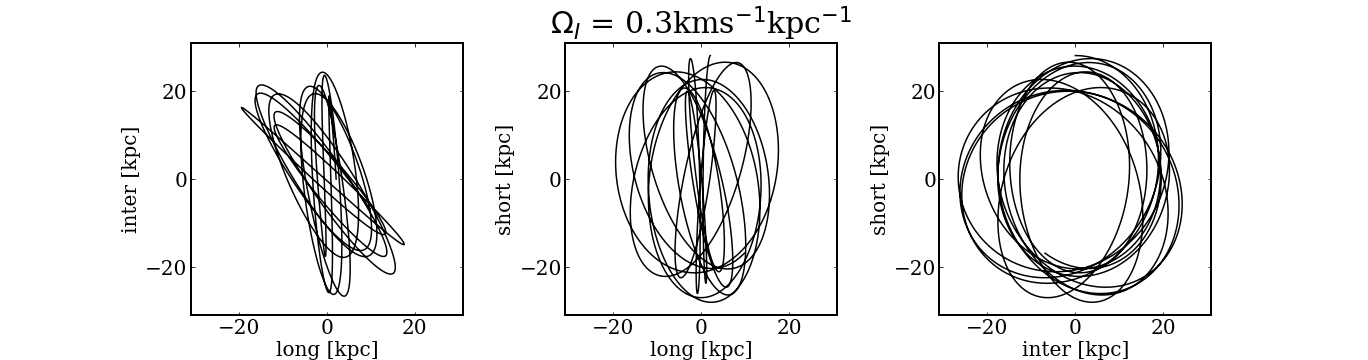} \\
\hspace{-0.25cm}\includegraphics[angle=0,trim=120. 0. 100. 0., clip, width=0.495\textwidth]{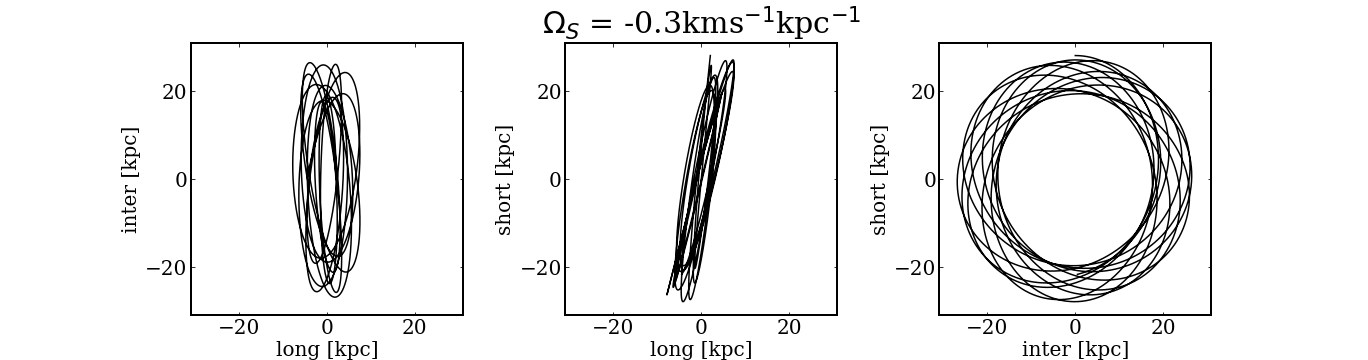} 
\includegraphics[angle=0,trim=120. 0. 100. 0., clip, width=0.495\textwidth]{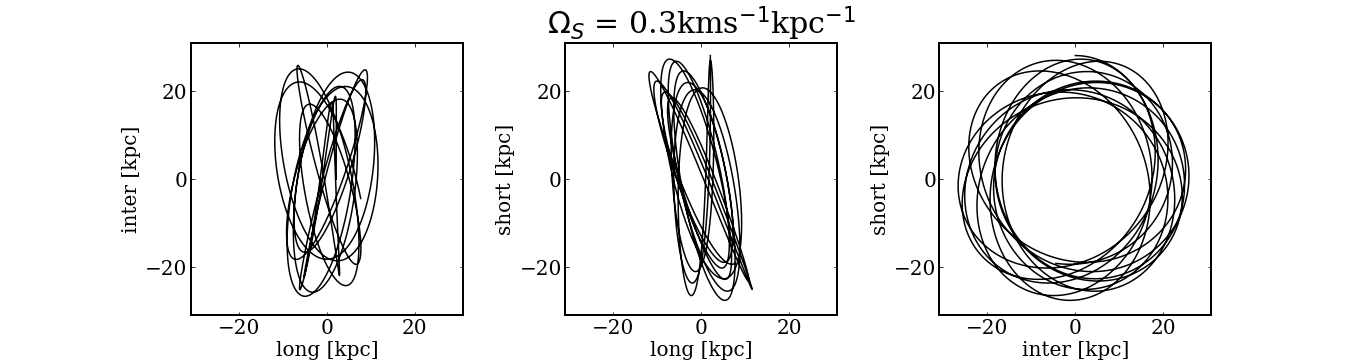} \\
\end{center}
    \caption{{Top row: a long-axis tube orbit in a static triaxial potential. The axes are marked `long', `inter', `short' to signify the long, intermediate and short axes of the potential respectively. This orbit has negative angular momentum about the `L'-axis (i.e. it is rotating clockwise in the top row-right panel). Rows 2-4 show the same orbit in the rotating frame with  clockwise (left 3 panels) and anti-clockwise (right 3 panels) rotation about the `L', `I', `S' axes respectively. The color in the top two rows denotes the orbital integration time and is provided to enable the reader to assess the rate of orbital precession. Figure rotation about the orbit's angular momentum axis (in this case the long-axis) alters the precession rate of the orbit (compare row 1 right-most panel with row 2, panels 3 \& 6). Clockwise/anti-clockwise rotation about the `I' axis (3rd row, 1st and 4th panels) causes the orbit to be become tilted with respect to this axis.  Clockwise/anti-clockwise rotation about the `S' axis (4th row, 2nd and 5th panels) causes the orbit to be become tilted with respect to this axis. }
\label{fig:LAT}
}
\end{figure*}

Although simulated triaxial halos in cosmological hydrodynamical simulations have radially varying halo shapes \citep{kazantzidis_etal_04_shapes, debattista_etal_08, zemp_etal_12}, our DM halos have their mass stratified on concentric ellipsoidal shells of fixed axis length ratios $a:b:c$. The three models with NFW DM halo density distributions have the different shapes as described below. The oblate-triaxial model (here after \OT\ model) has  axis scale lengths $a=0.9, b=1, c=0.8$ (i.e. the long-axis and short axes are along Galactocentric  $y$ and $z$ axes respectively). The prolate halo model  \citep[hereafter the \Fm\ model,][modified]{fardal_etal_19} is similar to the one found by \citet{fardal_etal_19} with $a=0.95, b=1., c=1.106$ (for the density\footnote{\citet{fardal_etal_19} find their best triaxial NFW potential model has a potential axis lengths of 1: 1.1: 1.15. The density axis lengths we use give the same axis lengths for the derived potential.}) but with an NFW halo parameters ($v_c, r_s$), disk and bulge parameters as given in the paragraph above.  The last model is the \LMm\ model (``Law-Majewski modified'') which has the same axis scale lengths (for the density) as the \LM\ model  $(a=0.44, b=1, c=0.97)$, but with halo, disk and bulge parameters as defined in the paragraph above. 

The  streams were evolved in  each of the four models above, both in static halos ($\Omega_p=0$) and rotating halos with clockwise and anti-clockwise rotation about each of the three Galactocentric principal axes ($x,y,z$) with $|\Omega_p| = 0.1, 0.3, 0.6, 0.8, 1.0, 1.4\kmskpc$. { In the rest of this paper the pattern frequency will either be denoted by a 3-vector ${\mathbf \Omega_p} =(\Omega_{x},\Omega_{y}, \Omega_{z})$ or just by the non-zero component of the pattern speed with the units always in \kmskpc .} After some initial exploration (not presented) we kept potential parameters, the duration of evolution, the initial mass and phase space coordinates of the Sgr dwarf fixed to the values stated above.  Our simulations do not include dynamical friction and the mass of the halo/disk does not change with time. While previous authors \citep{law_majewski_10, deg_widrow_12, belokurov_etal_14, fardal_etal_19} have done far more exhaustive searches of parameter space with the goal of constraining the shape of the halo and other Galactic parameters, we defer such an exploration to the future. In  Section~\ref{sec:results} we present a limited set of results to illustrate the broad {\it qualitative effects} of halo figure rotation on the properties of the stream. 

 \citetalias{bailin_steinmetz_04} showed that 85\% of DM halos  in cosmological simulations rotate about an axis that is within 25$^{\circ}$ of  the minor axis of the halo, but \citet{bryan_cress_07} found that less than half of their halos rotate about the minor axis. These authors and Valluri, Hofer et al. (in prep) find that recent or ongoing interactions can induce figure rotation over a short duration of time.  The MW is currently undergoing an interaction with the LMC, which if  massive enough could itself have induced halo rotation. As it is beyond the scope of this paper to determine the axis about which the LMC would induce figure rotation, for the sake of completeness, we consider rotation about each of the three principal axes of the Galactic potential.

{
\section{Effects of figure rotation on orbits and streams in triaxial halos}
}
\label{sec:orbits}

\subsection{Orbits in simple triaxial potentials with figure rotation}
\label{sec:orbits_triaxial}

The effects of figure rotation  on the main families of orbits in triaxial potentials \mdash\, box orbits, short axis tubes, and  long axis tubes \mdash\, have been reported in previous works \citep{schwarzschild_82,heisler_etal_82,dezeeuw_merritt_83,udry_pfenniger_88, udry_91,deibel_etal_11,valluri_etal_16}. However, as far as we are aware none of these studies have investigated the effects on orbits of figure rotation about any axis other than the short-axis of the potential.
 { In this section we briefly summarize previous results and discuss the behavior of tube orbits under rotation about the intermediate and long axes. As we will show the results, of rotation about each of the three axes are fairly similar and some general behaviors can be inferred. We  focus our discussion on orbits in a triaxial density distribution of NFW form with  axis ratios ${\rm intermediate}/{\rm long} = 0.8$ and ${\rm short}/{\rm long}=0.6$ where.  The potential is defined by a circular velocity $v_c = 200\kms$ and a scale radius $r_s =  20$~kpc.  }

In a rotating potential, the energy of an orbit is not an integral of motion but the Jacobi integral ($E_J$) is a conserved quantity:
\begin{equation}
E_J = {\frac{1}{2}}|\dot{\mathbf{x}}|^2+\Phi-{\frac{1}{2}}|{\mathbf \Omega_p} \times {\mathbf{x}}|^2,
\end{equation}
where $\mathbf{x}$ and $\mathbf{\dot{x}}$ are three dimensional spatial and velocity vectors, respectively.  The equation of motion in the rotation frame is given by the vector differential equation:

{\begin{eqnarray}
\ddot{\mathbf{x}} & = & - {\mathbf{\nabla\Phi}} - 2{\mathbf{\Omega_p \times \dot{x}}}  -{\mathbf \Omega_p} \times ({\mathbf\Omega_p}\times {\mathbf x})  \\ 
& = & - {\mathbf{\nabla\Phi}} - 2{\mathbf{\Omega_p \times \dot{x}}}  +|\Omega_p|^2{\mathbf x} - {\mathbf \Omega_p}({\mathbf \Omega_p}\cdot {\mathbf x}) \nonumber
\end{eqnarray} }
where  $- 2{\mathbf{\Omega_p \times \dot{x}}}$ is the Coriolis acceleration  (hereafter \vecaco) and ${\mathbf \Omega_p} \times ({\mathbf\Omega_p}\times {\mathbf x})$
 is the centrifugal acceleration  (hereafter \vecacf)  \citepalias[see \S~3.3.2, eq. 3.116][]{BT08}. In what follows we refer to the gravitational acceleration as \vecag. Table~\ref{tab:f_pseudo} gives exact expressions for the Coriolis and centrifugal accelerations for figure rotation about each of the three principal axes. In all the figures in this paper orbits in rotating potentials are plotted in the rotating frame and not in an external inertial frame. 

In triaxial potentials there are two families of tube-like orbits: long-axis tubes and short-axis tubes, both of which are affected by rotation about the short axis of the potential. \citet{heisler_etal_82} showed that closed periodic orbits rotating about the long-axis of the halo are stable to figure rotation but the Coriolis force tips them about the intermediate axis. Two stable periodic orbit families exist, which rotate clockwise and anti-clockwise about the long-axis. The Coriolis force, causes orbits with positive angular momentum to be tipped clockwise about the intermediate axis while orbits with negative angular momentum to be tipped anti-clockwise about the intermediate axis. In other words the Coriolis force causes these orbits to become misaligned with the figure rotation axis. Since these two periodic orbits ``parent'' the clockwise and anti-clockwise rotating long-axis tube orbit families are similarly tipped  about the intermediate axis (tilted relative to the short axis) \citep{deibel_etal_11,valluri_etal_16}.

\begin{figure*}
\begin{center}
\includegraphics[angle=0,trim=120. 0. 100. 0., clip, width=0.49\textwidth]{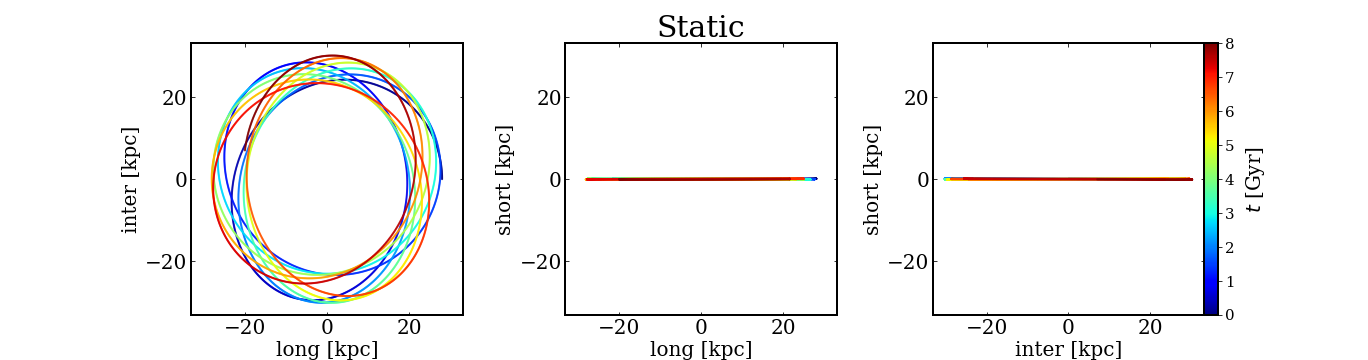} \\
\hspace{-0.25cm}\includegraphics[angle=0,trim=120. 0. 100. 0., clip, width=0.49\textwidth]{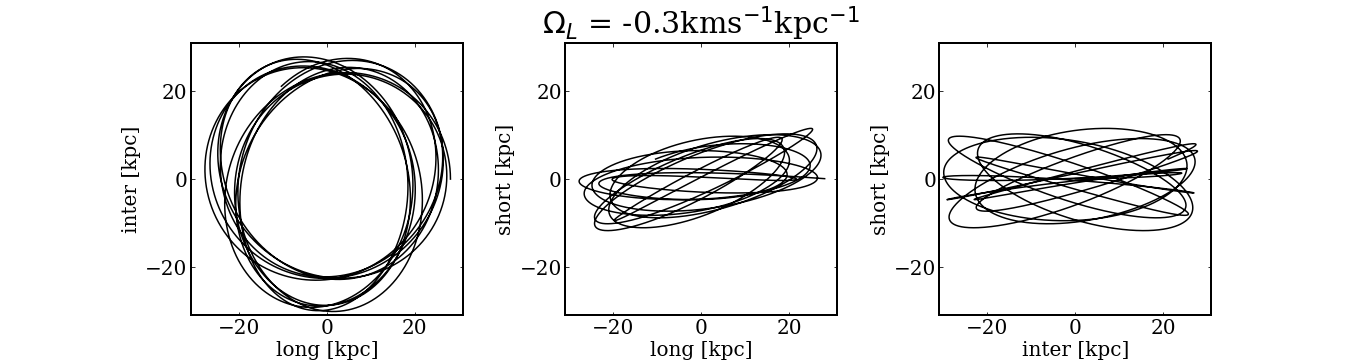} 
\includegraphics[angle=0,trim=120. 0. 100. 0., clip, width=0.49\textwidth]{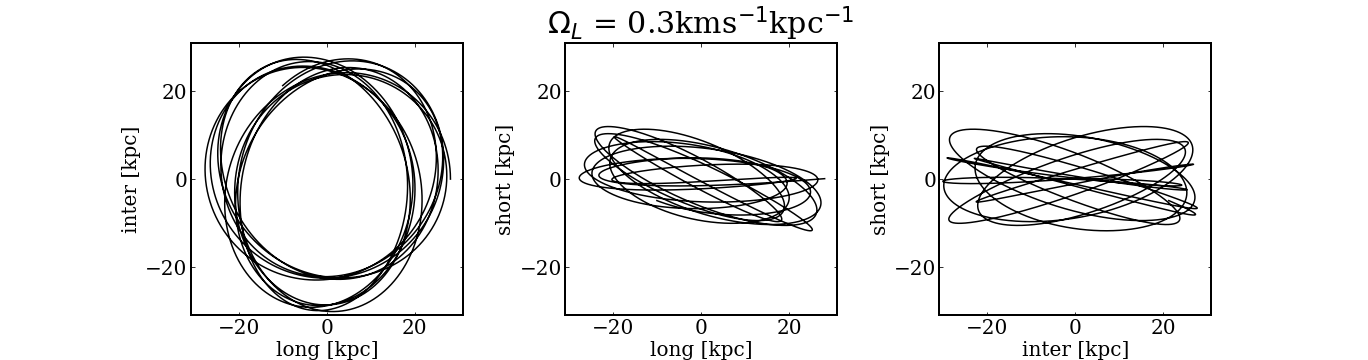} \\
\hspace{-0.25cm}\includegraphics[angle=0,trim=120. 0. 100. 0., clip, width=0.49\textwidth]{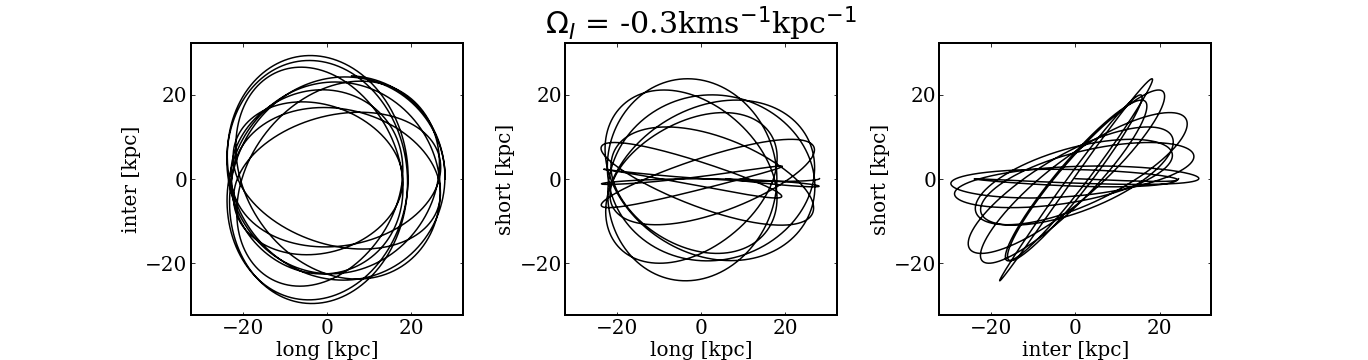} 
\includegraphics[angle=0,trim=120. 0. 100. 0., clip, width=0.49\textwidth]{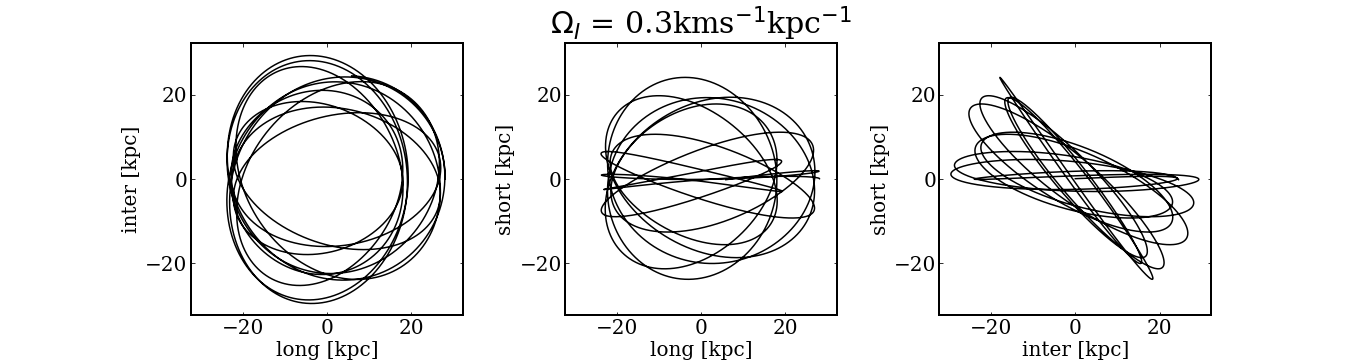} \\
\hspace{-0.25cm}\includegraphics[angle=0,trim=120. 0. 100. 0., clip, width=0.49\textwidth]{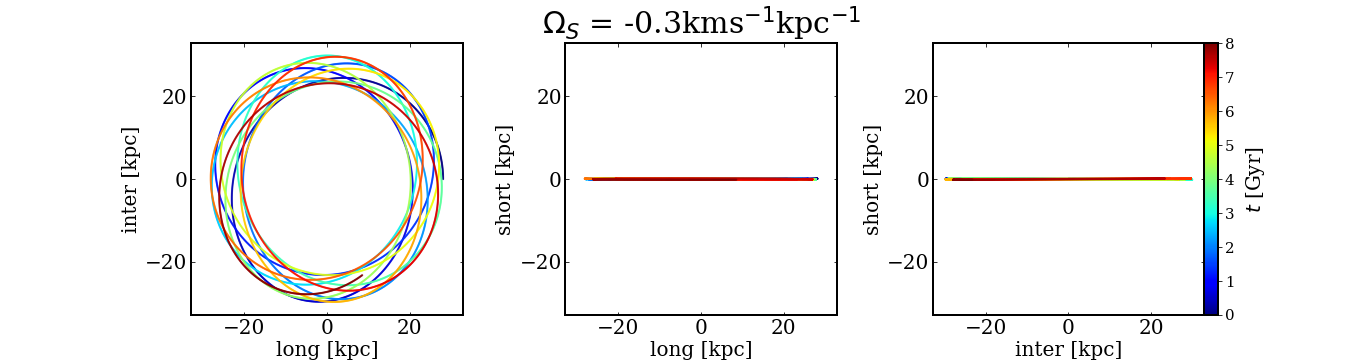} 
\includegraphics[angle=0,trim=120. 0. 100. 0., clip, width=0.49\textwidth]{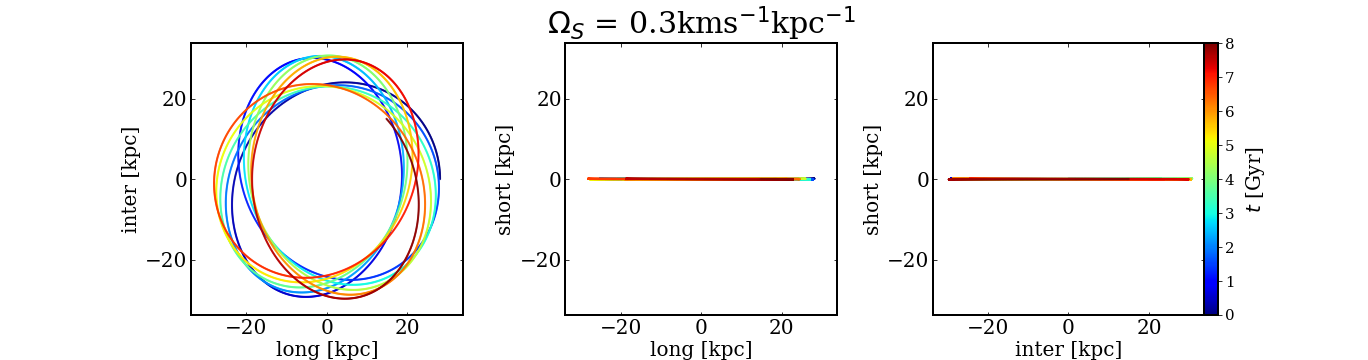} \\
\end{center}
\caption{{Similar to Fig.~\ref{fig:LAT} for a short-axis tube orbit in a static triaxial potential. This orbit has positive angular momentum about the  short 'S' axis (i.e. it is rotating anti-clockwise in top-left panel). Rows 2-4: the same orbit in the rotating frame with  clockwise (left) and anti-clockwise (right) rotation about the long (L), intermediate (I), and short (S) axes respectively. Again we see that figure rotation about the orbital angular momentum (`S') axis causes a change in orbital precession rate (4th row). Figure rotation about the other two axes cause the orbit to become tilted with respect to that axis (rows 2,3).}
\label{fig:SAT}
}
\end{figure*}

\citet{binney_81} showed that figure rotation about the short-axis of a triaxial potential destabilizes some loop orbits that circulate retrograde about the rotation axis close to the equatorial plane, making them unstable to perturbations perpendicular to that plane. Orbits are destabilized if they lie in an annular region (called the ``Binney instability strip''). This instability results from a resonant coupling that can cause orbits to become unstable to oscillations perpendicular to the equatorial plane resulting in their being tipped out of the $x-y$ plane. Since this is an instability that depends on resonant coupling it only affects orbits in a small region called the ``instability strip'', and  is of limited interest.

{
Figure~\ref{fig:LAT} and Figure~\ref{fig:SAT} show a long axis tube and short axis tube respectively in a static potential (top rows) and subject to  rotation about the long, intermediate and short axes of the potential (rows 2-4). The left (right) three panels in each row show three projections of the orbit when the potential is subjected to  clockwise (anti-clockwise) figure rotation with pattern speed $|\Omega_p|=0.3$\kmskpc. Detailed descriptions of the behavior of each type of tube orbit are provided in the captions to Figures~\ref{fig:LAT} and ~\ref{fig:SAT}, however the response of each orbit to figure rotation can be summarized by two basic results. For any tube orbit in a triaxial potential: 
\begin{itemize}
\item{figure rotation about an axis {\it parallel} to the angular momentum vector of the orbit will cause a change in the angular precession rate and will therefore alter the angle between successive apocenters of the orbit in the plane perpendicular to the angular momentum vector;}
\item{figure rotation about an axis {\it perpendicular} to the angular momentum vector of the tube will result in the orbit being tipped or misaligned with the figure rotation axis.}
\end{itemize}
}

{
The change in the precession rate of a tube orbit due to figure rotation about its angular momentum axis can be seen by comparing the top-right panel of Figures~\ref{fig:LAT} with the 3rd and 6th panels of the 2nd row and  in Figure~ \ref{fig:SAT} by comparing the top-left panel with the 1st and 4th panels of the 4th row. The change in the angle between successive apocenters and the shape of the lobes of the rosette is more clearly seen in Figure~\ref{fig:stream_forces} (2nd row, panels 2 \& 3).
The tilting of the orbit relative to the rotation axis and the change in the precession rate of the orbit about the angular momentum axis are both consequences of the Coriolis force which depends on the {\it sign} of the velocity vector and acts in a direction perpendicular to both the velocity vector and the figure rotation vector. This will be further illustrated in Fig.~\ref{fig:stream_forces} which shows various components of the Coriolis force  acting along a Sagittarius-like stream.
}

In self-consistent (equilibrium) potentials no tube orbits are found to circulate around the intermediate axis of a triaxial potential since it has been shown that intermediate axis tubes are unstable \citep{heiligman_schwarzschild_79,adams_etal_07}. Since it is impossible to generate stable intermediate axis tubes in a manner similar to the generation of long and short axis tubes, we do not consider them further.

Finally, \citet{schwarzschild_82} showed that when a triaxial potential is rotated about the short axis, the linear long-axis orbits acquire prograde rotation about the short axis as a result of the Coriolis force. Since this orbit is the ``parent'' of the box orbit family, many such orbits acquire small prograde rotation and also experience ``envelope doubling'' \citep{valluri_etal_16} which causes some resonant (``boxlet'') and non-resonant box orbits to acquire a small net angular momentum in the rotating frame (frequently called `x1' orbits in bars). Since our focus in the rest  of this paper is on the Sgr stream, which, based on the observed multiple wraps of the stream is on a tube-like orbit, we do not consider tidal streams on box orbits further in this paper.

\subsection{Sagittarius-like orbit and stream}
\label{sec:Sgr-stream}

We now  describe some theoretical principles which govern the behavior of the orbit of a Sagittarius-like dwarf satellite and the tidal stream it produces. We assume that the dark matter halo of the MW galaxy experiences steady figure rotation over a duration of 4~Gyr. As was seen in the previous section the effects of figure rotation are most easily seen in the rotating frame of the halo. While strictly speaking, the Sun  cannot  be regarded as being in the rotating frame of the dark halo, at the solar position a pattern speed of $0.15-0.6\kmskpc$ (the range of figure rotation values explored) translates to a velocity of only $1-5\kms$, which is smaller than the random velocities of stars in the solar neighborhood and smaller than the velocity of the sun relative to the LSR. Thus, we assume that the heliocentric view of the stream is (almost) identical to what it would be in the rotating frame of the dark halo (although it would be straightforward to make the transformation to a heliocentric frame if necessary).

\begin{table*}[t]
\begin{centering}
\caption{Pseudo forces for a polar Sgr-like orbit in rotating frame}
\begin{tabular}{ll|l|l}
\hline
 &
\multicolumn{1}{c|}{Rotation about $z$} &
\multicolumn{1}{c|}{Rotation about $x$}&
\multicolumn{1}{c}{Rotation about $y$} \\
\hline 
\vecaco &  
= $\hat{i}(2\Omega_p v_y) - \hat{j}(2\Omega_p v_x)$ &  
= $\hat{j}(2\Omega_p v_z) - \hat{k}(2\Omega_p v_y)$ & 
= $\hat{i}(-2\Omega_p v_z) + \hat{k}(2\Omega_p v_x)$\\
                                 & 
 $\approx - \hat{j}(2\Omega_p v_x)$ \; $[\because |v_y|\approx 0]$ &  
 $\approx \hat{j}(2\Omega_p v_z) $ \; $[\because |v_y|\approx 0]$  & 
  \\
  \hline
\vecacf& 
$= \Omega_p^2\sqrt{x^2+y^2}(\hat{i}+\hat{j})$&
$= \Omega_p^2\sqrt{y^2+z^2}(\hat{j}+\hat{k})$ &
$= \Omega_p^2\sqrt{x^2+z^2}(\hat{i}+\hat{k})$  \\
   & $ \approx \Omega_p^2|x|\hat{i}$\; $[\because |y|\approx 0]$ &
       $ \approx \Omega_p^2|z|\hat{k}$\; $[\because |y|\approx 0]$ &
       \\
\hline
\end{tabular}
 \label{tab:f_pseudo}
 \end{centering}
\end{table*}

\begin{figure*}
\begin{center}
\includegraphics[angle=0,trim=10. 0. 70. 20., clip,width=0.515\textwidth]{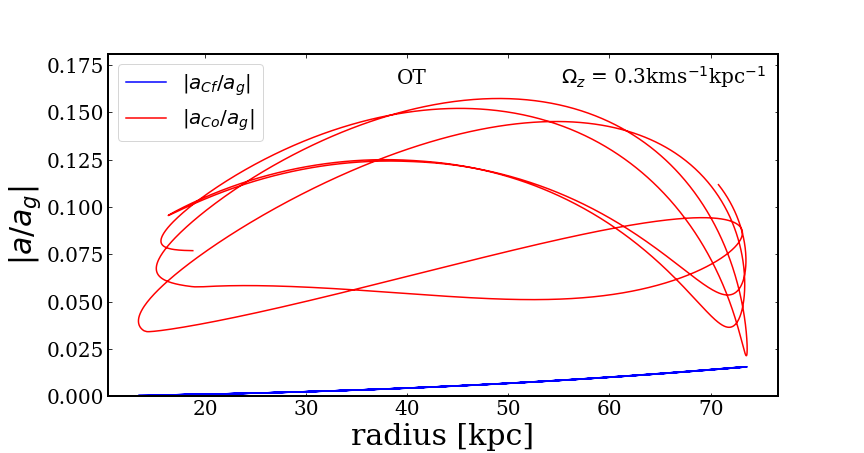} 
\includegraphics[angle=0,trim=105. 0. 40. 20., clip,width=0.475\textwidth]{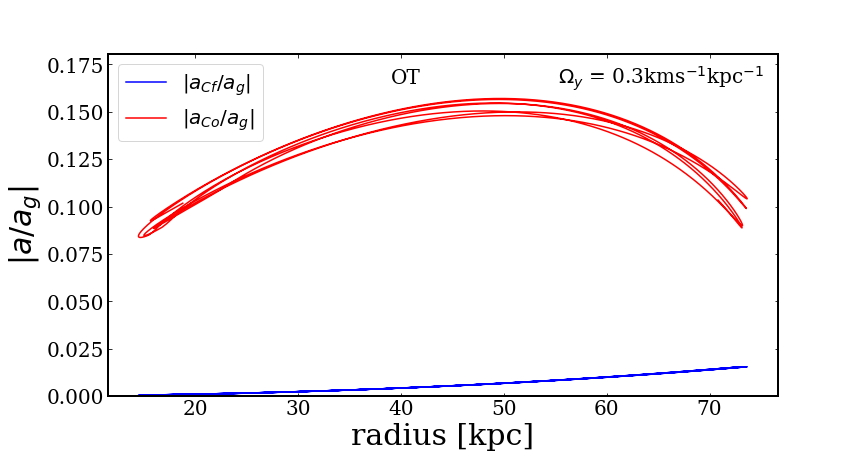} 
\end{center}
\caption{The ratio of the magnitude of $|$\vecacf/\vecag$|$, and  $|$\vecaco/\vecag$|$ along an orbit in a halo potential  {with rotation about the $z$  axis with pattern speed $\Omega_z = 0.3$\kmskpc (left) and rotation about the  $y$ axis with pattern speed $\Omega_y = 0.3$\kmskpc (right) (in all plots that follow the labels gives pattern speed in units of \kmskpc, if not specified). }While the centrifugal acceleration (blue) is small everywhere ($<$2\% of the gravitational acceleration) and increases monotonically with radius, the Coriolis acceleration (which depends on orbital velocity) changes through out the orbit and can be as high as $\sim$15\% of gravitational acceleration. 
\label{fig:orb_forces}
}
\end{figure*}

In this section we quantify the magnitude of the centrifugal acceleration (\vecacf) and Coriolis acceleration (\vecaco) and compare them with the gravitational acceleration (\vecag) for a satellite on a Sgr-like orbit in one of the MW-like potentials described in Section~\ref{sec:method}. Figure~\ref{fig:orb_forces} shows  the ratios of the magnitudes of the Coriolis and centrifugal accelerations to the gravitational acceleration: $|$\vecaco/\vecag$|$  (red curve) and $|$\vecacf/\vecag$|$  (blue curve) as a function of Galactocentric radius along such an orbit. The orbit shown was evolved in an {\it OT} model, with angular velocity vectors as indicated by labels in each panel. For a Sgr-like stream that lies approximately in the $x-z$-plane, rotation about the $x$ axis produces pseudo forces of similar magnitude to rotation about the $z$ axis and is not shown. The figures show that \vecacf\ (blue curves) is never more than $\sim 2$\% of \vecag\, and changes monotonically with radius. In contrast, the Coriolis acceleration, changes non-monotonically along the entire orbit (since the velocity changes along the orbit) and can be as large as $\sim 15$\% of the gravitational acceleration along this Sgr-like orbit. Since the Coriolis acceleration is a significant fraction of the gravitational acceleration even for $|\Omega_p| =0.3$ it should  alter an orbit of a Sgr-like dwarf, and the tidal stream generated by it.

\begin{figure*}
\begin{center}
\includegraphics[angle=0,trim=30. 0. 125. 0., clip, width=0.63\textwidth]{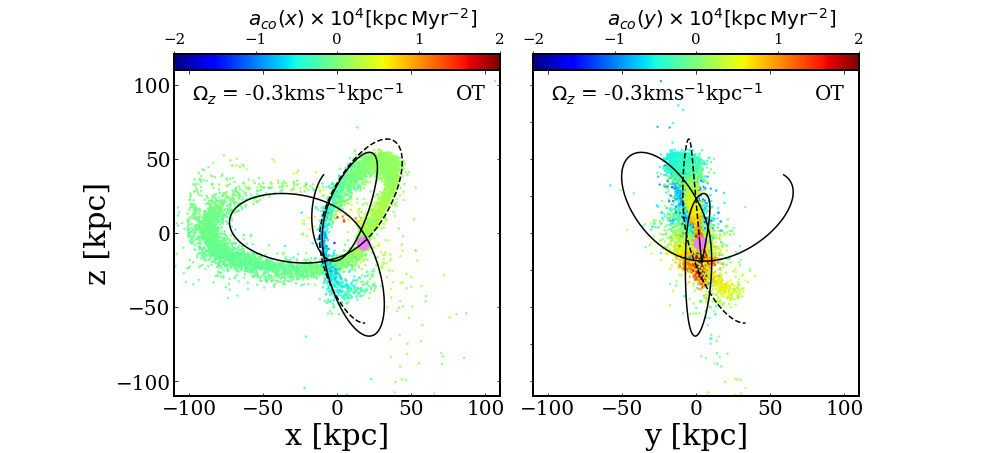} 
\includegraphics[angle=0,trim=530. 0. 40. 0., clip, width=0.325\textwidth]{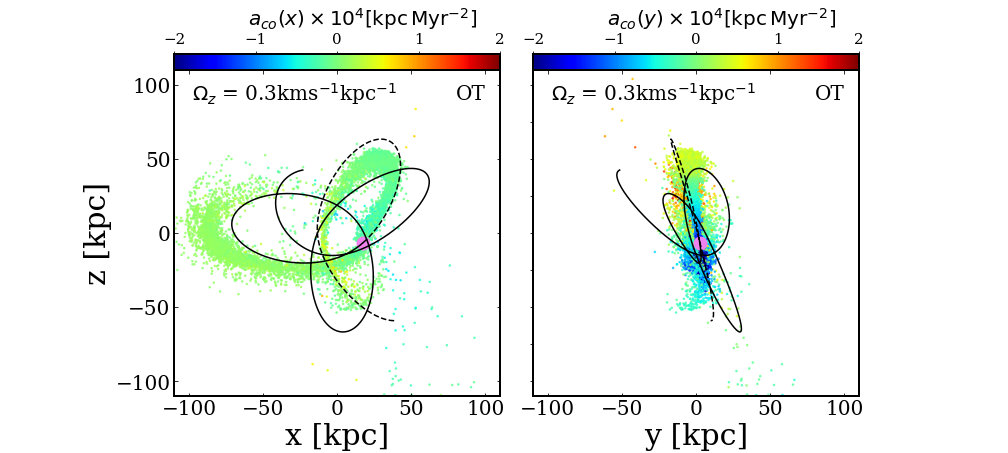} 
\includegraphics[angle=0,trim=30. 0. 125. 0., clip, width=0.63\textwidth]{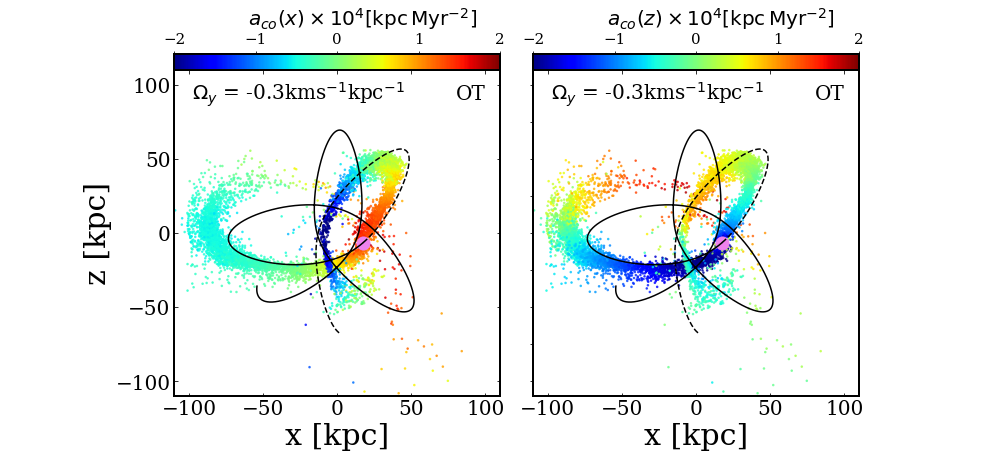} 
\includegraphics[angle=0,trim=530. 0. 40. 0., clip, width=0.325\textwidth]{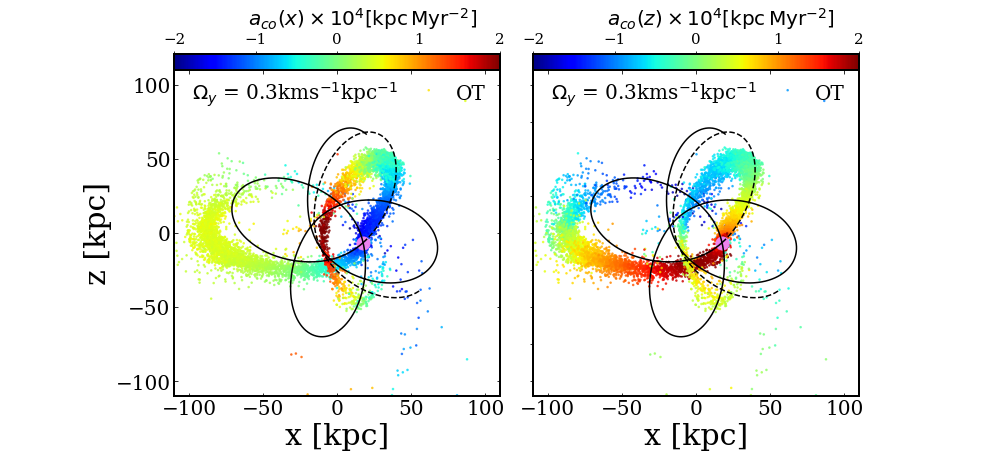} 
\end{center}
\caption{Galactocentric Cartesian frame projections of a Sgr-like tidal stream with points colored  by  Coriolis acceleration \aco\, along different axes as indicated by the color bar labels. Pattern speed as shown in legend. Solid (dashed) black lines show past (future) orbit of the Sgr dwarf whose current position is shown as mauve dot. Top row: left and middle panels show two projections of the stream and orbit for clockwise figure rotation; right panel shows $y-z$ projection for anti-clockwise rotation of the same magnitude. Top left panel: for rotation about  $z$-axis  \aco$(x) \sim 0$ everywhere.  Top middle  (right) panel shows the effect of figure rotation on the warping of the stream in the $y-z$ plane for clockwise (anti-clockwise) rotation about the $z$-axis. The southern-tip of the leading arm is warped to  positive (negative) $y$ values due to \aco$(y) <0$ ( \aco$(y) >0$). Bottom row:  left and middle panels show $x-z$ projection of orbit and stream with particles colored by $x$ and $z$ components of Coriolis force with clockwise rotation about $y$-axis. Bottom row, right shows the effect of anti-clockwise rotation on \aco$(z)$ . 
\label{fig:stream_forces}
}
\end{figure*}

In the rotating frame the pseudo forces (\vecacf\, and \vecaco)  alter the trajectory of an object relative to the trajectory in a static (non-rotating) potential. We focus here on the effects on orbits in triaxial potentials but note that orbits in other potentials (e.g. axisymmetric potentials -- oblate or prolate) rotated about a axis perpendicular to the angular momentum axis would also be altered by the Coriolis force and appear tilted in the rotating frame relative to orbits in static potentials in exactly the same manner as described in Section~\ref{sec:orbits_triaxial} (Figs.~\ref{fig:LAT} \& \ref{fig:SAT}).

The Sgr stream is on a polar orbit with net angular momentum about an axis approximately lying in the Galactic equatorial plane \citep{majewski_etal_03,law_majewski_16}.  In the current best fit models for the Sgr stream \citep{law_majewski_10, deg_widrow_12} the long-axis of the triaxial halo lies  about 7$^{\circ}$ away from the Galactocentric $y$-axis. The short-axis of the triaxial halo lies roughly along the Galactocentric $x$-axis (along the sun-Galactic center line) and the intermediate axis is aligned with Galactocentric $z$-axis perpendicular to the disk plane. This would put the Sgr dwarf on a long-axis tube orbit -- a  stable orbital configuration. However, since the Sgr stream has been evolving for less than 10 orbital periods, considerations of orbital stability under halo figure rotation are likely to primarily affect the coherence of the stream \citep{price_whelan_etal_16}.

Since the Sgr dwarf and its resultant tidal stream are on an almost planar orbit that lies approximately in the Galactocentric $x-z$ plane, we can simplify the discussion of the expected effects of figure rotation on the appearance of the Sgr stream by considering an orbit that lies  {\em exactly} in the $x-z$ plane in the stationary potential. (For such an orbit, $y \approx v_y \approx 0$.)  Table~\ref{tab:f_pseudo} gives explicit equations for the Coriolis and centrifugal accelerations (\vecaco, \vecacf) for rotation about each of the three principal axes and the approximate expressions for accelerations on an orbit lying in the $x-z$ plane.  The simulations however, compute  the exact orbits for the progenitor and stream particles assuming the current position of the Sgr-dwarf from \gaia\ DR2 \citep{gaiacollab_helmi_18,vasiliev_belokurov_20}.

From Table~\ref{tab:f_pseudo} we see that rotation about the $z$ or $x$ axes gives rise to \vecaco\ with non-zero components primarily along the $y$ axis, while rotation about the $y$ axis gives rise to both  $x$ and $z$ components of \vecaco. The effect of the centrifugal acceleration \vecacf\ 
is to push the stream away from the axis of rotation. Table~\ref{tab:f_pseudo} shows that rotation about the $z$ ($x$) axis causes the stream to be pushed outwards to larger $|x|$ ($|z|$), while rotation about the $y$ axis causes the entire stream to experience an outward radial centrifugal acceleration whose magnitude is linearly proportional to the distance from the Galactic center. Since the pattern speeds being considered in this paper are tiny ($\Omega_p = \pm0.3\kmskpc$) \vecacf\,  is much smaller than \vecaco. As will be shown later, these qualitative predictions for a strictly planar orbit are in general agreement with the behavior  seen for the simulated streams, which are not confined to the $x-z$ plane.

From Table~\ref{tab:f_pseudo} one sees that rotation about  the $z$ (or  $x$) axis results in significant Coriolis forces only along the $y$ axis. Therefore it primarily causes  {\em tilting}  of the orbit and consequently the plane of the tidal stream (which is approximately perpendicular to the Galactocentric $y$ axis). The top row, middle  panel of Figure~\ref{fig:stream_forces} shows that the southernmost tip of the stream experiences a negative (colored light blue)  \aco$(y)$ and is therefore pushed to negative $y$ values, while changing the sign of figure rotation (top row, right panel) causes this part of the  stream to experience positive (yellow-orange) \aco$(y)$ and hence it is pushed to positive $y$ values. These results are consistent with expectations from the analysis of the  tube orbits in Figs.~\ref{fig:LAT} \& ~\ref{fig:SAT}.

The Galactocentric $y$-axis is the angular momentum axis of the orbit in this figure. Rotation about this axis (bottom row of Fig.~\ref{fig:stream_forces}) results in significant Coriolis forces in the orbital plane of the stream. This is evident from the strongly varying colors of the stream particles in the 2nd row which show the Coriolis force in the $x$ direction (left panel) and $z$ direction (middle and right). As we showed in Section~\ref{sec:orbits_triaxial}, figure rotation about the angular momentum axis of the orbit changes the precession rate of the orbit and the angle between the lobes. It is clear from this figure that the Coriolis force in the $x-z$ plane has a strong effect on the shape of the {\em orbit} of the progenitor, particularly the angles between the  lobes of the rosette and the widths of the lobes.   In  the bottom row, the middle (right) panel shows that a positive (negative)  \aco$(z)$ pushes the orbit to more positive (negative) $z$ values. Similarly, the positive (negative)  \aco$(x)$ pushes the orbit to more positive (negative) $x$ values. While the stream approximately follows the orbit of the progenitor, we  can see from  Fig.~\ref{fig:stream_forces} that figure rotation causes clear and strong changes in the shape of the orbit. 

\section{Results}
\label{sec:results}

We now present results of simulations designed to study the effects of figure rotation on a Sgr-like stream evolved for 4~Gyrs in each of the four models described in  Section~\ref{sec:method}. For completeness we discuss rotation about each of the three principal axes of the Galaxy.

\begin{figure*}
\begin{center}
\includegraphics[trim=10. 60. 80. 60., clip, angle=0,width=0.364\textwidth]{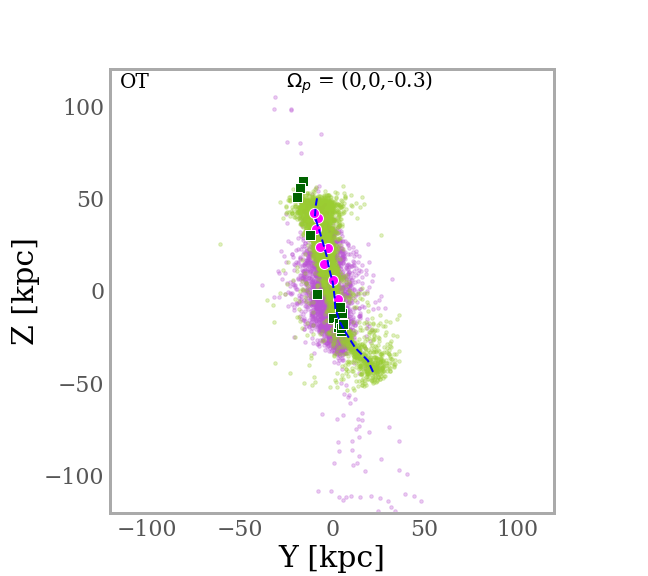} 
\includegraphics[trim=110. 60. 80. 60., clip, angle=0,width=0.30\textwidth]{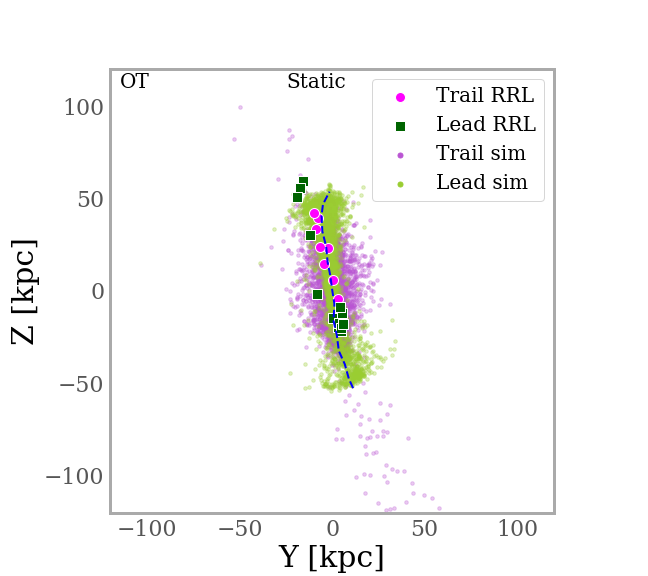} 
\includegraphics[trim=110. 60. 80. 60., clip, angle=0,width=0.30\textwidth]{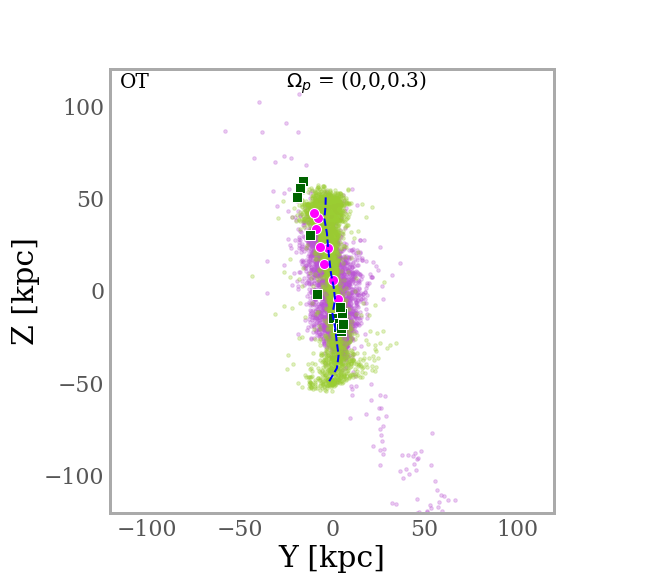}  
\includegraphics[trim=10. 60. 80. 60., clip, angle=0,width=0.367\textwidth]{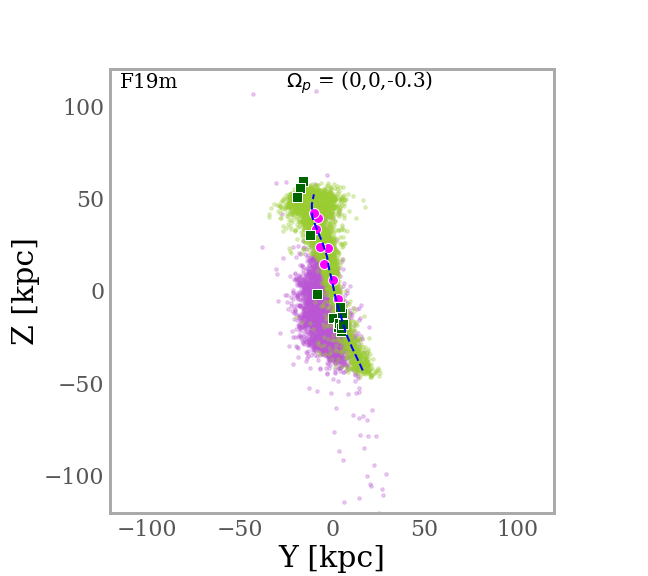} 
\includegraphics[trim=110. 60. 80. 60., clip, angle=0,width=0.30\textwidth]{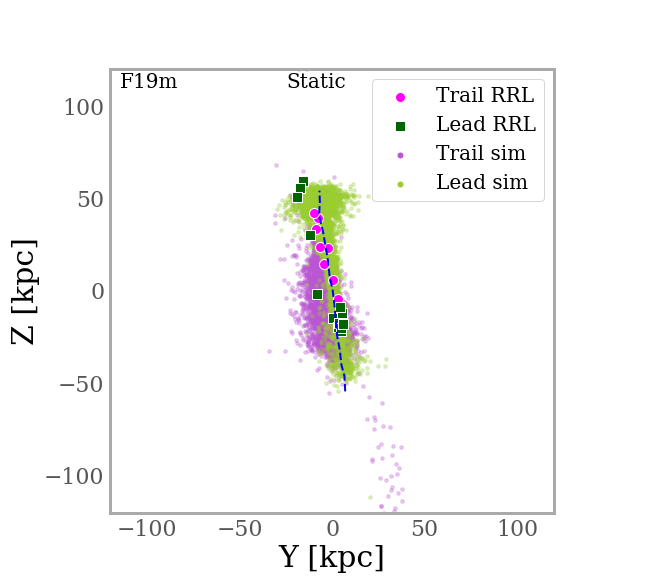} 
\includegraphics[trim=110. 60. 80. 60., clip, angle=0,width=0.30\textwidth]{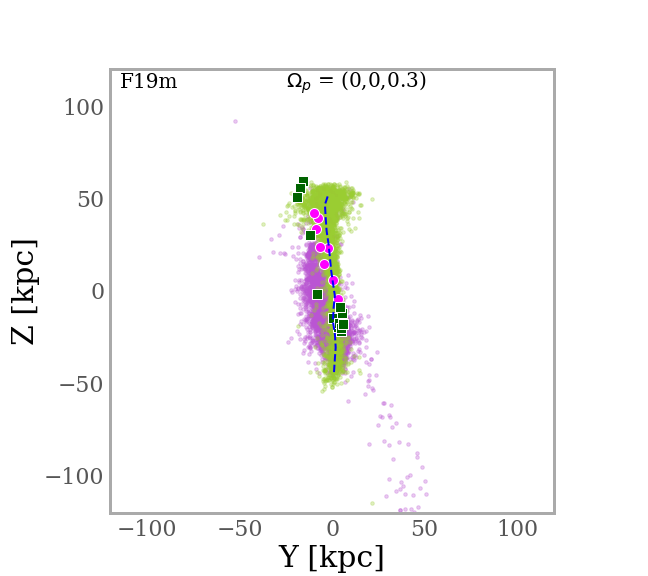} 
\includegraphics[trim=10. 0. 80. 60., clip, angle=0,  width=0.367\textwidth]{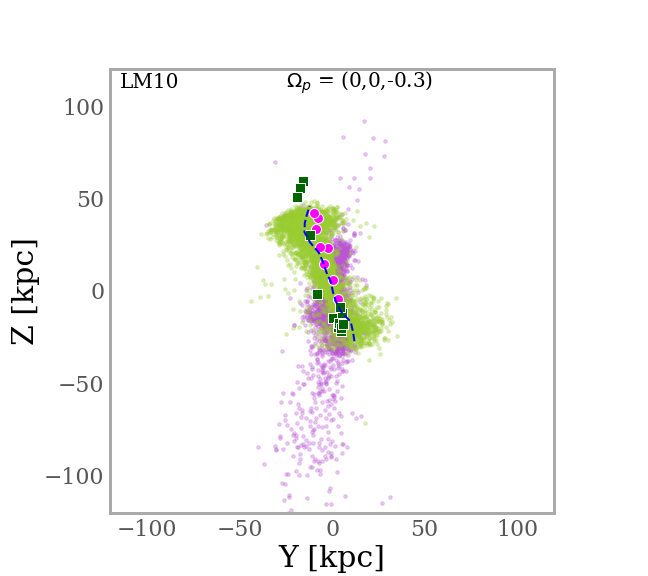} 
\includegraphics[trim=110. 0. 80. 60., clip, angle=0,  width=0.30\textwidth]{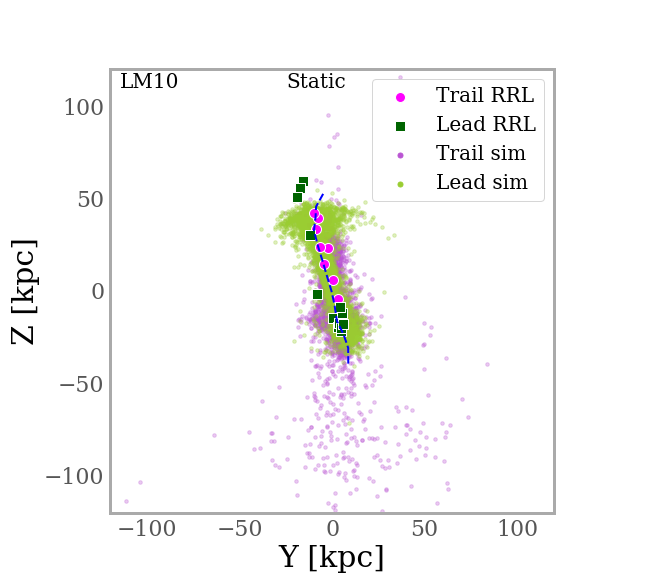} 
\includegraphics[trim=110. 0. 80. 60., clip, angle=0,  width=0.30\textwidth]{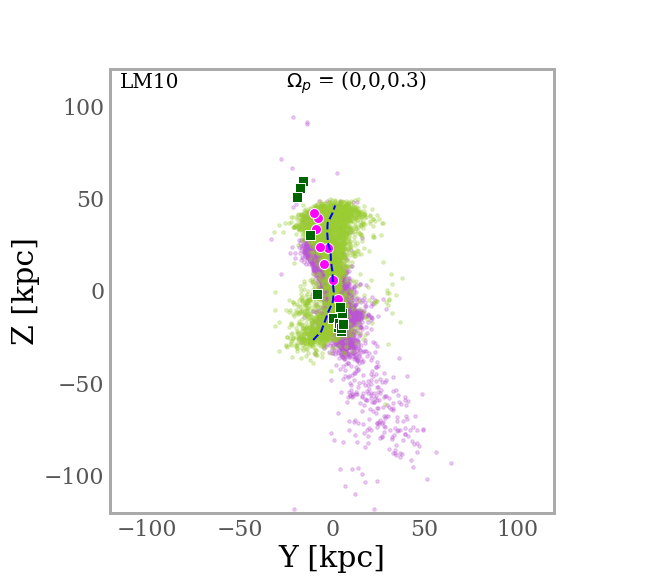} 
\end{center}
\caption{Cartesian ($y-z$) projection of the stream for static models (middle column) or rotating models with pattern speed as in labels, for \OT, \Fm, \LM\ models. {The black dashed curve in each plot shows the median $y$-coordinate of the {\it leading arm } of the simulated stream in each panel. The black curve guides the eye showing the warp of the southern-most end of the stream is warped in opposite directions when halo rotation is reversed. } Median positions of observed Sgr stream RRLyrae stars from PanSTARRS \citep{hernitschek_etal_17} are shown for the leading arm (large squares) and  trailing arm (large circles).
\label{fig:YZ_OT_F19_z}
}
\end{figure*}

The primary effects of rotation about the $z$ or $x$ axes arise from the Coriolis force in the $y$ direction (see Table~1) which causes the warping of the northern most and southern most tips of the stream, as previously shown in Figure~\ref{fig:stream_forces} (top row middle and right panels). Similar effects are seen in all the models. Figure~\ref{fig:YZ_OT_F19_z} shows  $yz$-projections of Sgr-like streams evolved in the \OT\ model (top row), the \Fm\ model (middle row) and the \LM\ model (bottom row). The middle column  shows the  stream in a static halo for each model while the left and right columns show streams in halos with pattern speeds as indicated by ${\mathbf\Omega_p}$. {Small  dots show simulated stream stars in the leading  and trailing arms of the simulated stream with colors as indicated in the legend in the ``Static'' panels. The median location of the leading arm of the simulated stream is shown by the black dashed curve in each panel. As can be seen from the median stream positions, clockwise rotation of the halo (left column) in all 3 models pushes the southern end of the stream to positive $y$ values while anti-clockwise rotation (right column) does the opposite.} For reference the median positions of RRLyrae stars of the Sgr stream from PanSTARRS \citep{hernitschek_etal_17} are shown for both the leading arm (large squares) and trailing arms (large circles). In all three models the stream in the static halo is slightly tilted relative to the $z$-axis. Although the effects of rotation are subtle it is clear that clockwise rotation about $z$ (left column) in all three halos causes simulated stars in the leading arm at negative $z$ values to be shifted towards positive $y$ values while anti-clockwise rotation (right column) causes the same stars to be pushed towards negative $y$ values. The warping in the plane of the leading arm is a result of the $y$-component of the Coriolis force being greatest  at the point where the $v_x$ is largest. As expected from Table 1 (first column) the direction of the warping is reversed when the sense of rotation is reversed. 

In all the models we see that the planes of the leading and trailing arms become slightly  misaligned  (especially for clockwise rotation, see left column). This is because most of the leading arm stars (except for those at the northern and southern tips) are moving along the $z$ axis during which time they experience no Coriolis acceleration. In contrast, trailing arm stars (see Fig~\ref{fig:Pol_LMm_rxyz}) are moving primarily along the $x$ axis and therefore experience a larger Coriolis acceleration along the $y$ axis. This is the primary cause of the misalignment of the planes which contain the leading and trailing arms. Thus it is clear from this figure that figure rotation can result in subtle, but predictable, changes to the morphology of a Sgr-like stream. 

We now examine the effect of figure rotation about each of the three principal axes in the \LMm\ model. As mentioned previously this model has the same halo shape as the \LM\ model, but the masses of the disk and halo are lower, resulting in a trailing arm that extends to a much larger Galactocentric radius, and therefore gives a better match to  observations of BHB stars \citep{belokurov_etal_14} and RR-Lyrae stars from PanSTARRS \citep{hernitschek_etal_17} at the trailing apocenter.

\begin{figure*}
\begin{center}
\includegraphics[trim=10. 60. 80. 50., clip, angle=0,width=0.364\textwidth]{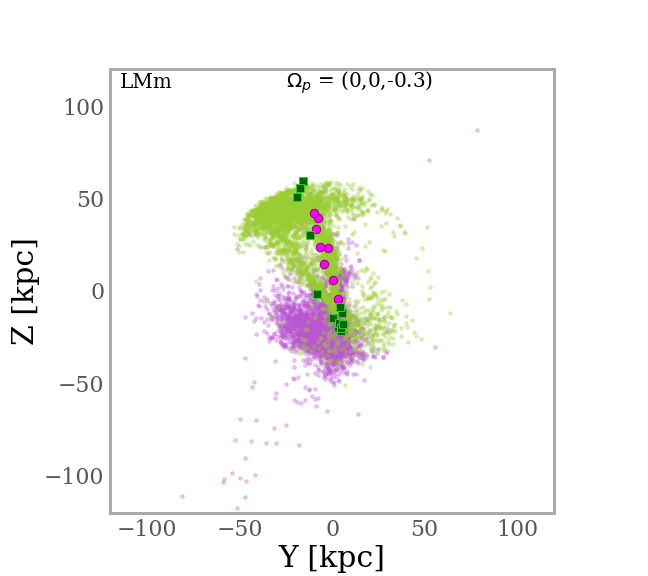} 
\hspace{0.24\textwidth}
\includegraphics[trim=10. 60. 80. 50., clip, angle=0,width=0.37\textwidth]{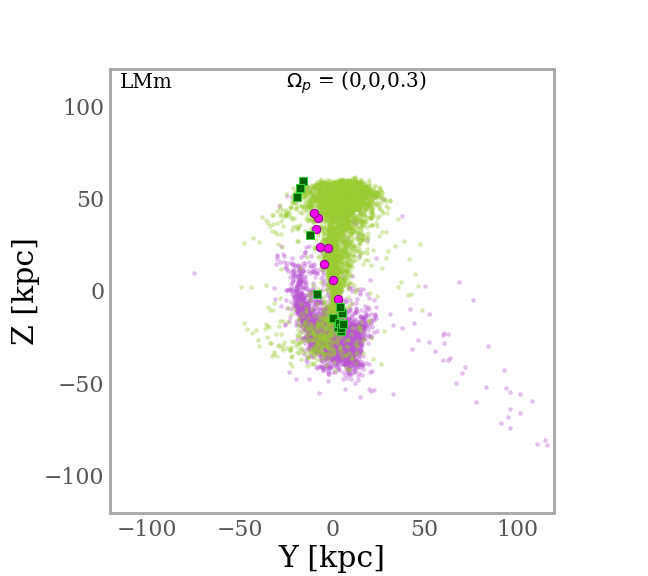}  
\includegraphics[trim=10. 60. 80. 50., clip, angle=0,width=0.364\textwidth]{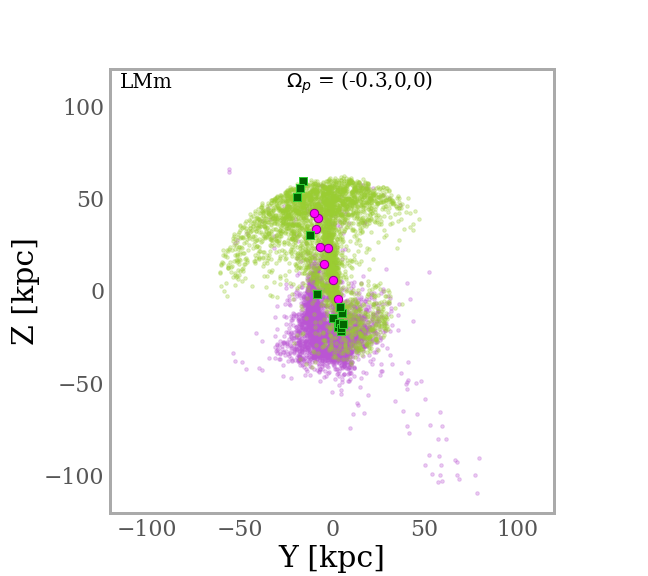} 
\hspace{0.24\textwidth}
\includegraphics[trim=10. 60. 80. 50., clip, angle=0,width=0.37\textwidth]{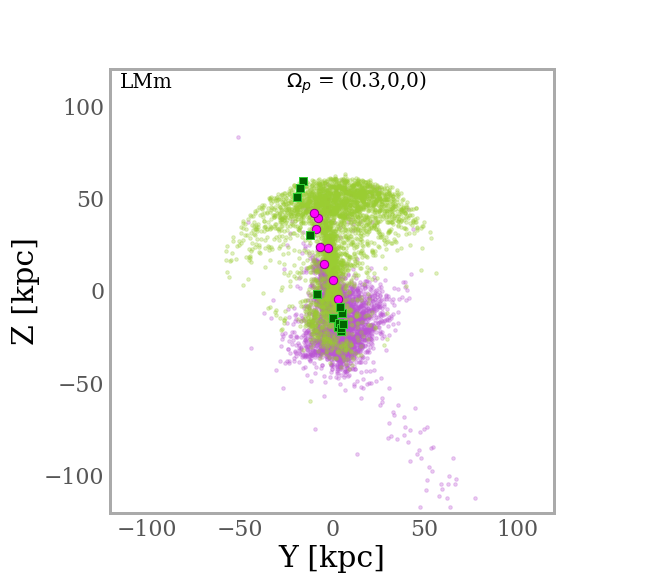} 
\includegraphics[trim=10. 0. 80. 50., clip, angle=0,width=0.370\textwidth]{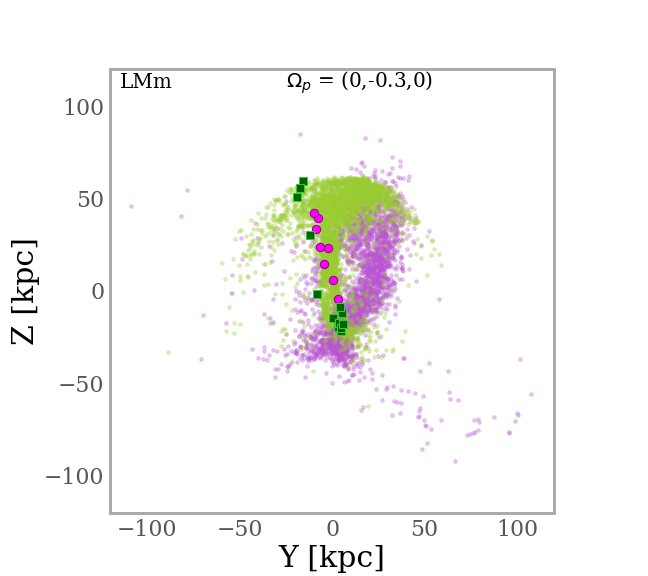} 
\includegraphics[trim=110. 0. 80. 50., clip, angle=0,width=0.305\textwidth]{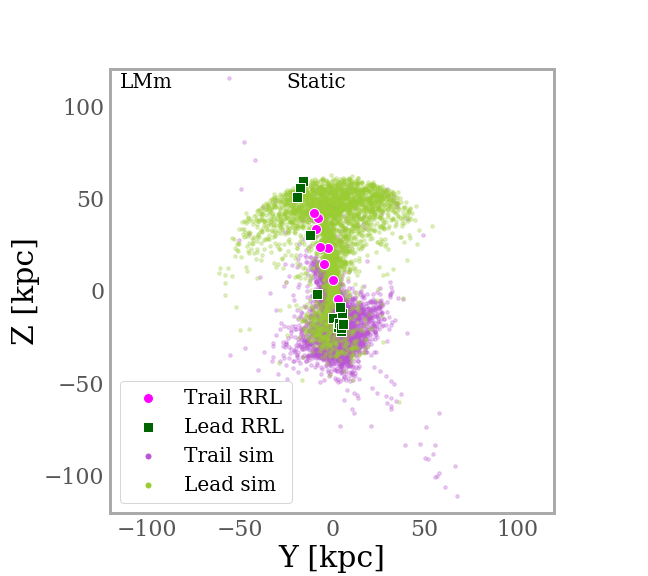}  
\includegraphics[trim=110. 0. 80. 50., clip, angle=0,width=0.305\textwidth]{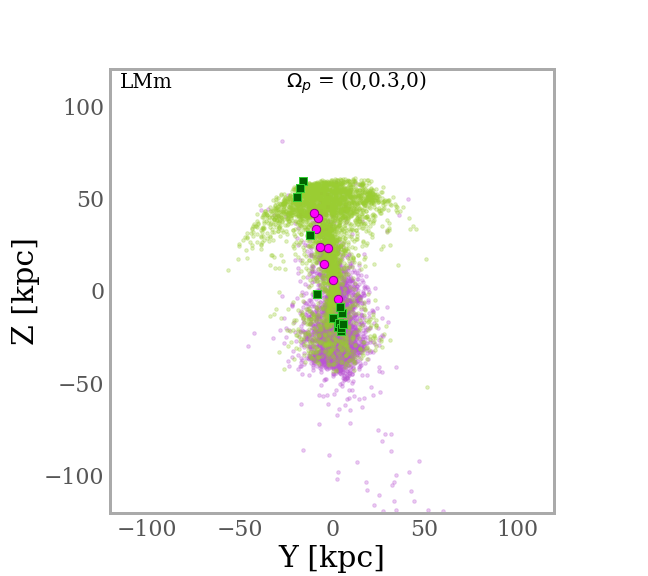}  
\end{center}
\caption{Cartesian plot in $y-z$ plane for streams in the  \LMm\ model rotated about each of the three principle axes: $z$-axis (top), $x$-axis (middle row), $y$-axis (bottom row). Rotation axes and magnitudes as shown by the labels.
\label{fig:YZ_LMm_allax}
} 
\end{figure*}

Figure~\ref{fig:YZ_LMm_allax} shows $yz$-projections of the \LMm\ model for clockwise (anti-clockwise) rotation about three different axes shown in the left (right) columns, with the static model shown in the middle of the bottom row. The top row  shows that rotation about the $z$ axis causes dramatic warping and misalignment of the leading arm and trailing arm of the stream with the direction of the tilting of the leading arm reversing when the sign of rotation flips. Rotation about the $x$ axis  (middle row), which is perpendicular to the plane of the figure, causes  fanning of the  northern-most end of the leading arm of the stream, but less misalignment between the leading and trailing arm planes.  Recall that the halo of the \LMm\ model has its short axis along this ($x$) axis. The greatest asymmetry between clockwise and anti-clockwise rotation is seen in the bottom row which shows rotation about the $y$ axis (long-axis of the \LMm\ and \LM\ halos). It is particularly striking that clockwise rotation (bottom row, left) cause the plane of the trailing arm to deviate very strongly from that of the leading arm.  

Once again we see that the primary signatures of the effect of the Coriolis force on a Sgr-like stream are to warp the stream and cause more significant misalignment (precession) between the planes containing the leading and trailing arms of the stream, regardless of the axis of rotation.

\begin{figure*}
\begin{center} 
\includegraphics[trim=40. 80. 130. 0., clip, angle=0,width=0.29\textwidth]{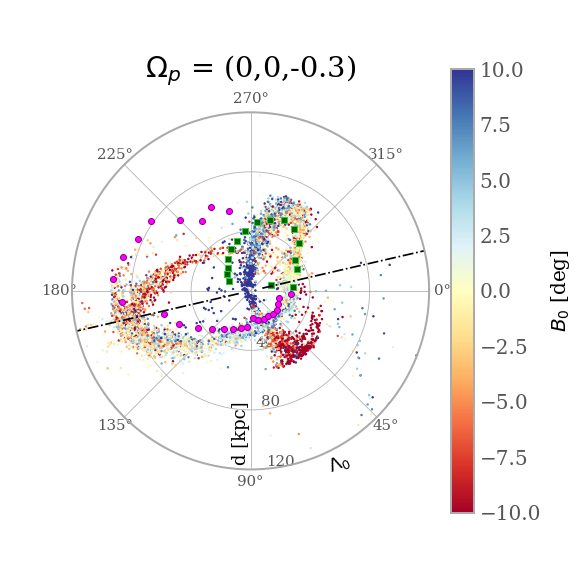} 
\includegraphics[trim=40. 80. 130. 0., clip, angle=0,width=0.29\textwidth]{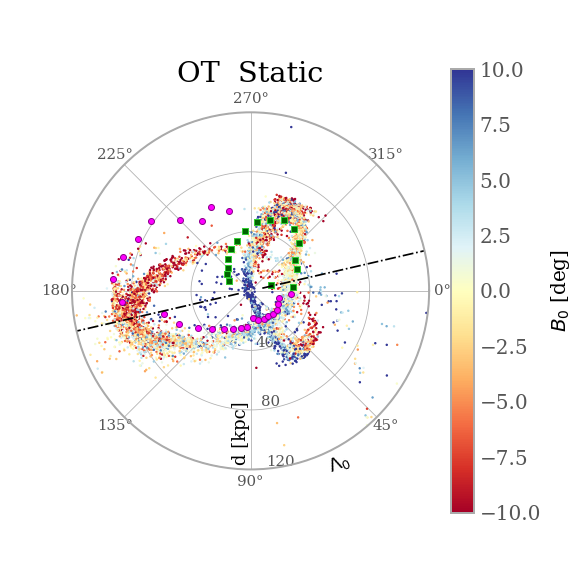} 
\includegraphics[trim=40. 80. 0. 0., clip, angle=0,width=0.39\textwidth]{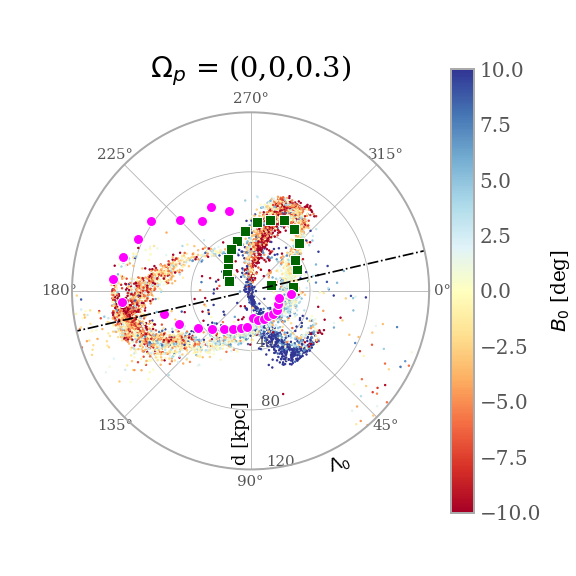} 
\includegraphics[trim=40. 80. 130. 0., clip, angle=0,width=0.29\textwidth]{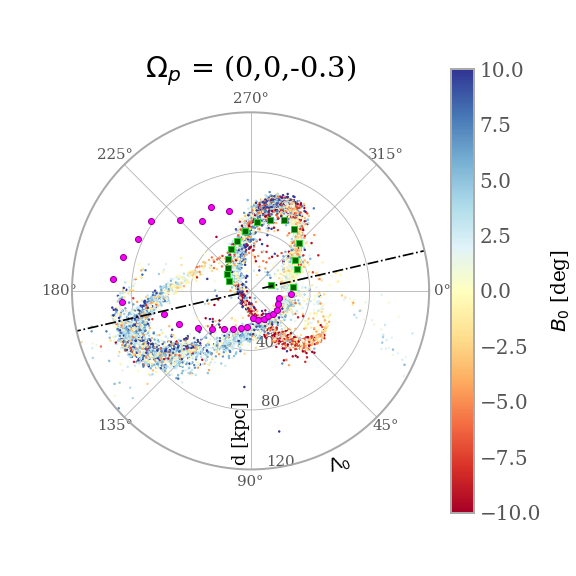}
\includegraphics[trim=40. 80. 130. 0., clip, angle=0,width=0.29\textwidth]{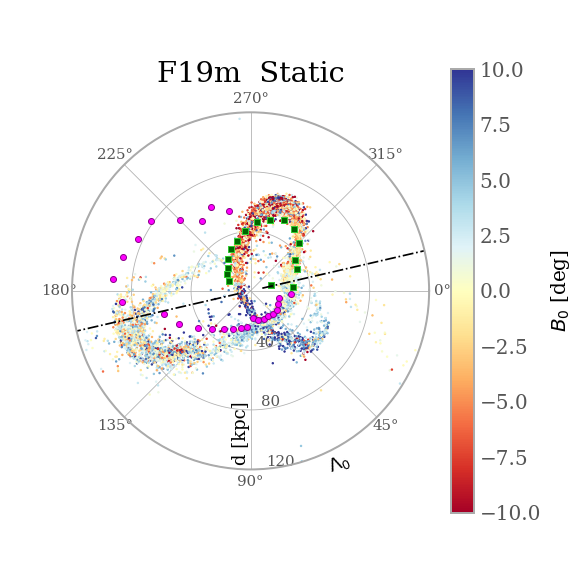} 
\includegraphics[trim=40. 80. 0. 0., clip, angle=0,width=0.39\textwidth]{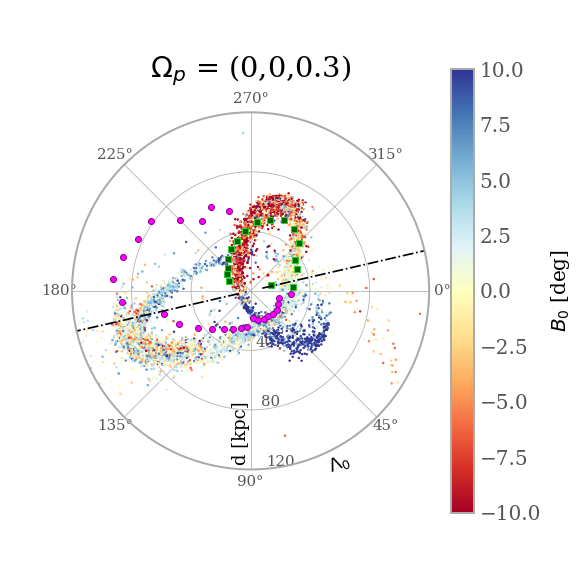} 
\includegraphics[trim=40. 80. 130. 0., clip, angle=0,width=0.29\textwidth]{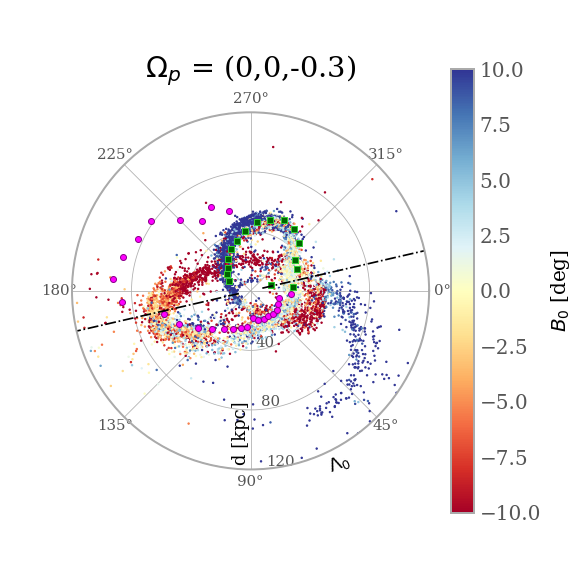} 
\includegraphics[trim=40. 80. 130. 0., clip, angle=0,width=0.29\textwidth]{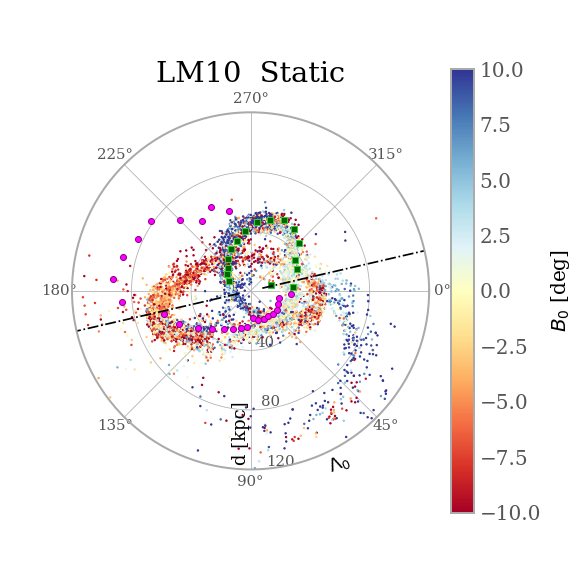} 
\includegraphics[trim=40. 80. 0. 0., clip, angle=0,width=0.39\textwidth]{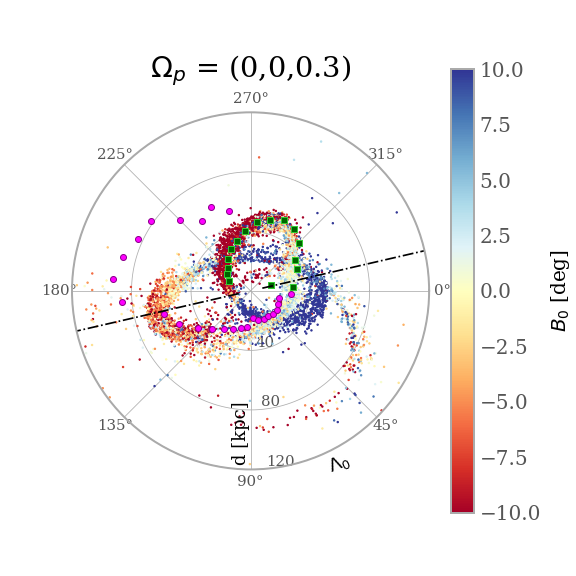} 
\end{center}
\vspace{-.4cm}
\caption{Polar plot for models \OT (top row), \Fm (middle), \LM (bottom) with polar angle showing Sgr-coordinate $\Lambda_0$ clockwise from the current position of the Sgr dwarf ($\Lambda_0=0$), and radial coordinate showing heliocentric distance in kpc. The stream is shown for clockwise rotation (right), static (middle), anti-clockwise rotation about the $z$ axis. Stream particles are colored by their angular distance $B_0$ [ degrees] from the Sgr-stream great-circle plane. Median positions of observed RRLyrae stars \citep{hernitschek_etal_17} are shown by large squares (leading arm) and large  circles (trailing arm).The Galactic plane is marked by a dot-dashed line. 
\label{fig:Pol_all_rxyz}
}
\end{figure*}
\begin{figure*}
\begin{center} 
\includegraphics[trim=40. 80. 130. 0., clip, angle=0,width=0.29\textwidth]{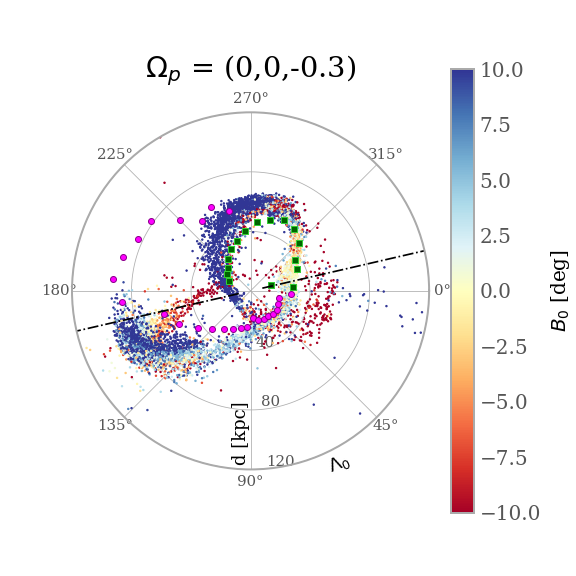} 
\hspace{0.29\textwidth}
\includegraphics[trim=40. 80. 0. 0., clip, angle=0,width=0.39\textwidth]{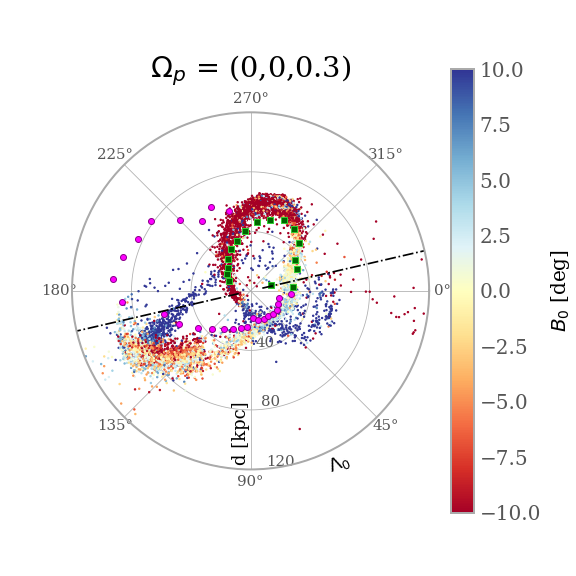} 
\includegraphics[trim=40. 80. 130. 0., clip, angle=0,width=0.29\textwidth]{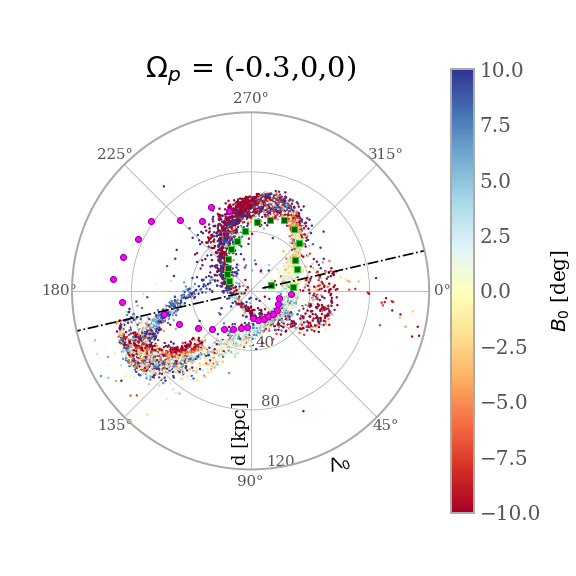}
\includegraphics[trim=40. 80. 130. 0., clip, angle=0,width=0.29\textwidth]{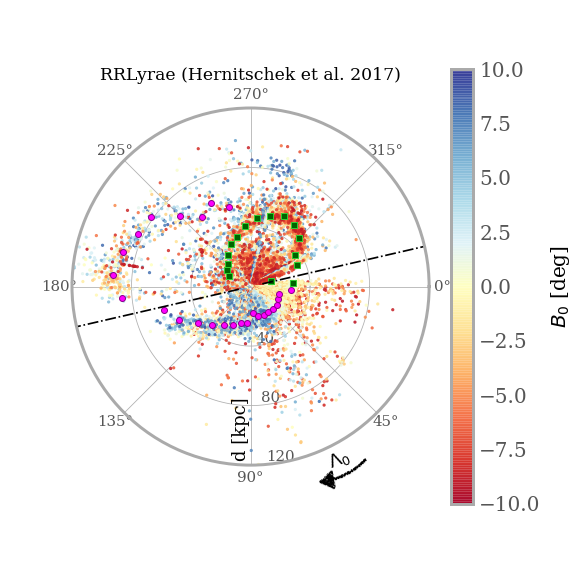} 
\includegraphics[trim=40. 80. 0. 0., clip, angle=0,width=0.39\textwidth]{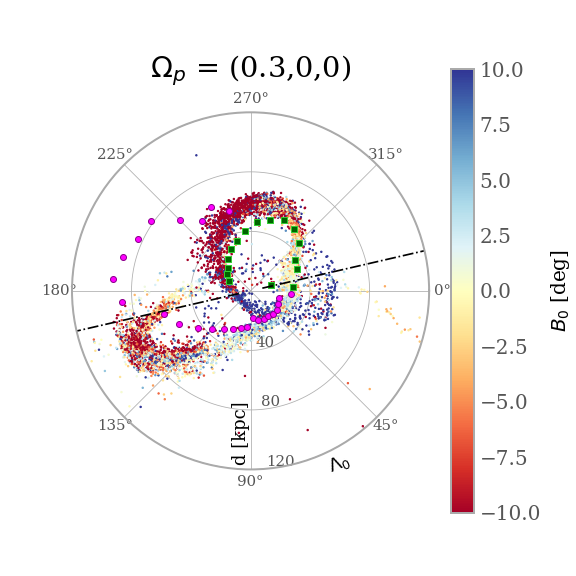} 
\includegraphics[trim=40. 80. 130. 0., clip, angle=0,width=0.29\textwidth]{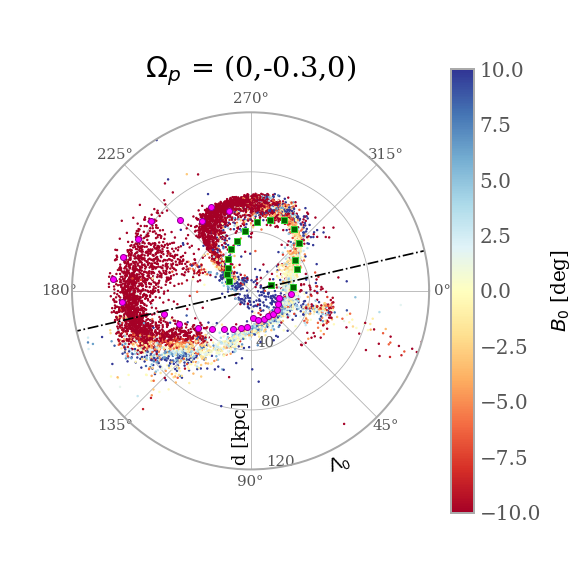} 
\includegraphics[trim=40. 80. 130. 0., clip, angle=0,width=0.29\textwidth]{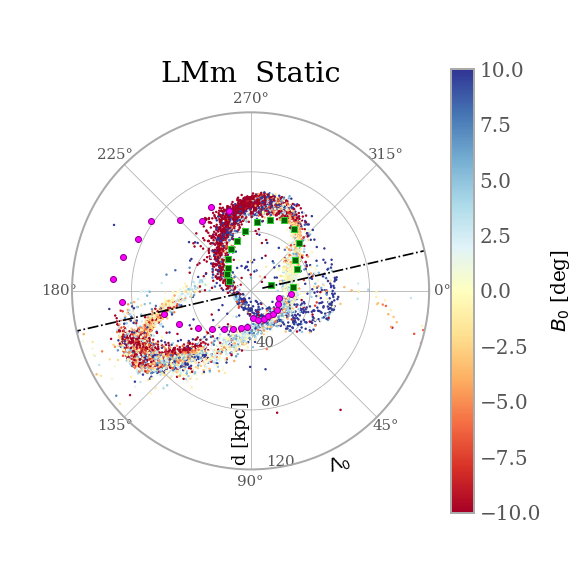} 
\includegraphics[trim=40. 80. 0. 0., clip, angle=0,width=0.39\textwidth]{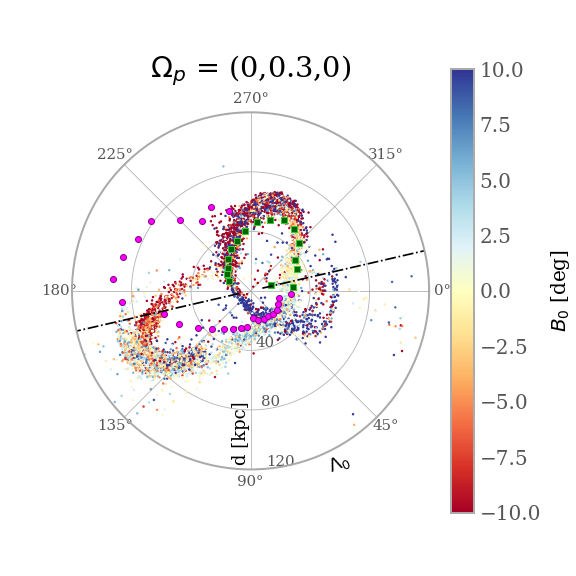} 
\end{center}
\vspace{-.4cm}
\caption{
Similar to Fig.~\ref{fig:Pol_all_rxyz} for \LMm\ model for rotation about each of the three principle axes: $z$ (top), $x$ (left and right of 2nd row), $y$ (left and right of bottom row. The stream in the static \LMm\ model is shown in the bottom middle panel. The central panel shows 15050 individual observed RR-Lyrae stars from \citet{hernitschek_etal_17}.
\label{fig:Pol_LMm_rxyz}
}
\end{figure*}

Figure~\ref{fig:Pol_all_rxyz} shows the simulated Sgr streams in Figure~\ref{fig:YZ_OT_F19_z} in a polar plot with Sgr great-circle coordinate \citep[$\Lambda_0$,][]{majewski_etal_03} in the angular direction and heliocentric distance in the radial direction. The dot-dashed line in each panel marks the orientation of the Galactic plane. Following \citet{majewski_etal_03} the angular Sgr stream coordinate $\Lambda_0$ increases clockwise from $\Lambda_0=0$ (which marks the position of the Sgr dwarf) and is offset by 13$^\circ$ clockwise from the Galactic plane (shown by the dot-dashed line).  Simulated stream stars (small dots) are colored by their Sgr coordinate $B_0$, which is the angle in degrees perpendicular to the Sgr great-circle plane defined by $B_0=0$. Coloring the stream by $B_0$ provides a 3D view of how figure rotation alters the warping and misalignment (or precession) in the planes containing the leading and trailing arms.  To further aid comparison with the observed Sgr stream we also show observed median positions of RRLyrae stars along the leading arm (large squares) and trailing arm (large circles) from PanSTARRS \citep{hernitschek_etal_17}. This polar plot is close to what would be observed if the stream was plotted in the $x-z$ plane. As this figure shows, all the simulated streams do a reasonably good job of matching parts of the observed stream but none of them match the the RR-Lyrae data points precisely. In particular we see that the \LM\, model (bottom row) produces a trailing arm with too small an apocenter, although it does the best job of matching the leading arm. 

In Figure~\ref{fig:Pol_LMm_rxyz}, the simulated stream in the \LMm\ model is shown for rotation about each of the three principal axes $z$, $x$, $y$ (from top to bottom) with  pattern speeds (left to right) $\Omega_p =  -0.3, 0, 0.3$\kmskpc. The middle panel in the 2nd row shows  15,050 individual RRLyrae  observed with PanSTARRS from the catalog published by \citet{hernitschek_etal_17}. The observed and simulated stream stars  are both shown on  polar plots as in Figure~\ref{fig:Pol_all_rxyz}.

The heliocentric distance of the apocenter of the leading arm in all the models is greater than the observed apocenter distance (marked by large squares), most likely because we do not include the effects of dynamical friction from the Milky Way's dark matter halo on the Sgr dwarf \citet{fardal_etal_19} {or the reflex motion of the Milky Way's center-of-mass in response to the gravitational field of the LMC \citep{Vasiliev_tango}. }

Nonetheless, the middle panel shows that, like the simulated streams, there are substantial gradients in $B_0$ across the observed stream stars, with the leading arm lying primarily at negative $B_0$ values (the observed RRLyrae stars of this arm are marked by large squares) and the trailing arm (marked by large circles) lying primarily at positive $B_0$, except at trailing apocenter which is at $B_0\lesssim0$. It is clear (from the colors of the points) that rotation about the $z$-axis (top row) causes the plane of the leading arm  ($225^\circ <\Lambda_0 < 315^\circ$, marked by large squares)  to  be warped so that stars at the leading apocenter lie at  $B_0>0$ for negative figure rotation (left column) and $B_0>0$ for positive figure rotation (right column).

None of the simulated streams matches the observed angle between the leading and trailing apocenters. As discussed in Section~\ref{sec:method} it has previously been shown that this angle is determined by the radial density profile of the dark matter halo \citep{belokurov_etal_14, fardal_etal_19} and that cored dark matter halos with larger scale lengths are needed to produce the observed angle of $\sim 95^{\circ}$ between the apocenters. We have kept $r_s=28$~kpc fixed for all of our models. Therefore, none of our models give the correct angle between the apocenters. 

We showed in Figures~\ref{fig:LAT} (2nd row), \ref{fig:SAT} (4th row) \& \ref{fig:stream_forces} (2nd row) that figure rotation about the angular momentum axis of a tube orbit can alter the precession rate and the angle between successive apocenters of orbits. The bottom left panel  shows that this causes the angle between apocenters  of the leading/trailing arms  of the tidal stream to also  be altered by figure rotation. While the static model (bottom row, middle)  fails to produce streams that match the observed positions of stars near trailing apocenter ($135^\circ <\Lambda_0 < 225^\circ$, marked by large circles), rotation about the $y$ axis (bottom row) with $\Omega_p = -0.3$ (left column) results in a Coriolis force that push the southern part of the trailing arm  towards Galactocentric north resulting in a slightly closer match to the observed stream.  The colors of the trailing stream stars (bright red) in this panel show that stars at trailing apocenter and beyond are pushed to negative $B_0$, which is not observed for the PanSTARRS RRLyrae (middle panel, 2nd row).  While this implies that rotation about the $y$-axis may not be adequate to change the angle between the apocenters, it certainly has a strong enough effect that it should be considered in future models, since it could allow for a halo with less extreme values of $r_s$ than found by \citet{fardal_etal_19}.

\begin{figure*}
\begin{center}
\includegraphics[trim=30. 45. 0. 40., clip, angle=0, width=0.45\textwidth]{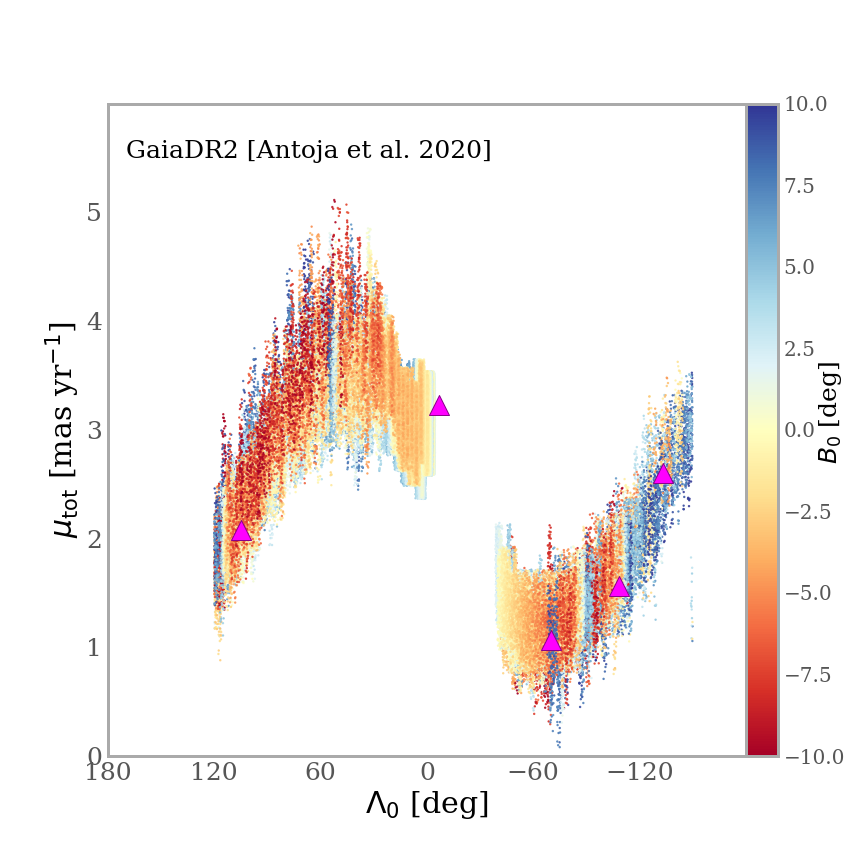} \\
\includegraphics[trim=30. 90. 116. 75., clip, angle=0, width=0.39\textwidth]{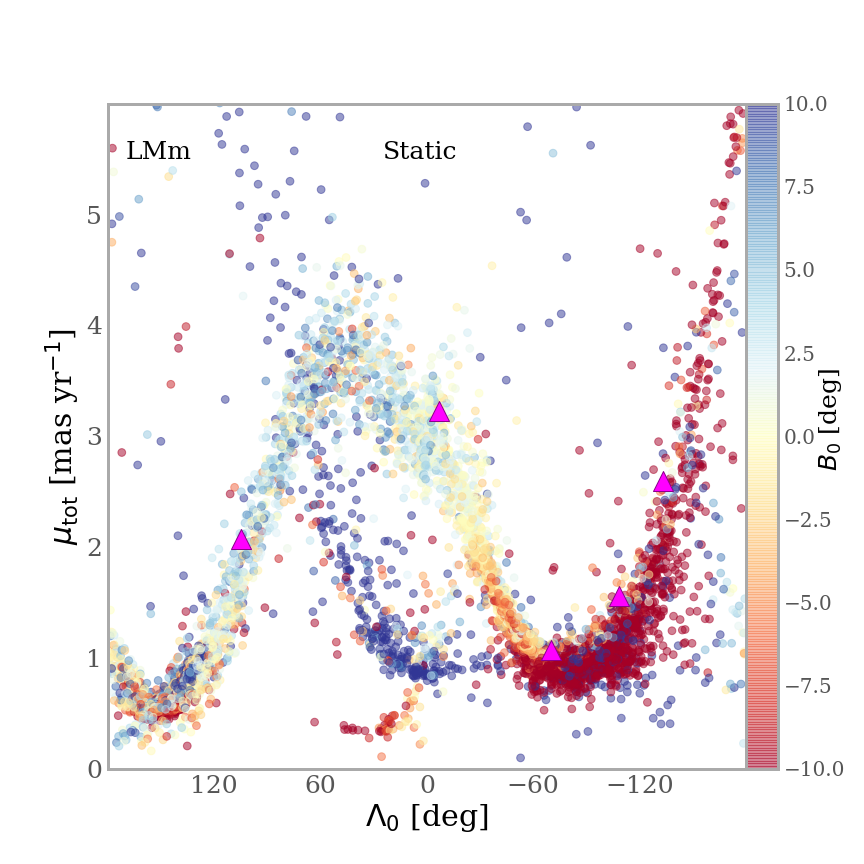} 
\includegraphics[trim=100. 90. 116. 75., clip, angle=0, width=0.352\textwidth]{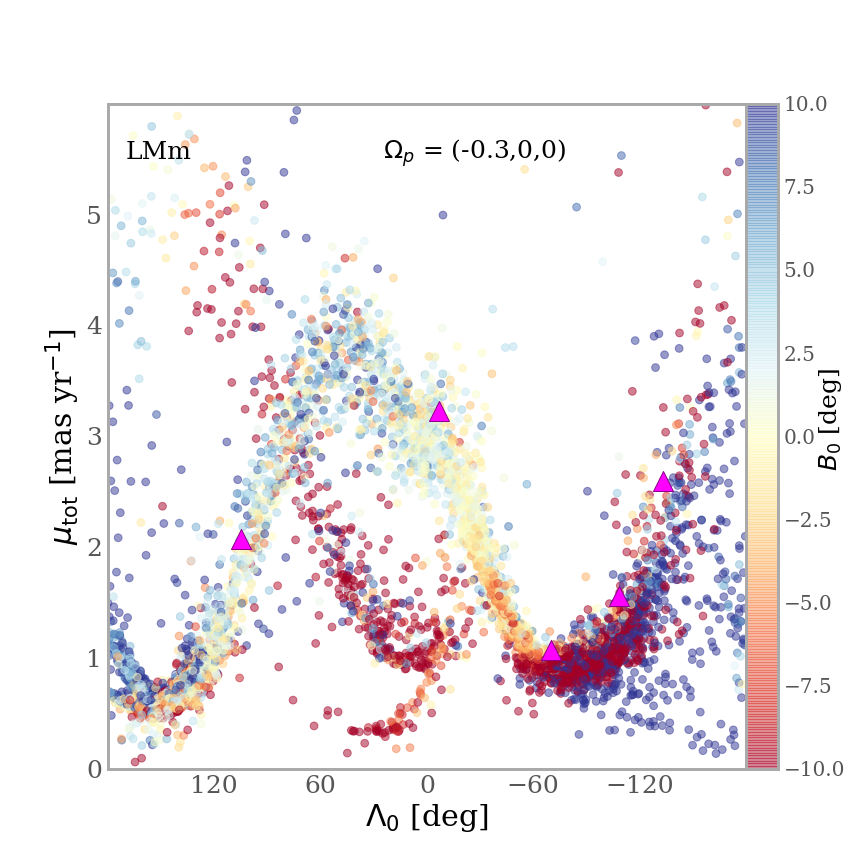} \\
\vspace{-0.05cm}
\includegraphics[trim=30. 33. 116. 85., clip, angle=0,width=0.39\textwidth]{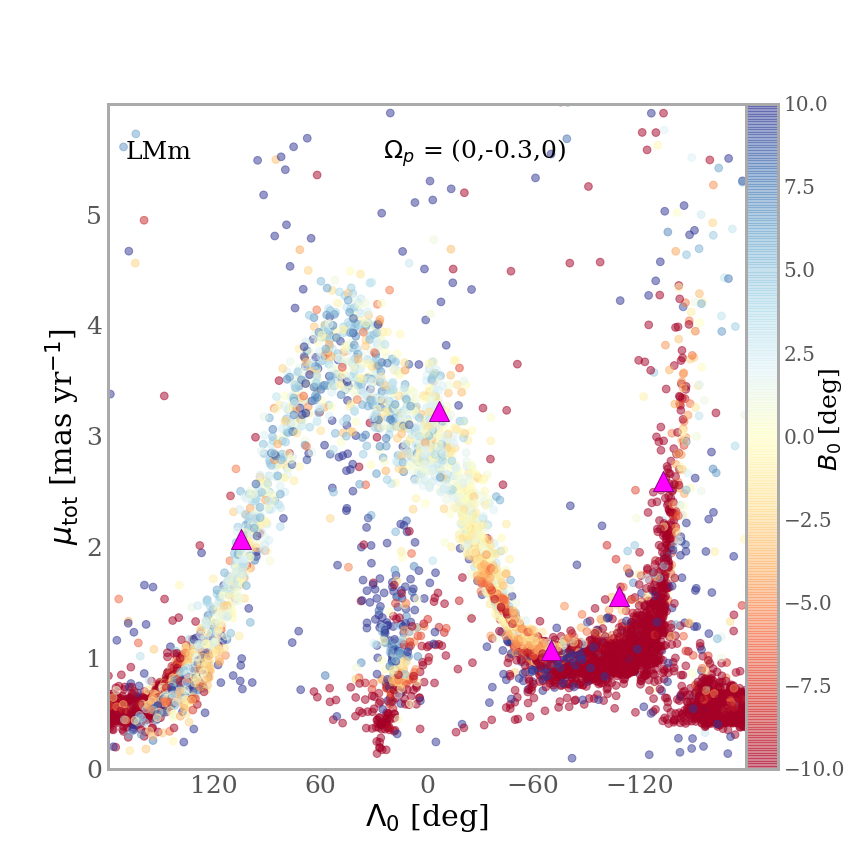} 
\includegraphics[trim=100. 33. 116. 85., clip, angle=0,width=0.352\textwidth]{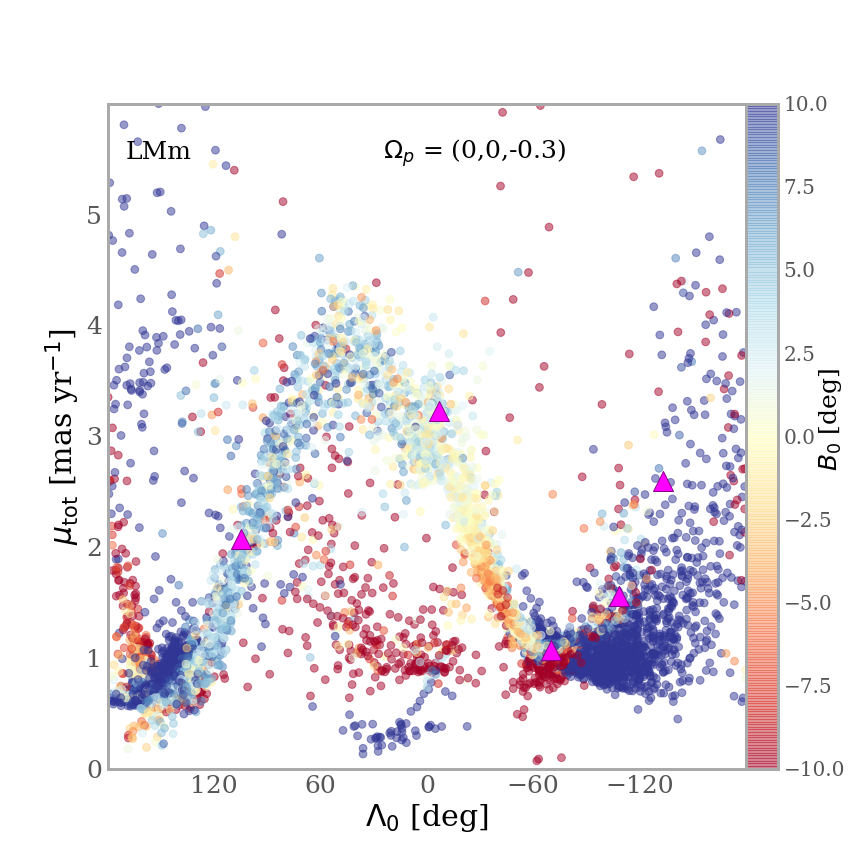} 
\caption{$\mu_{\rm tot}$ vs. $\Lambda_0$, with color bar signifying $B_0$. Top row shows observed total proper motion as a function of $\Lambda_0$  from  \gaia\ DR2 (Table~E.1. of \citet{antoja_etal_20}).  2nd row left panel shows simulated stream in a static \LMm\, model. The other three panels show streams in \LMm\ model with clockwise rotation about $x$, $y$, $z$ axes respectively. $\Lambda_0$ is plotted from positive to negative values to enable easier comparison with $\tilde{\Lambda}_\odot$ \citep{belokurov_etal_14}. The five triangles mark HST proper motions \citep{sohn_etal_15, sohn_etal_16}.
\label{fig:Lam_pm}}
\end{center}
\end{figure*}

Figure~\ref{fig:Lam_pm}  
shows the total proper motion of  stream stars vs. $\Lambda_0$ for the observed Sgr stream  from \gaia\ DR2 observations \citep{antoja_etal_20}\footnote{These authors only provide what they consider reliable Sgr stream data between $-150^\circ < \Lambda_0 <120^\circ$.} (top row). The color bar shows $B_0$ in degrees from the Sgr great-circle plane. The other panels  in Figure~\ref{fig:Lam_pm}  show the stream in the \LMm\ model with rotation axes and pattern speed as indicated by the labels. In this figure $\Lambda_0$ is plotted in reverse to match $\tilde{\Lambda}_0$ \citep{belokurov_etal_14} which increase anti-clockwise per standard convention. The color bar shows $B_0$ on the same scale for both simulated Sgr stream  and the observed stream. In each panel the five triangles  show the proper motions in five fields from \citep{sohn_etal_15, sohn_etal_16,fardal_etal_19} based on Hubble Space Telescope observations.

While none of the simulated streams shown produce both the amplitude and gradient of deviation of the stream from the $B_0=0$ great circle plane that is observed, rotation about each of the three principal axes  changes the gradient in $B_0$ over some parts of the stream. Nonetheless two facts are immediately clear. First, the overall sinusoidal shape of the observed total proper motions as a function of $\Lambda_0$ along the stream is broadly in agreement with all of the simulated streams (the simulated streams show both wraps of the stream which are not shown for the observations).   \citet{antoja_etal_20} found similar broad agreement between the observed proper motions for Sgr stream stars and  the N-body simulation from \citet{law_majewski_10}.  Second, the observed streams show a substantial gradient in $B_0$ along the stream especially in the range $-60 > \Lambda_0 > -120$, starting at fairly negative values of $B_0$ (red/oranges at $\Lambda_0 \sim -60$) at  and increasing to positive values of $B_0$ (blues at $\Lambda_0 \sim -120$).  While this precise gradient in $B_0$ is not seen over this stretch of the stream in any of the simulations shown, it is clear that the static model (top row, right panel) does not show such a gradient and is strictly at negative $B_0$. Only clockwise rotation about the $y$-axis causes this part of the steam to show a gradient going from negative $B_0$ to positive $B_0$.

\begin{figure*}
\begin{center} 
\includegraphics[trim=40. 80. 130. 0., clip, angle=0,width=0.386\textwidth]{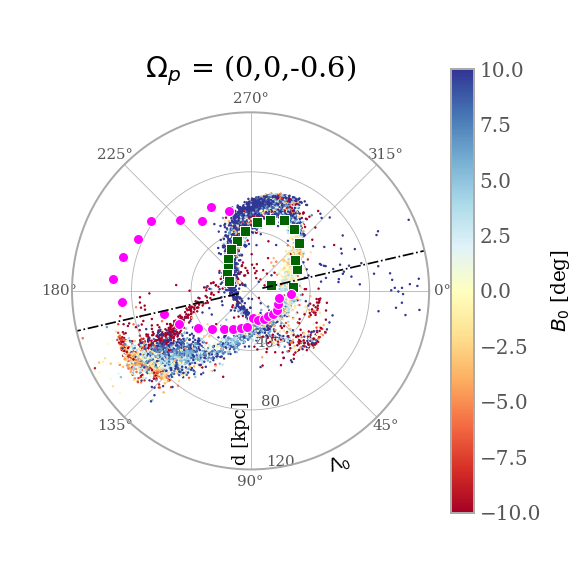}
\includegraphics[trim=40. 80.  0. 0., clip, angle=0,width=0.524\textwidth]{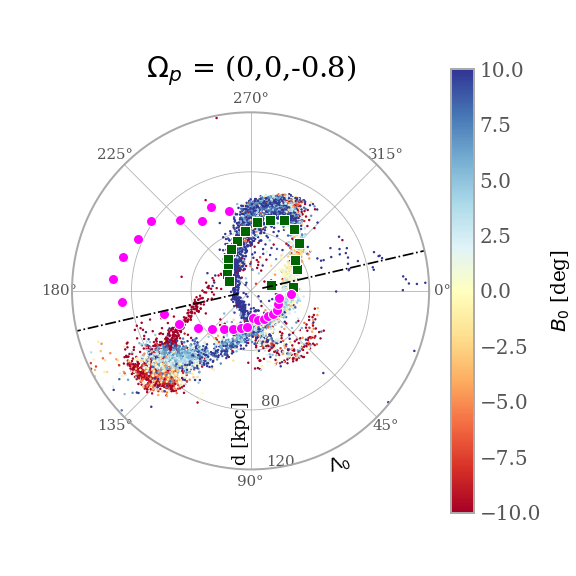}
\includegraphics[trim=40. 80. 130. 40., clip, angle=0,width=0.386\textwidth]{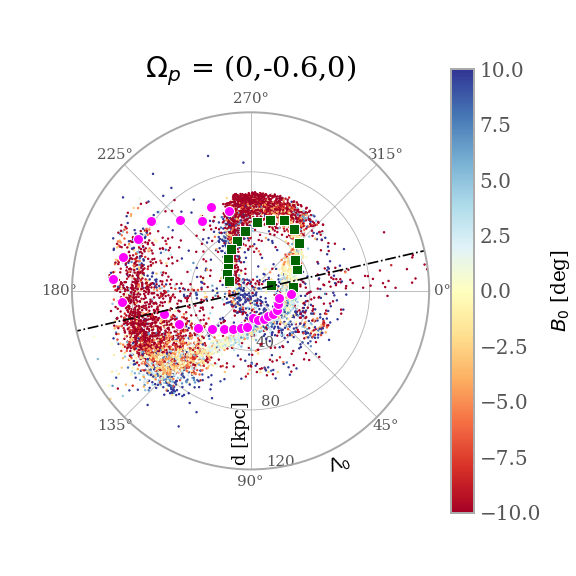}
\includegraphics[trim=40. 80. 0. 40., clip, angle=0,width=0.524\textwidth]{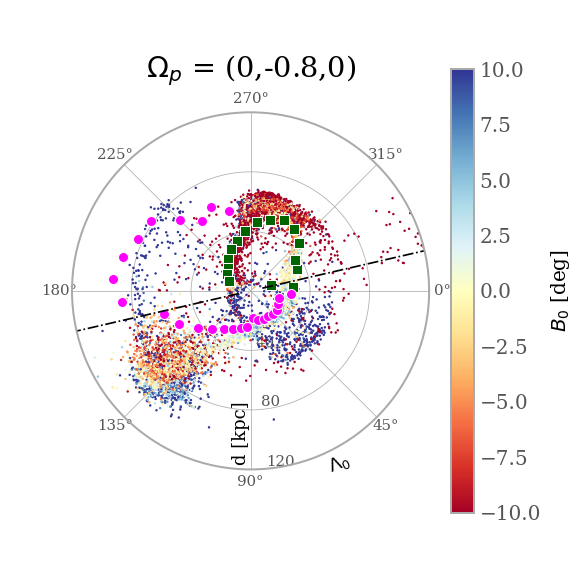} 
\end{center}
\vspace{-.4cm}
\caption{
Effect of changing pattern speed in the \LMm\ model  for rotation about $z$-axis and $y$-axis. Increasing $|\Omega_p|$ to  0.6\kmskpc or greater produces stronger distortions in the stream.  
\label{fig:Pol_LMm_mult_omega}
}
\end{figure*}

In Figure~\ref{fig:Pol_LMm_mult_omega} we show streams in the \LMm\ model for two  values of pattern speed ($|\Omega_p| = 0.6, 0.8$) that are larger than the value used in previous figures. The top row shows rotation about the $z$-axis and 2nd row shows rotation about the $y$-axis. As pattern speed increases we see that the coherence of the stream  decreases and the distortions to the stream increase dramatically. Increasing  $|\Omega_p|$ increases the angle between the apocenters of the leading and trailing arms in this model. Our examination of the other models confirms that even a pattern speed of $|\Omega_p|=0.6$ produces a significantly larger distortions than $|\Omega_p|=0.3$ (see Fig.~\ref{fig:Pol_LMm_rxyz}). The pattern speeds in this figure are at the high end of the pattern speed distribution expected from cosmological simulations. Based on the work of \citep{bailin_steinmetz_04} we infer that less than 5\% of DM halos have such large
pattern speeds. Nonetheless this is further indication that the Sgr stream is a sensitive probes of the sign, axis and
pattern speed of figure rotation, in the range of pattern speeds values predicted by cosmological simulations.

\section{ Conclusions and Discussion}
\label{sec:conclusions}
It has been nearly thirty years since it was first demonstrated using cosmological simulations that triaxial dark matter halos could have figure rotation \citep{dubinski_92}.  The predicted pattern speeds of simulated halos have a log-normal distribution with median $\Omega_p \sim 0.15h$\kmskpc\ and a width of 0.83$h$\kmskpc.  We show for the first time that if the dark matter halo of the Milky Way galaxy is triaxial  and maintains a steady pattern speed over a duration of 3-4~Gyr (as predicted by cosmological simulations), even small pattern speeds will produce significant Coriolis forces on a Sagittarius-like tidal stream which will alter its morphology and kinematics in ways that should already be detectable with current data. 

{Previous studies of the effects of figure rotation on tube orbits have been limited to studying the effects of rotation about the short-axis of the potential. We show that these results can be generalized into two simple principles that affect all tube orbits (Figs.~\ref{fig:LAT} \& \ref{fig:SAT}):
\begin{itemize}
\item{figure rotation about an axis {\it parallel} to the angular momentum vector of the tube orbit alters the angular precession rate and the angle between successive apocenters of the orbit in the plane perpendicular to the angular momentum vector;}
\item{figure rotation about an axis {\it perpendicular} to the angular momentum vector of the tube will result in the orbit being tipped or misaligned with the rotation axis.}
\end{itemize}
Both these effects arise due to the Coriolis force which alters both the orbit of a satellite and the debris stream that arises from it.}

We do not attempt to model the observed properties of the Sgr stream in this paper, rather we show that rotation with a pattern speed even as small as $|\Omega_p|= 0.3$\kmskpc\ about any of the three principal axes generates a Coriolis acceleration that varies between $\sim$ 2--15\% of the gravitational acceleration  along the stream (see Fig.~\ref{fig:orb_forces}). The Coriolis forces in the direction perpendicular to the stream plane (for rotation about any axis) and in the stream plane for rotation about the $y$ axis result  in detectable differences in the progenitor orbit and therefore the morphology and kinematics of the tidal stream.  Our main results are listed below:
\begin{itemize}
\item Our simulations suggest that figure rotation of the Milky Way's dark matter halo would warp the Sagittarius stream at its northern- and southern-most Galactic positions and produce a misalignment (or relative precession) between the instantaneous orbital planes of the leading and trailing arms (see Figures~\ref{fig:YZ_OT_F19_z} and \ref{fig:YZ_LMm_allax}).

\item Figure rotation about the Galactocentric $y$ axis (the angular momentum axis of the Sgr-stream) produces significant Coriolis forces throughout the halo in the $x-z$ plane (Figure~\ref{fig:stream_forces}), that change the angle between apocenters of the progenitor orbit even at fixed radial density profile of the halo. This in turn can result in a change in the angle between the apocenters of the leading and trailing tidal arms (see the bottom left panel of Figure~\ref{fig:Pol_LMm_rxyz}), which suggests that this angle encodes information both about the density profile of the halo \citep[e.g.,][]{belokurov_etal_14} and the rotation of the halo.

\item The recently observed, sinusoidal form of the total proper motion as a function of $\Lambda_0$ is qualitatively seen in the \LMm models (as well as other models). However, if the halo has the shape inferred by \citet{law_majewski_10}, the observed values and gradient in $B_0$ along the stream (especially for $120^\circ<\Lambda_0<20^\circ$) are not produced in either a static halo or in any of the rotating models. 

\item Based on our simulations and within the context of the Sagittarius stream, we find that the southern-most portion of the leading tidal arm will likely provide the strongest constraints on the pattern speed of the halo. This part of the stream is warped in significantly different ways depending on the sign (and magnitude) of figure rotation almost independent of the type of potential used. Therefore, mapping the locations (especially $B_0$) of Sgr stream stars in this region could help to constrain the magnitude of figure rotation of the halo. Unfortunately, as can be seen in the middle panel of Figure~\ref{fig:Pol_LMm_rxyz}, the leading arm of the Sgr stream is difficult to trace after it passes through the disk plane.

\item Pattern speeds of $|\Omega_p| \gtrsim 0.6\kmskpc$ can cause sever distortions to the Sgr-stream, altering both its coherence and morphology. Although fewer than 5\% of DM halos  in cosmological simulations are expected to have such large pattern speeds \citep{bailin_steinmetz_04} this implies the the mere fact that the Sgr stream is quite coherent can be used to set a realistic upper limit on the pattern speed of figure rotation of the Milky Way's DM halo.
\end{itemize}

Although we have not shown results for the line-of-sight velocities of stream stars, we found that streams in the \LM, \LMm\, and \Fm\, models provide a reasonably good match to the line of sight velocities of the leading arm, consistent with \citet{law_majewski_10} while these line-of-sight velocities in the \OT\, model are much too negative (as found by previous authors).  Line-of-sight velocities of stars in the trailing arm are  well fitted by most models. Our test particle simulations have shown that figure rotation of the halo has a negligible effect on heliocentric distances of stream stars.  

While it is not yet well established how halos acquire figure rotation in cosmological simulations, it was found in early studies \citep{bailin_steinmetz_04, bryan_cress_07} that it frequently arises after a close tidal interaction with  a massive galaxies or satellites. The LMC, is known to be on its first infall towards the Milky Way \citep{besla_07, besla_10} and is probably massive  enough ($\sim 10^{11}\msun$) to have moved the center of the Milky Way such that the pair of galaxies is orbiting their common center of mass \citep{gomez_besla_15}. This motion of the center of the Milky Way would also affect the Sgr stream and is not simulated here. {Following the submission of this paper \citet{Vasiliev_tango} presented detailed N-body models of the Sgr stream that account for the gravitational effect of the LMC on both the time-dependent shape of the Milky Way's dark matter halo  and the motion of the center-of-mass of the Milky Way. } If the Milky Way's halo is triaxial as determined by many previous models of the Sgr stream \citep{law_majewski_10,deg_widrow_12,Vasiliev_tango}, a massive satellite like the LMC is probably  capable of inducing figure rotation in the MW halo. A more detailed study of figure rotation in cosmological and controlled simulations is needed to understand exactly how figure rotation is induced and what determines the axis of rotation and its magnitude and direction. 

An alternative method for generating an effective rotation of the halo (relative to our viewpoint in the disk) is if the disk is currently tilting relative to the Milky Way halo. As shown by \citet{debattista_etal_13} a halo with shape determined  by \citep{law_majewski_10}, which has the intermediate axis of the halo perpendicular to the plane of the disk would be violently unstable and would result in the disk tilting relative to the halo such that it evolved to an orientation with the short axis of the halo (currently approximately along the Galactocentric $x$ axis) becoming aligned with the rotation axis of the disk.  If this is currently occurring in the Milky Way, it would result in rotation about the $y$ axis (which causes some of the most significant effects on the Sgr stream). \citet{debattista_etal_13} showed that  this instability-induced tilting of the disk would occur fairly rapidly and estimated a rate of $\sim 20^\circ$Gyr$^{-1}$ which is comparable to the values we have considered. \citet{earp_etal_17} have shown that if the disk of the Milky Way is tilting, the angular speed of this tilting would be observable with \gaia. They recently showed using  state-of-the-art cosmological hydrodynamical simulations, that the minimum tilting rate of disks is high enough  \citep{earp_etal_17, earp_etal_19} to be detectable with the astrometric precision of the \gaia\ reference frame \citep{perryman_etal_14}, which will have an end-of-mission accuracy better than 1$\mu$ arcsec\,yr$^{-1}$ (0.28$^{\circ}$Gyr$^{-1}$). 
At the present time, too little is known about the circumstances giving rise to figure rotation in cosmological halos or the circumstances that would produce a halo with its intermediate axis in the unstable condition where it is perpendicular to the disk. However, with the large number of publicly available cosmological hydrodynamical simulations currently available both for $\Lambda$CDM and with other types of DM candidates (WDM, SIDM), these questions can be answered in the near future and can lead to improved simulations of the Sgr stream.

{Our study has been restricted to rotation about the three principal axes of four specific halo models. In addition to other possible shapes and density profile parameters for the dark matter halo, the problem has many other parameters (such as the mass and potential of the Sgr dwarf progenitor, the evolution time), which affect other properties of the stream but have been held fixed in this study. A future study that allows for rotation about an arbitrary axis and carries out a systematic search over other model parameters could result in streams that match the observed gradients in $B_0$ as well as other properties of the stream possibly allowing us to constrain both the gravitational potential and the pattern speed and axis of figure rotation. We have shown that the morphology of the stream is sensitive to all properties of figure rotation of the halo: the rotation axis, the magnitude of pattern speed and the sign of rotation. At first glance it would appear that there are so many parameters in the problem and that numerous degeneracies might exist. While this could be true the recent model of the Sgr stream by \citet{Vasiliev_tango} provides an excellent match to all the currently observed data and demonstrated that some of these were impossible to produce unless their model considered the effects of {\it time dependent distortions of the Milky Way potential} due to the gravitational field of the LMC and the effect of the not holding the center-of-mass of the Milk Way fixed. While it may be difficult to distinguish between the tide due to the LMC and steady figure rotation with just a single stream, numerous shorter streams have recently been discovered by \gaia\, and photometric surveys like DES \citep{Shipp_2018}. A model that aims to jointly fit several streams simultaneously could in principle distinguish between these options and will be the effort of future work.}

With the wealth of existing and upcoming data from large photometric and astrometric surveys (\gaia, WFIRST, LSST), and large spectroscopic surveys \citep[DESI, WEAVE, 4MOST,][]{DESI_2016a, DESI_2016b,WEAVE2014,4MOST2012}  it will soon be possible to construct much more accurate models for the Sgr stream and other streams, and potentially to constrain not only the halo shape and density profile but also, for the first time, its pattern speed and axis of figure rotation. 

\acknowledgments

We thank the referee for a constructive report that improved the quality of this manuscript. MV thanks members of the stellar halos group at the University of Michigan and Eugene Vasiliev for stimulating discussion and continued camaraderie especially during the COVID-19 pandemic. MV is supported by NASA-ATP awards NNX15AK79G and  80NSSC20K0509 and a Catalyst Grant from the Michigan Institute for Computational Discovery and Engineering at the University of Michigan. This work has made use of data from the European Space Agency (ESA) mission \gaia\ 
(\url{http://www.cosmos.esa.int/gaia}),  processed by the \gaia\ Data Processing and Analysis Consortium (DPAC,
\url{http://www.cosmos.esa.int/web/gaia/dpac/consortium}).  Funding for the DPAC has been provided by national institutions, in particular the institutions participating in the {\it Gaia} Multilateral Agreement.
This research has also made use of NASA's Astrophysics Data System Bibliographic Services;
the arXiv pre-print server operated by Cornell University.

\software{
Astropy \citep{astropy, astropy:2018},
gala \citep{gala:JOSS, gala},
IPython \citep{IPython},
matplotlib \citep{hunter07},
numpy \citep{vanderwalt11},
scipy \citep{jones01}.}

\medskip


\bibliographystyle{aasjournal}
\bibliography{rotate}

\begin{thebibliography}{}
\expandafter\ifx\csname natexlab\endcsname\relax\def\natexlab#1{#1}\fi
\providecommand{\url}[1]{\href{#1}{#1}}
\providecommand{\dodoi}[1]{doi:~\href{http://doi.org/#1}{\nolinkurl{#1}}}
\providecommand{\doeprint}[1]{\href{http://ascl.net/#1}{\nolinkurl{http://ascl.net/#1}}}
\providecommand{\doarXiv}[1]{\href{https://arxiv.org/abs/#1}{\nolinkurl{https://arxiv.org/abs/#1}}}

\bibitem[{{Adams} {et~al.}(2007){Adams}, {Bloch}, {Butler}, {Druce}, \&
  {Ketchum}}]{adams_etal_07}
{Adams}, F.~C., {Bloch}, A.~M., {Butler}, S.~C., {Druce}, J.~M., \& {Ketchum},
  J.~A. 2007, \apj, 670, 1027, \dodoi{10.1086/522581}

\bibitem[{{Antoja} {et~al.}(2020){Antoja}, {Ramos}, {Mateu}, {Helmi}, {Anders},
  {Jordi}, \& {Carballo-Bello}}]{antoja_etal_20}
{Antoja}, T., {Ramos}, P., {Mateu}, C., {et~al.} 2020, \aap, 635, L3,
  \dodoi{10.1051/0004-6361/201937145}

\bibitem[{{Astropy Collaboration} {et~al.}(2013){Astropy Collaboration},
  {Robitaille}, {Tollerud}, {Greenfield}, {Droettboom}, {Bray}, {Aldcroft},
  {Davis}, {Ginsburg}, {Price-Whelan}, {Kerzendorf}, {Conley}, {Crighton},
  {Barbary}, {Muna}, {Ferguson}, {Grollier}, {Parikh}, {Nair}, {Unther},
  {Deil}, {Woillez}, {Conseil}, {Kramer}, {Turner}, {Singer}, {Fox}, {Weaver},
  {Zabalza}, {Edwards}, {Azalee Bostroem}, {Burke}, {Casey}, {Crawford},
  {Dencheva}, {Ely}, {Jenness}, {Labrie}, {Lim}, {Pierfederici}, {Pontzen},
  {Ptak}, {Refsdal}, {Servillat}, \& {Streicher}}]{astropy}
{Astropy Collaboration}, {Robitaille}, T.~P., {Tollerud}, E.~J., {et~al.} 2013,
  \aap, 558, A33, \dodoi{10.1051/0004-6361/201322068}

\bibitem[{{Astropy Collaboration} {et~al.}(2018){Astropy Collaboration},
  {Price-Whelan}, {Sip{\H{o}}cz}, {G{\"u}nther}, {Lim}, {Crawford}, {Conseil},
  {Shupe}, {Craig}, {Dencheva}, {Ginsburg}, {Vand erPlas}, {Bradley},
  {P{\'e}rez-Su{\'a}rez}, {de Val-Borro}, {Aldcroft}, {Cruz}, {Robitaille},
  {Tollerud}, {Ardelean}, {Babej}, {Bach}, {Bachetti}, {Bakanov}, {Bamford},
  {Barentsen}, {Barmby}, {Baumbach}, {Berry}, {Biscani}, {Boquien}, {Bostroem},
  {Bouma}, {Brammer}, {Bray}, {Breytenbach}, {Buddelmeijer}, {Burke},
  {Calderone}, {Cano Rodr{\'\i}guez}, {Cara}, {Cardoso}, {Cheedella}, {Copin},
  {Corrales}, {Crichton}, {D'Avella}, {Deil}, {Depagne}, {Dietrich}, {Donath},
  {Droettboom}, {Earl}, {Erben}, {Fabbro}, {Ferreira}, {Finethy}, {Fox},
  {Garrison}, {Gibbons}, {Goldstein}, {Gommers}, {Greco}, {Greenfield},
  {Groener}, {Grollier}, {Hagen}, {Hirst}, {Homeier}, {Horton}, {Hosseinzadeh},
  {Hu}, {Hunkeler}, {Ivezi{\'c}}, {Jain}, {Jenness}, {Kanarek}, {Kendrew},
  {Kern}, {Kerzendorf}, {Khvalko}, {King}, {Kirkby}, {Kulkarni}, {Kumar},
  {Lee}, {Lenz}, {Littlefair}, {Ma}, {Macleod}, {Mastropietro}, {McCully},
  {Montagnac}, {Morris}, {Mueller}, {Mumford}, {Muna}, {Murphy}, {Nelson},
  {Nguyen}, {Ninan}, {N{\"o}the}, {Ogaz}, {Oh}, {Parejko}, {Parley}, {Pascual},
  {Patil}, {Patil}, {Plunkett}, {Prochaska}, {Rastogi}, {Reddy Janga},
  {Sabater}, {Sakurikar}, {Seifert}, {Sherbert}, {Sherwood-Taylor}, {Shih},
  {Sick}, {Silbiger}, {Singanamalla}, {Singer}, {Sladen}, {Sooley},
  {Sornarajah}, {Streicher}, {Teuben}, {Thomas}, {Tremblay}, {Turner},
  {Terr{\'o}n}, {van Kerkwijk}, {de la Vega}, {Watkins}, {Weaver}, {Whitmore},
  {Woillez}, {Zabalza}, \& {Astropy Contributors}}]{astropy:2018}
{Astropy Collaboration}, {Price-Whelan}, A.~M., {Sip{\H{o}}cz}, B.~M., {et~al.}
  2018, \aj, 156, 123, \dodoi{10.3847/1538-3881/aabc4f}

\bibitem[{{Bailin} \& {Steinmetz}(2004)}]{bailin_steinmetz_04}
{Bailin}, J., \& {Steinmetz}, M. 2004, \apj, 616, 27, \dodoi{10.1086/424912}

\bibitem[{{Bekki} \& {Freeman}(2002)}]{bekki_freeman_02}
{Bekki}, K., \& {Freeman}, K.~C. 2002, \apjl, 574, L21, \dodoi{10.1086/342262}

\bibitem[{{Belokurov} {et~al.}(2014){Belokurov}, {Koposov}, {Evans},
  {Pe{\~n}arrubia}, {Irwin}, {Smith}, {Lewis}, {Gieles}, {Wilkinson},
  {Gilmore}, {Olszewski}, \& {Niederste-Ostholt}}]{belokurov_etal_14}
{Belokurov}, V., {Koposov}, S.~E., {Evans}, N.~W., {et~al.} 2014, \mnras, 437,
  116, \dodoi{10.1093/mnras/stt1862}

\bibitem[{{Besla} {et~al.}(2007){Besla}, {Kallivayalil}, {Hernquist},
  {Robertson}, {Cox}, {van der Marel}, \& {Alcock}}]{besla_07}
{Besla}, G., {Kallivayalil}, N., {Hernquist}, L., {et~al.} 2007, \apj, 668,
  949, \dodoi{10.1086/521385}

\bibitem[{{Besla} {et~al.}(2010){Besla}, {Kallivayalil}, {Hernquist}, {van der
  Marel}, {Cox}, \& {Kere{\v{s}}}}]{besla_10}
---. 2010, \apj, 721, L97, \dodoi{10.1088/2041-8205/721/2/L97}

\bibitem[{{Binney}(1981)}]{binney_81}
{Binney}, J. 1981, \mnras, 196, 455

\bibitem[{{Binney} \& {Tremaine}(2008)}]{BT08}
{Binney}, J., \& {Tremaine}, S. 2008, {Galactic Dynamics: Second Edition}
  (Princeton University Press)

\bibitem[{{Bland-Hawthorn} \& {Gerhard}(2016)}]{bland_hawthorn_gerhard_16}
{Bland-Hawthorn}, J., \& {Gerhard}, O. 2016, \araa, 54, 529,
  \dodoi{10.1146/annurev-astro-081915-023441}

\bibitem[{{Bose} {et~al.}(2016){Bose}, {Hellwing}, {Frenk}, {Jenkins},
  {Lovell}, {Helly}, \& {Li}}]{bose_frenk_16_WDM}
{Bose}, S., {Hellwing}, W.~A., {Frenk}, C.~S., {et~al.} 2016, \mnras, 455, 318,
  \dodoi{10.1093/mnras/stv2294}

\bibitem[{{Bryan} \& {Cress}(2007)}]{bryan_cress_07}
{Bryan}, S.~E., \& {Cress}, C.~M. 2007, \mnras, 380, 657,
  \dodoi{10.1111/j.1365-2966.2007.12096.x}

\bibitem[{{Bureau} {et~al.}(1999){Bureau}, {Freeman}, {Pfitzner}, \&
  {Meurer}}]{bureau_etal_99}
{Bureau}, M., {Freeman}, K.~C., {Pfitzner}, D.~W., \& {Meurer}, G.~R. 1999,
  \aj, 118, 2158, \dodoi{10.1086/301064}

\bibitem[{{Carlin} {et~al.}(2011){Carlin}, {Majewski}, {Casetti-Dinescu},
  {Law}, {Girard}, \& {Patterson}}]{carlin_etal_11}
{Carlin}, J.~L., {Majewski}, S.~R., {Casetti-Dinescu}, D.~I., {et~al.} 2011,
  ArXiv e-prints.
\newblock \doarXiv{1111.0014}

\bibitem[{{Carlin} {et~al.}(2012){Carlin}, {Majewski}, {Casetti-Dinescu},
  {Law}, {Girard}, \& {Patterson}}]{carlin_etal_12}
---. 2012, \apj, 744, 25, \dodoi{10.1088/0004-637X/744/1/25}

\bibitem[{{Chakrabarty} \& {Dubinski}(2011)}]{chakrabarty_dubinski_11}
{Chakrabarty}, D., \& {Dubinski}, J. 2011, Memorie della Societa Astronomica
  Italiana Supplementi, 18, 139

\bibitem[{{Chua} {et~al.}(2019){Chua}, {Pillepich}, {Vogelsberger}, \&
  {Hernquist}}]{Chua_2019}
{Chua}, K. T.~E., {Pillepich}, A., {Vogelsberger}, M., \& {Hernquist}, L. 2019,
  \mnras, 484, 476, \dodoi{10.1093/mnras/sty3531}

\bibitem[{{Dalton} {et~al.}(2014){Dalton}, {Trager}, {Abrams}, {Bonifacio},
  {L{\'o}pez Aguerri}, {Middleton}, {Benn}, {Dee}, {Say{\`e}de}, {Lewis},
  {Pragt}, {Pico}, {Walton}, {Rey}, {Allende Prieto}, {Pe{\~n}ate}, {Lhome},
  {Ag{\'o}cs}, {Alonso}, {Terrett}, {Brock}, {Gilbert}, {Ridings}, {Guinouard},
  {Verheijen}, {Tosh}, {Rogers}, {Steele}, {Stuik}, {Tromp}, {Jasko}, {Kragt},
  {Lesman}, {Mottram}, {Bates}, {Gribbin}, {Fernand o Rodriguez}, {Delgado},
  {Martin}, {Cano}, {Navarro}, {Irwin}, {Lewis}, {Gonzalez Solares},
  {O'Mahony}, {Bianco}, {Zurita}, {ter Horst}, {Molinari}, {Lodi}, {Guerra},
  {Vallenari}, \& {Baruffolo}}]{WEAVE2014}
{Dalton}, G., {Trager}, S., {Abrams}, D.~C., {et~al.} 2014, Society of
  Photo-Optical Instrumentation Engineers (SPIE) Conference Series, Vol. 9147,
  {Project overview and update on WEAVE: the next generation wide-field
  spectroscopy facility for the William Herschel Telescope} (SPIE), 91470L,
  \dodoi{10.1117/12.2055132}

\bibitem[{de~Jong {et~al.}(2012)de~Jong, Bellido-Tirado, Chiappini, Ãric
  Depagne, Haynes, Johl, Schnurr, Schwope, Walcher, Dionies, Haynes, Kelz,
  Kitaura, Lamer, Minchev, MÃŒller, Nuza, Olaya, Piffl, Popow, Steinmetz,
  Ural, Williams, Winkler, Wisotzki, Ansorge, Banerji, Solares, Irwin, Jr.,
  King, McMahon, Koposov, Parry, Sun, Walton, Finger, Iwert, Krumpe, Lizon,
  Vincenzo, Amans, Bonifacio, Cohen, Francois, Jagourel, Mignot, Royer,
  Sartoretti, Bender, Grupp, Hess, Lang-Bardl, Muschielok, BÃ¶hringer,
  Boller, Bongiorno, Brusa, Dwelly, Merloni, Nandra, Salvato, Pragt, Navarro,
  Gerlofsma, Roelfsema, Dalton, Middleton, Tosh, Boeche, Caffau, Christlieb,
  Grebel, Hansen, Koch, Ludwig, Quirrenbach, Sbordone, Seifert, Thimm,
  Trifonov, Helmi, Trager, Feltzing, Korn, \& Boland}]{4MOST2012}
de~Jong, R.~S., Bellido-Tirado, O., Chiappini, C., {et~al.} 2012, in Society of
  Photo-Optical Instrumentation Engineers (SPIE) Conference Series, Vol. 8446,
  Ground-based and Airborne Instrumentation for Astronomy IV, ed. I.~S. McLean,
  S.~K. Ramsay, \& H.~Takami, International Society for Optics and Photonics
  (SPIE), 252 -- 266, \dodoi{10.1117/12.926239}

\bibitem[{{de Zeeuw} \& {Merritt}(1983)}]{dezeeuw_merritt_83}
{de Zeeuw}, T., \& {Merritt}, D. 1983, \apj, 267, 571, \dodoi{10.1086/160894}

\bibitem[{{Debattista} {et~al.}(2008{\natexlab{a}}){Debattista}, {Moore},
  {Quinn}, {Kazantzidis}, {Maas}, {Mayer}, {Read}, \& {Stadel}}]{deb_etal_08}
{Debattista}, V.~P., {Moore}, B., {Quinn}, T., {et~al.} 2008{\natexlab{a}},
  \apj, 681, 1076, \dodoi{10.1086/587977}

\bibitem[{{Debattista} {et~al.}(2008{\natexlab{b}}){Debattista}, {Moore},
  {Quinn}, {Kazantzidis}, {Maas}, {Mayer}, {Read}, \&
  {Stadel}}]{debattista_etal_08}
---. 2008{\natexlab{b}}, \apj, 681, 1076, \dodoi{10.1086/587977}

\bibitem[{{Debattista} {et~al.}(2013){Debattista}, {Ro{\v s}kar}, {Valluri},
  {Quinn}, {Moore}, \& {Wadsley}}]{debattista_etal_13}
{Debattista}, V.~P., {Ro{\v s}kar}, R., {Valluri}, M., {et~al.} 2013, \mnras,
  434, 2971, \dodoi{10.1093/mnras/stt1217}

\bibitem[{{Deg} \& {Widrow}(2013)}]{deg_widrow_12}
{Deg}, N., \& {Widrow}, L. 2013, \mnras, 428, 912, \dodoi{10.1093/mnras/sts089}

\bibitem[{{Deibel} {et~al.}(2011){Deibel}, {Valluri}, \&
  {Merritt}}]{deibel_etal_11}
{Deibel}, A.~T., {Valluri}, M., \& {Merritt}, D. 2011, \apj, 728, 128,
  \dodoi{10.1088/0004-637X/728/2/128}

\bibitem[{{DESI Collaboration} {et~al.}(2016{\natexlab{a}}){DESI
  Collaboration}, {Aghamousa}, {Aguilar}, {Ahlen}, {Alam}, {Allen}, {Allende
  Prieto}, {Annis}, {Bailey}, {Balland}, \& et~al.}]{DESI_2016a}
{DESI Collaboration}, {Aghamousa}, A., {Aguilar}, J., {et~al.}
  2016{\natexlab{a}}, ArXiv e-prints, arXiv:1611.00036.
\newblock \doarXiv{1611.00036}

\bibitem[{{DESI Collaboration} {et~al.}(2016{\natexlab{b}}){DESI
  Collaboration}, {Aghamousa}, {Aguilar}, {Ahlen}, {Alam}, {Allen}, {Allende
  Prieto}, {Annis}, {Bailey}, {Balland}, \& et~al.}]{DESI_2016b}
---. 2016{\natexlab{b}}, ArXiv e-prints, arXiv:1611.00037.
\newblock \doarXiv{1611.00037}

\bibitem[{{Dierickx} \& {Loeb}(2017{\natexlab{a}})}]{dierickx_loeb_17a}
{Dierickx}, M. I.~P., \& {Loeb}, A. 2017{\natexlab{a}}, \apj, 836, 92,
  \dodoi{10.3847/1538-4357/836/1/92}

\bibitem[{{Dierickx} \& {Loeb}(2017{\natexlab{b}})}]{dierickx_loeb_17b}
---. 2017{\natexlab{b}}, \apj, 847, 42, \dodoi{10.3847/1538-4357/aa8767}

\bibitem[{{Dubinski}(1992)}]{dubinski_92}
{Dubinski}, J. 1992, \apj, 401, 441, \dodoi{10.1086/172076}

\bibitem[{{Dubinski} \& {Carlberg}(1991)}]{dubinski_carlberg_91}
{Dubinski}, J., \& {Carlberg}, R.~G. 1991, \apj, 378, 496,
  \dodoi{10.1086/170451}

\bibitem[{{Dubinski} \& {Chakrabarty}(2009)}]{dubinski_chakrabarty_09}
{Dubinski}, J., \& {Chakrabarty}, D. 2009, \apj, 703, 2068,
  \dodoi{10.1088/0004-637X/703/2/2068}

\bibitem[{{Dutton} \& {Macci{\`o}}(2014)}]{dutton_maccio_14}
{Dutton}, A.~A., \& {Macci{\`o}}, A.~V. 2014, \mnras, 441, 3359,
  \dodoi{10.1093/mnras/stu742}

\bibitem[{{Earp} {et~al.}(2017){Earp}, {Debattista}, {Macci{\`o}}, \&
  {Cole}}]{earp_etal_17}
{Earp}, S. W.~F., {Debattista}, V.~P., {Macci{\`o}}, A.~V., \& {Cole}, D.~R.
  2017, \mnras, 469, 4095, \dodoi{10.1093/mnras/stx1143}

\bibitem[{{Earp} {et~al.}(2019){Earp}, {Debattista}, {Macci{\`o}}, {Wang},
  {Buck}, \& {Khachaturyants}}]{earp_etal_19}
{Earp}, S. W.~F., {Debattista}, V.~P., {Macci{\`o}}, A.~V., {et~al.} 2019,
  \mnras, 488, 5728, \dodoi{10.1093/mnras/stz2109}

\bibitem[{{Eyre} \& {Binney}(2009)}]{eyre_binney_09}
{Eyre}, A., \& {Binney}, J. 2009, \mnras, 1338,
  \dodoi{10.1111/j.1365-2966.2009.15494.x}

\bibitem[{{Fardal} {et~al.}(2015){Fardal}, {Huang}, \&
  {Weinberg}}]{fardal_etal_15}
{Fardal}, M.~A., {Huang}, S., \& {Weinberg}, M.~D. 2015, \mnras, 452, 301,
  \dodoi{10.1093/mnras/stv1198}

\bibitem[{{Fardal} {et~al.}(2019){Fardal}, {van der Marel}, {Law}, {Sohn},
  {Sesar}, {Hernitschek}, \& {Rix}}]{fardal_etal_19}
{Fardal}, M.~A., {van der Marel}, R.~P., {Law}, D.~R., {et~al.} 2019, \mnras,
  483, 4724, \dodoi{10.1093/mnras/sty3428}

\bibitem[{{Fellhauer} {et~al.}(2006){Fellhauer}, {Belokurov}, {Evans},
  {Wilkinson}, {Zucker}, {Gilmore}, {Irwin}, {Bramich}, {Vidrih}, {Wyse},
  {Beers}, \& {Brinkmann}}]{fellhauer_etal_06}
{Fellhauer}, M., {Belokurov}, V., {Evans}, N.~W., {et~al.} 2006, \apj, 651,
  167, \dodoi{10.1086/507128}

\bibitem[{{Gaia Collaboration} {et~al.}(2018){Gaia Collaboration}, {Helmi},
  {van Leeuwen}, {McMillan}, {Massari}, {Antoja}, {Robin}, {Lindegren},
  {Bastian}, {Arenou}, {Babusiaux}, {Biermann}, {Breddels}, {Hobbs}, {Jordi},
  {Pancino}, {Reyl{\'e}}, {Veljanoski}, {Brown}, {Vallenari}, {Prusti}, {de
  Bruijne}, {Bailer-Jones}, {Evans}, {Eyer}, {Jansen}, {Klioner}, {Lammers},
  {Luri}, {Mignard}, {Panem}, {Pourbaix}, {Randich}, {Sartoretti}, {Siddiqui},
  {Soubiran}, {Walton}, {Cropper}, {Drimmel}, {Katz}, {Lattanzi}, {Bakker},
  {Cacciari}, {Casta{\~n}eda}, {Chaoul}, {Cheek}, {De Angeli}, {Fabricius},
  {Guerra}, {Holl}, {Masana}, {Messineo}, {Mowlavi}, {Nienartowicz}, {Panuzzo},
  {Portell}, {Riello}, {Seabroke}, {Tanga}, {Th{\'e}venin}, {Gracia-Abril},
  {Comoretto}, {Garcia-Reinaldos}, {Teyssier}, {Altmann}, {Andrae}, {Audard},
  {Bellas-Velidis}, {Benson}, {Berthier}, {Blomme}, {Burgess}, {Busso},
  {Carry}, {Cellino}, {Clementini}, {Clotet}, {Creevey}, {Davidson}, {De
  Ridder}, {Delchambre}, {Dell'Oro}, {Ducourant},
  {Fern{\'a}ndez-Hern{\'a}ndez}, {Fouesneau}, {Fr{\'e}mat}, {Galluccio},
  {Garc{\'\i}a-Torres}, {Gonz{\'a}lez-N{\'u}{\~n}ez}, {Gonz{\'a}lez-Vidal},
  {Gosset}, {Guy}, {Halbwachs}, {Hambly}, {Harrison}, {Hern{\'a}ndez},
  {Hestroffer}, {Hodgkin}, {Hutton}, {Jasniewicz}, {Jean-Antoine-Piccolo},
  {Jordan}, {Korn}, {Krone-Martins}, {Lanzafame}, {Lebzelter}, {L{\"o}ffler},
  {Manteiga}, {Marrese}, {Mart{\'\i}n-Fleitas}, {Moitinho}, {Mora}, {Muinonen},
  {Osinde}, {Pauwels}, {Petit}, {Recio-Blanco}, {Richards}, {Rimoldini},
  {Sarro}, {Siopis}, {Smith}, {Sozzetti}, {S{\"u}veges}, {Torra}, {van Reeven},
  {Abbas}, {Abreu Aramburu}, {Accart}, {Aerts}, {Altavilla}, {{\'A}lvarez},
  {Alvarez}, {Alves}, {Anderson}, {Andrei}, {Anglada Varela}, {Antiche},
  {Arcay}, {Astraatmadja}, {Bach}, {Baker}, {Balaguer-N{\'u}{\~n}ez}, {Balm},
  {Barache}, {Barata}, {Barbato}, {Barblan}, {Barklem}, {Barrado}, {Barros},
  {Barstow}, {Bartholom{\'e} Mu{\~n}oz}, {Bassilana}, {Becciani}, {Bellazzini},
  {Berihuete}, {Bertone}, {Bianchi}, {Bienaym{\'e}}, {Blanco-Cuaresma}, {Boch},
  {Boeche}, {Bombrun}, {Borrachero}, {Bossini}, {Bouquillon}, {Bourda},
  {Bragaglia}, {Bramante}, {Bressan}, {Brouillet}, {Br{\"u}semeister},
  {Brugaletta}, {Bucciarelli}, {Burlacu}, {Busonero}, {Butkevich}, {Buzzi},
  {Caffau}, {Cancelliere}, {Cannizzaro}, {Cantat-Gaudin}, {Carballo},
  {Carlucci}, {Carrasco}, {Casamiquela}, {Castellani}, {Castro-Ginard},
  {Charlot}, {Chemin}, {Chiavassa}, {Cocozza}, {Costigan}, {Cowell}, {Crifo},
  {Crosta}, {Crowley}, {Cuypers}, {Dafonte}, {Damerdji}, {Dapergolas}, {David},
  {David}, {de Laverny}, {De Luise}, {De March}, {de Martino}, {de Souza}, {de
  Torres}, {Debosscher}, {del Pozo}, {Delbo}, {Delgado}, {Delgado}, {Di
  Matteo}, {Diakite}, {Diener}, {Distefano}, {Dolding}, {Drazinos},
  {Dur{\'a}n}, {Edvardsson}, {Enke}, {Eriksson}, {Esquej}, {Eynard Bontemps},
  {Fabre}, {Fabrizio}, {Faigler}, {Falc{\~a}o}, {Farr{\`a}s Casas}, {Federici},
  {Fedorets}, {Fernique}, {Figueras}, {Filippi}, {Findeisen}, {Fonti},
  {Fraile}, {Fraser}, {Fr{\'e}zouls}, {Gai}, {Galleti}, {Garabato},
  {Garc{\'\i}a-Sedano}, {Garofalo}, {Garralda}, {Gavel}, {Gavras}, {Gerssen},
  {Geyer}, {Giacobbe}, {Gilmore}, {Girona}, {Giuffrida}, {Glass}, {Gomes},
  {Granvik}, {Gueguen}, {Guerrier}, {Guiraud}, {Guti{\'e}rrez-S{\'a}nchez},
  {Hofmann}, {Holland}, {Huckle}, {Hypki}, {Icardi}, {Jan{\ss}en}, {Jevardat de
  Fombelle}, {Jonker}, {Juh{\'a}sz}, {Julbe}, {Karampelas}, {Kewley}, {Klar},
  {Kochoska}, {Kohley}, {Kolenberg}, {Kontizas}, {Kontizas}, {Koposov},
  {Kordopatis}, {Kostrzewa-Rutkowska}, {Koubsky}, {Lambert}, {Lanza}, {Lasne},
  {Lavigne}, {Le Fustec}, {Le Poncin-Lafitte}, {Lebreton}, {Leccia}, {Leclerc},
  {Lecoeur-Taibi}, {Lenhardt}, {Leroux}, {Liao}, {Licata}, {Lindstr{\o}m},
  {Lister}, {Livanou}, {Lobel}, {L{\'o}pez}, {Managau}, {Mann}, {Mantelet},
  {Marchal}, {Marchant}, {Marconi}, {Marinoni}, {Marschalk{\'o}}, {Marshall},
  {Martino}, {Marton}, {Mary}, {Matijevi{\v{c}}}, {Mazeh}, {Messina},
  {Michalik}, {Millar}, {Molina}, {Molinaro}, {Moln{\'a}r}, {Montegriffo},
  {Mor}, {Morbidelli}, {Morel}, {Morris}, {Mulone}, {Muraveva}, {Musella},
  {Nelemans}, {Nicastro}, {Noval}, {O'Mullane}, {Ord{\'e}novic},
  {Ord{\'o}{\~n}ez-Blanco}, {Osborne}, {Pagani}, {Pagano}, {Pailler},
  {Palacin}, {Palaversa}, {Panahi}, {Pawlak}, {Piersimoni}, {Pineau}, {Plachy},
  {Plum}, {Poggio}, {Poujoulet}, {Pr{\v{s}}a}, {Pulone}, {Racero}, {Ragaini},
  {Rambaux}, {Ramos-Lerate}, {Regibo}, {Riclet}, {Ripepi}, {Riva}, {Rivard},
  {Rixon}, {Roegiers}, {Roelens}, {Romero-G{\'o}mez}, {Rowell}, {Royer},
  {Ruiz-Dern}, {Sadowski}, {Sagrist{\`a} Sell{\'e}s}, {Sahlmann}, {Salgado},
  {Salguero}, {Sanna}, {Santana-Ros}, {Sarasso}, {Savietto}, {Schultheis},
  {Sciacca}, {Segol}, {Segovia}, {S{\'e}gransan}, {Shih}, {Siltala}, {Silva},
  {Smart}, {Smith}, {Solano}, {Solitro}, {Sordo}, {Soria Nieto}, {Souchay},
  {Spagna}, {Spoto}, {Stampa}, {Steele}, {Steidelm{\"u}ller}, {Stephenson},
  {Stoev}, {Suess}, {Surdej}, {Szabados}, {Szegedi-Elek}, {Tapiador}, {Taris},
  {Tauran}, {Taylor}, {Teixeira}, {Terrett}, {Teyssand ier}, {Thuillot},
  {Titarenko}, {Torra Clotet}, {Turon}, {Ulla}, {Utrilla}, {Uzzi}, {Vaillant},
  {Valentini}, {Valette}, {van Elteren}, {Van Hemelryck}, {van Leeuwen},
  {Vaschetto}, {Vecchiato}, {Viala}, {Vicente}, {Vogt}, {von Essen}, {Voss},
  {Votruba}, {Voutsinas}, {Walmsley}, {Weiler}, {Wertz}, {Wevems},
  {Wyrzykowski}, {Yoldas}, {{\v{Z}}erjal}, {Ziaeepour}, {Zorec}, {Zschocke},
  {Zucker}, {Zurbach}, \& {Zwitter}}]{gaiacollab_helmi_18}
{Gaia Collaboration}, {Helmi}, A., {van Leeuwen}, F., {et~al.} 2018, \aap, 616,
  A12, \dodoi{10.1051/0004-6361/201832698}

\bibitem[{{Garavito-Camargo} {et~al.}(2019){Garavito-Camargo}, {Besla},
  {Laporte}, {Johnston}, {G{\'o}mez}, \& {Watkins}}]{garavito-camargo_etal_19}
{Garavito-Camargo}, N., {Besla}, G., {Laporte}, C. F.~P., {et~al.} 2019, \apj,
  884, 51, \dodoi{10.3847/1538-4357/ab32eb}

\bibitem[{{Gibbons} {et~al.}(2017){Gibbons}, {Belokurov}, \&
  {Evans}}]{gibbons_etal_17}
{Gibbons}, S.~L.~J., {Belokurov}, V., \& {Evans}, N.~W. 2017, \mnras, 464, 794,
  \dodoi{10.1093/mnras/stw2328}

\bibitem[{{G{o}mez} {et~al.}(2015){G{o}mez}, {Besla}, {Carpintero},
  {Villalobos}, {O'Shea}, \& {Bell}}]{gomez_besla_15}
{G{o}mez}, F.~A., {Besla}, G., {Carpintero}, D.~D., {et~al.} 2015, \apj, 802,
  128, \dodoi{10.1088/0004-637X/802/2/128}

\bibitem[{{Hattori} {et~al.}(2018){Hattori}, {Valluri}, {Bell}, \&
  {Roederer}}]{hattori_etal_18}
{Hattori}, K., {Valluri}, M., {Bell}, E.~F., \& {Roederer}, I.~U. 2018, \apj,
  866, 121, \dodoi{10.3847/1538-4357/aadee5}

\bibitem[{{Heiligman} \& {Schwarzschild}(1979)}]{heiligman_schwarzschild_79}
{Heiligman}, G., \& {Schwarzschild}, M. 1979, \apj, 233, 872,
  \dodoi{10.1086/157449}

\bibitem[{{Heisler} {et~al.}(1982){Heisler}, {Merritt}, \&
  {Schwarzschild}}]{heisler_etal_82}
{Heisler}, J., {Merritt}, D., \& {Schwarzschild}, M. 1982, \apj, 258, 490,
  \dodoi{10.1086/160100}

\bibitem[{{Helmi}(2004)}]{helmi_04}
{Helmi}, A. 2004, \mnras, 351, 643, \dodoi{10.1111/j.1365-2966.2004.07812.x}

\bibitem[{{Hernitschek} {et~al.}(2017){Hernitschek}, {Sesar}, {Rix},
  {Belokurov}, {Martinez-Delgado}, {Martin}, {Kaiser}, {Hodapp}, {Chambers}, \&
  {Wainscoat}}]{hernitschek_etal_17}
{Hernitschek}, N., {Sesar}, B., {Rix}, H.-W., {et~al.} 2017, \apj, 850, 96,
  \dodoi{10.3847/1538-4357/aa960c}

\bibitem[{{Hernquist}(1990)}]{hernquist_90}
{Hernquist}, L. 1990, \apj, 356, 359, \dodoi{10.1086/168845}

\bibitem[{{Hunter}(2007)}]{hunter07}
{Hunter}, J.~D. 2007, Computing in Science and Engineering, 9, 90,
  \dodoi{10.1109/MCSE.2007.55}

\bibitem[{{Jiang} \& {Binney}(2000)}]{jiang_binney_00}
{Jiang}, I.-G., \& {Binney}, J. 2000, \mnras, 314, 468,
  \dodoi{10.1046/j.1365-8711.2000.03311.x}

\bibitem[{{Jing} \& {Suto}(2000)}]{jing_suto_00}
{Jing}, Y.~P., \& {Suto}, Y. 2000, \apjl, 529, L69, \dodoi{10.1086/312463}

\bibitem[{{Johnston} {et~al.}(2005){Johnston}, {Law}, \&
  {Majewski}}]{johnston_etal_05}
{Johnston}, K.~V., {Law}, D.~R., \& {Majewski}, S.~R. 2005, \apj, 619, 800,
  \dodoi{10.1086/426777}

\bibitem[{{Johnston} {et~al.}(1999){Johnston}, {Zhao}, {Spergel}, \&
  {Hernquist}}]{johnston_etal_99}
{Johnston}, K.~V., {Zhao}, H., {Spergel}, D.~N., \& {Hernquist}, L. 1999,
  \apjl, 512, L109, \dodoi{10.1086/311876}

\bibitem[{{Jones} {et~al.}(2001){Jones}, {Oliphant}, \& {Peterson}}]{jones01}
{Jones}, E., {Oliphant}, T., \& {Peterson}, P., e.~a. 2001, {SciPy}: Open
  source scientific tools for {Python}.
\newblock \url{http://www.scipy.org/}

\bibitem[{{Kazantzidis} {et~al.}(2004){Kazantzidis}, {Kravtsov}, {Zentner},
  {Allgood}, {Nagai}, \& {Moore}}]{kazantzidis_etal_04_shapes}
{Kazantzidis}, S., {Kravtsov}, A.~V., {Zentner}, A.~R., {et~al.} 2004, \apjl,
  611, L73, \dodoi{10.1086/423992}

\bibitem[{{Klypin} {et~al.}(2016){Klypin}, {Yepes}, {Gottl{\"o}ber}, {Prada},
  \& {He{\ss}}}]{klypin_yepes_16}
{Klypin}, A., {Yepes}, G., {Gottl{\"o}ber}, S., {Prada}, F., \& {He{\ss}}, S.
  2016, \mnras, 457, 4340, \dodoi{10.1093/mnras/stw248}

\bibitem[{{Koposov} {et~al.}(2012){Koposov}, {Belokurov}, {Evans}, {Gilmore},
  {Gieles}, {Irwin}, {Lewis}, {Niederste-Ostholt}, {Pe{\~n}arrubia}, {Smith},
  {Bizyaev}, {Malanushenko}, {Malanushenko}, {Schneider}, \&
  {Wyse}}]{koposov_etal_12}
{Koposov}, S.~E., {Belokurov}, V., {Evans}, N.~W., {et~al.} 2012, \apj, 750,
  80, \dodoi{10.1088/0004-637X/750/1/80}

\bibitem[{{K{\"u}pper} {et~al.}(2012){K{\"u}pper}, {Lane}, \&
  {Heggie}}]{Kupper_etal_12}
{K{\"u}pper}, A. H.~W., {Lane}, R.~R., \& {Heggie}, D.~C. 2012, \mnras, 420,
  2700, \dodoi{10.1111/j.1365-2966.2011.20242.x}

\bibitem[{{Laporte} {et~al.}(2018){Laporte}, {Johnston}, {G{\'o}mez},
  {Garavito-Camargo}, \& {Besla}}]{laporte_etal_18}
{Laporte}, C. F.~P., {Johnston}, K.~V., {G{\'o}mez}, F.~A., {Garavito-Camargo},
  N., \& {Besla}, G. 2018, \mnras, 481, 286, \dodoi{10.1093/mnras/sty1574}

\bibitem[{{Law} \& {Majewski}(2010)}]{law_majewski_10}
{Law}, D.~R., \& {Majewski}, S.~R. 2010, \apj, 714, 229,
  \dodoi{10.1088/0004-637X/714/1/229}

\bibitem[{{Law} \& {Majewski}(2016)}]{law_majewski_16}
{Law}, D.~R., \& {Majewski}, S.~R. 2016, in Astrophysics and Space Science
  Library, Vol. 420, Tidal Streams in the Local Group and Beyond, ed. H.~J.
  {Newberg} \& J.~L. {Carlin}, 31, \dodoi{10.1007/978-3-319-19336-6_2}

\bibitem[{{Lee} \& {Suto}(2003)}]{lee_suto_03}
{Lee}, J., \& {Suto}, Y. 2003, \apj, 585, 151, \dodoi{10.1086/345931}

\bibitem[{{Majewski} {et~al.}(2003){Majewski}, {Skrutskie}, {Weinberg}, \&
  {Ostheimer}}]{majewski_etal_03}
{Majewski}, S.~R., {Skrutskie}, M.~F., {Weinberg}, M.~D., \& {Ostheimer}, J.~C.
  2003, \apj, 599, 1082, \dodoi{10.1086/379504}

\bibitem[{{Majewski} {et~al.}(2004){Majewski}, {Kunkel}, {Law}, {Patterson},
  {Polak}, {Rocha-Pinto}, {Crane}, {Frinchaboy}, {Hummels}, {Johnston}, {Rhee},
  {Skrutskie}, \& {Weinberg}}]{majewski_etal_04}
{Majewski}, S.~R., {Kunkel}, W.~E., {Law}, D.~R., {et~al.} 2004, \aj, 128, 245,
  \dodoi{10.1086/421372}

\bibitem[{{Masset} \& {Bureau}(2003)}]{masset_bureau_03}
{Masset}, F.~S., \& {Bureau}, M. 2003, \apj, 586, 152, \dodoi{10.1086/367550}

\bibitem[{{Mateo} {et~al.}(1996){Mateo}, {Mirabal}, {Udalski}, {Szymanski},
  {Kaluzny}, {Kubiak}, {Krzeminski}, \& {Stanek}}]{mateo_etal_96}
{Mateo}, M., {Mirabal}, N., {Udalski}, A., {et~al.} 1996, \apjl, 458, L13,
  \dodoi{10.1086/309919}

\bibitem[{{Mateo} {et~al.}(1998){Mateo}, {Olszewski}, \&
  {Morrison}}]{mateo_etal_98_sgrstream}
{Mateo}, M., {Olszewski}, E.~W., \& {Morrison}, H.~L. 1998, \apjl, 508, L55,
  \dodoi{10.1086/311720}

\bibitem[{{Milgrom}(1983)}]{MOND}
{Milgrom}, M. 1983, \apj, 270, 365, \dodoi{10.1086/161130}

\bibitem[{{Milgrom}(2019)}]{milgrom_2019}
---. 2019, arXiv e-prints, arXiv:1910.04368.
\newblock \doarXiv{1910.04368}

\bibitem[{{Miyamoto} \& {Nagai}(1975)}]{miyamoto_nagai_75}
{Miyamoto}, M., \& {Nagai}, R. 1975, PASJ, 27, 533

\bibitem[{{Navarro} {et~al.}(1997){Navarro}, {Frenk}, \& {White}}]{NFW}
{Navarro}, J.~F., {Frenk}, C.~S., \& {White}, S.~D.~M. 1997, \apj, 490, 493

\bibitem[{{Pe{\~n}arrubia} {et~al.}(2010){Pe{\~n}arrubia}, {Belokurov},
  {Evans}, {Mart{\'\i}nez-Delgado}, {Gilmore}, {Irwin}, {Niederste-Ostholt}, \&
  {Zucker}}]{penarrubia_etal_10}
{Pe{\~n}arrubia}, J., {Belokurov}, V., {Evans}, N.~W., {et~al.} 2010, \mnras,
  408, L26, \dodoi{10.1111/j.1745-3933.2010.00921.x}

\bibitem[{{Peebles}(1969)}]{peebles_69}
{Peebles}, P.~J.~E. 1969, \apj, 155, 393, \dodoi{10.1086/149876}

\bibitem[{{Perez} \& {Granger}(2007)}]{IPython}
{Perez}, F., \& {Granger}, B.~E. 2007, Computing in Science and Engineering, 9,
  21, \dodoi{10.1109/MCSE.2007.53}

\bibitem[{{Perryman} {et~al.}(2014){Perryman}, {Spergel}, \&
  {Lindegren}}]{perryman_etal_14}
{Perryman}, M., {Spergel}, D.~N., \& {Lindegren}, L. 2014, \apj, 789, 166,
  \dodoi{10.1088/0004-637X/789/2/166}

\bibitem[{{Peter} {et~al.}(2013){Peter}, {Rocha}, {Bullock}, \&
  {Kaplinghat}}]{peter_13_SIDM}
{Peter}, A. H.~G., {Rocha}, M., {Bullock}, J.~S., \& {Kaplinghat}, M. 2013,
  \mnras, 430, 105, \dodoi{10.1093/mnras/sts535}

\bibitem[{{Price-Whelan} {et~al.}(2019){Price-Whelan}, {Sipocz}, {Lenz},
  {Greco}, {Major}, {Oh}, \& {Lim}}]{gala}
{Price-Whelan}, A., {Sipocz}, B., {Lenz}, D., {et~al.} 2019, {adrn/gala: v1.0},
  v1.0,  Zenodo, \dodoi{10.5281/zenodo.2638307}

\bibitem[{{Price-Whelan}(2017)}]{gala:JOSS}
{Price-Whelan}, A.~M. 2017, The Journal of Open Source Software, 2, 388,
  \dodoi{10.21105/joss.00388}

\bibitem[{{Price-Whelan} {et~al.}(2016){Price-Whelan}, {Johnston}, {Valluri},
  {Pearson}, {K{\"u}pper}, \& {Hogg}}]{price_whelan_etal_16}
{Price-Whelan}, A.~M., {Johnston}, K.~V., {Valluri}, M., {et~al.} 2016, \mnras,
  455, 1079, \dodoi{10.1093/mnras/stv2383}

\bibitem[{{Schwarzschild}(1982)}]{schwarzschild_82}
{Schwarzschild}, M. 1982, \apj, 263, 599, \dodoi{10.1086/160531}

\bibitem[{{Shipp} {et~al.}(2018){Shipp}, {Drlica-Wagner}, {Balbinot},
  {Ferguson}, {Erkal}, {Li}, {Bechtol}, {Belokurov}, {Buncher}, {Carollo},
  {Carrasco Kind}, {Kuehn}, {Marshall}, {Pace}, {Rykoff}, {Sevilla-Noarbe},
  {Sheldon}, {Strigari}, {Vivas}, {Yanny}, {Zenteno}, {Abbott}, {Abdalla},
  {Allam}, {Avila}, {Bertin}, {Brooks}, {Burke}, {Carretero}, {Castander},
  {Cawthon}, {Crocce}, {Cunha}, {D'Andrea}, {da Costa}, {Davis}, {De Vicente},
  {Desai}, {Diehl}, {Doel}, {Evrard}, {Flaugher}, {Fosalba}, {Frieman},
  {Garc{\'\i}a-Bellido}, {Gaztanaga}, {Gerdes}, {Gruen}, {Gruendl}, {Gschwend},
  {Gutierrez}, {Hartley}, {Honscheid}, {Hoyle}, {James}, {Johnson}, {Krause},
  {Kuropatkin}, {Lahav}, {Lin}, {Maia}, {March}, {Martini}, {Menanteau},
  {Miller}, {Miquel}, {Nichol}, {Plazas}, {Romer}, {Sako}, {Sanchez},
  {Santiago}, {Scarpine}, {Schindler}, {Schubnell}, {Smith}, {Smith},
  {Sobreira}, {Suchyta}, {Swanson}, {Tarle}, {Thomas}, {Tucker}, {Walker},
  {Wechsler}, \& {DES Collaboration}}]{Shipp_2018}
{Shipp}, N., {Drlica-Wagner}, A., {Balbinot}, E., {et~al.} 2018, \apj, 862,
  114, \dodoi{10.3847/1538-4357/aacdab}

\bibitem[{{Slater} {et~al.}(2013){Slater}, {Bell}, {Schlafly}, {Juri{\'c}},
  {Martin}, {Rix}, {Bernard}, {Burgett}, {Chambers}, {Finkbeiner}, {Goldman},
  {Kaiser}, {Magnier}, {Morganson}, {Price}, \& {Tonry}}]{slater_etal_13}
{Slater}, C.~T., {Bell}, E.~F., {Schlafly}, E.~F., {et~al.} 2013, \apj, 762, 6,
  \dodoi{10.1088/0004-637X/762/1/6}

\bibitem[{{Sohn} {et~al.}(2015){Sohn}, {van der Marel}, {Carlin}, {Majewski},
  {Kallivayalil}, {Law}, {Anderson}, \& {Siegel}}]{sohn_etal_15}
{Sohn}, S.~T., {van der Marel}, R.~P., {Carlin}, J.~L., {et~al.} 2015, \apj,
  803, 56, \dodoi{10.1088/0004-637X/803/2/56}

\bibitem[{{Sohn} {et~al.}(2016){Sohn}, {van der Marel}, {Kallivayalil},
  {Majewski}, {Besla}, {Carlin}, {Law}, {Siegel}, \& {Anderson}}]{sohn_etal_16}
{Sohn}, S.~T., {van der Marel}, R.~P., {Kallivayalil}, N., {et~al.} 2016, \apj,
  833, 235, \dodoi{10.3847/1538-4357/833/2/235}

\bibitem[{{Udry}(1991)}]{udry_91}
{Udry}, S. 1991, \aap, 245, 99

\bibitem[{{Udry} \& {Pfenniger}(1988)}]{udry_pfenniger_88}
{Udry}, S., \& {Pfenniger}, D. 1988, \aap, 198, 135

\bibitem[{{Valluri} {et~al.}(2016){Valluri}, {Shen}, {Abbott}, \&
  {Debattista}}]{valluri_etal_16}
{Valluri}, M., {Shen}, J., {Abbott}, C., \& {Debattista}, V.~P. 2016, \apj,
  818, 141, \dodoi{10.3847/0004-637X/818/2/141}

\bibitem[{{van Albada} {et~al.}(1982){van Albada}, {Kotanyi}, \&
  {Schwarzschild}}]{vanalbada_etal_82}
{van Albada}, T.~S., {Kotanyi}, C.~G., \& {Schwarzschild}, M. 1982, \mnras,
  198, 303

\bibitem[{{van der Walt} {et~al.}(2011){van der Walt}, {Colbert}, \&
  {Varoquaux}}]{vanderwalt11}
{van der Walt}, S., {Colbert}, S.~C., \& {Varoquaux}, G. 2011, Computing in
  Science Engineering, 13, 22, \dodoi{10.1109/MCSE.2011.37}

\bibitem[{{Vasiliev} \& {Belokurov}(2020)}]{vasiliev_belokurov_20}
{Vasiliev}, E., \& {Belokurov}, V. 2020, arXiv e-prints, arXiv:2006.02929.
\newblock \doarXiv{2006.02929}

\bibitem[{{Vasiliev} {et~al.}(2020){Vasiliev}, {Belokurov}, \&
  {Erkal}}]{Vasiliev_tango}
{Vasiliev}, E., {Belokurov}, V., \& {Erkal}, D. 2020, \mnras,
  \dodoi{10.1093/mnras/staa3673}

\bibitem[{{Vera-Ciro} \& {Helmi}(2013)}]{vera-ciro_helmi_13}
{Vera-Ciro}, C., \& {Helmi}, A. 2013, \apjl, 773, L4,
  \dodoi{10.1088/2041-8205/773/1/L4}

\bibitem[{{Vogelsberger} {et~al.}(2014){Vogelsberger}, {Genel}, {Springel},
  {Torrey}, {Sijacki}, {Xu}, {Snyder}, {Nelson}, \&
  {Hernquist}}]{vogelsberger_etal_14}
{Vogelsberger}, M., {Genel}, S., {Springel}, V., {et~al.} 2014, \mnras, 444,
  1518, \dodoi{10.1093/mnras/stu1536}

\bibitem[{{Zemp} {et~al.}(2012){Zemp}, {Gnedin}, {Gnedin}, \&
  {Kravtsov}}]{zemp_etal_12}
{Zemp}, M., {Gnedin}, O.~Y., {Gnedin}, N.~Y., \& {Kravtsov}, A.~V. 2012, \apj,
  748, 54, \dodoi{10.1088/0004-637X/748/1/54}

\end{thebibliography}


\end{document}